\documentclass[12pt,preprint]{aastex}

\usepackage{graphicx}
\usepackage{amsmath}

\shorttitle{Light Curves from GRMHD Simulations}
\shortauthors{Schnittman, Krolik, \& Hawley}

\begin{document}

\title{Light Curves from an MHD Simulation of a Black Hole Accretion
  Disk}

\author{Jeremy D.\ Schnittman}
\affil{Department of Physics,
University of Maryland\\
82 Regents Drive, College Park, MD 20742}
\email{schnittm@umd.edu}

\author{Julian H.\ Krolik}
\affil{Department of Physics and Astronomy,
Johns Hopkins University\\
Baltimore, MD 21218}
\email{jhk@pha.jhu.edu}

\and
\author{John F.\ Hawley}
\affil{Astronomy Department,
University of Virginia\\
P.O. Box 400325, Charlottesville, VA 22904-4325}
\email{jh8h@virginia.edu}

\begin{abstract}
We use a relativistic ray-tracing code to calculate the light
curves observed from a global general relativistic magneto-hydrodynamic
simulation of an accretion flow onto a Schwarzschild black hole. We apply
three basic emission models to sample different properties of the
time-dependent accretion disk. With one of these models, which assumes
thermal blackbody emission and free-free absorption, we can predict
qualitative features of 
the high-frequency power spectrum from stellar-mass black holes in the
``Thermal Dominant'' state.  
The simulated power spectrum is characterized by a power law of
index $\Gamma \approx 3$ and total rms fractional variance of
$\lesssim 2\%$ above 10 Hz.  
For each emission model, we find that the variability 
amplitude should increase with increasing inclination angle.
On the basis of a newly-developed formalism
for quantifying the significance of quasi-periodic oscillations (QPOs)
in simulation data, we find that these simulations are able to
identify any such features with (rms/mean) amplitudes
$\gtrsim 1 \%$ near the orbital frequency at the inner-most stable
orbit. Initial results indicate the existence of transient QPO peaks
with frequency
ratios of nearly 2:3 at a $99.9\%$ confidence limit, but they are
not generic features because at any given time they are seen only from
certain observer
directions.  Additionally, we present detailed analysis of
the azimuthal structure of the accretion disk and the evolution of
density perturbations in the inner disk. These ``hot spot'' structures appear
to be roughly self-similar over a range of disk radii, with a single
characteristic size $\delta\phi=25^\circ$ and $\delta r/r=0.3$, and
typical lifetimes $T_l\approx 0.3T_{\rm orb}$. 
\end{abstract}


\section{INTRODUCTION}\label{intro}
Our understanding of accretion onto black holes has grown
substantially in recent years. Correlated magneto-hydrodynamic
(MHD) turbulence,
stirred by an underlying magneto-rotational instability, has now been
well-established as the fundamental mechanism of angular momentum
transfer in accretion disks \citep{balbu98}. With that
achievement in hand, detailed studies of global accretion disk
dynamics have been undertaken via large-scale numerical simulations in
both the pseudo-Newtonian \citep{hawle01,hawle02,machi03,armit03}
and general relativistic \citep{devil03a,devil03b,gammi03} frameworks.

At the same time, we have also come into possession of a wealth of
observational data.  High signal-to-noise spectra of active galactic
nuclei (AGN; see e.g., the SDSS
composite: \citet{vande01}) as well as of black hole binaries
in a variety of spectral states are now available \citep{mccli05}; so, too,
are detailed light curves and power-density spectra, particularly of black
hole binaries \citep{mccli05,vande05}, but also for Seyfert 1 galaxies
and other AGN
\citep{marko04}. Fluorescent Fe K$\alpha$ profiles have been used to
infer detailed diagnostics of the disk surface in its innermost
portions \citep{fabia95,reyno97,done00,mille02}. 

One particularly exciting area of research has been the discovery of
high-frequency quasi-periodic oscillations (QPOs) in accreting black
hole binaries \citep{stroh01a,stroh01b}. In a growing number of
sources, these QPOs appear in
pairs with integer frequency ratios of 2:3
\citep{mille01,remil02,homan05}. The high frequencies of these QPOs
(near the frequency of the inner-most stable circular orbit; ISCO)
suggests a strongly relativistic origin, but there exists a very
broad range of theoretical models, none of which at this point appear
overwhelmingly convincing
\citep{abram01,rezzo03,rebus04,schni05,petri05}. At the same time,
there have not yet been any clear identifications of QPOs in the
MHD simulations, much less pairs of QPOs with integer frequency ratios, making
it difficult to choose one imperfect theoretical model over another.

%Unfortunately, we have as yet no direct way of associating light
%output with numerical simulations of accretion disk dynamics.  For
%that, it will be necessary to develop computer codes that faithfully
%track the heating of gas in the accretion flow, its photon production,
%and the radiation transfer of those photons as they escape the disk.
%Consequently, we cannot yet directly compare the predictions of our best
%theoretical disk models with any spectral or timing observations.

Without a physically consistent way of associating light output with
numerical simulations of accretion dynamics, we cannot directly compare
the predictions of our best theoretical disk models with any spectral
or timing observations. Therefore
it is the goal of this paper to move toward the objective of linking
dynamical simulations to observational diagnostics.   We will do so by
applying several phenomenological models to detailed simulation
data in order to predict the light that would be produced.  From these
models, we generate images and light curves for accretion disks
as they might be seen by distant observers.  Statistical analysis of
this derived data will lead us to several interesting generalizations
about the nature of variability in the light output of accreting black
holes.  Although the particular models we employ here for the
emissivity and opacity of disks are not entirely realistic, the
formalism we create can be readily applied to more realistic models in
the future. 

\section{PRESENT WORK}\label{present}

To be specific, in this paper we will combine the results from general
relativistic (GR) MHD simulations similar to those of
\citet{devil03b} with the ray-tracing and radiative transfer
calculations described in \citet{schni04} and \citet{schni06} to produce
light curves
and power spectra of accreting black holes. This coupling of
ray-tracing with the output data from an independent simulation is
often referred to as a ``post-processor,'' a technique commonly used
in other branches of physics, particularly when attempting to compare
a simulation directly with experiment. The ray-tracing post-processor
described in this paper is closely related to a number of
radiative transfer codes used in modeling laboratory plasma physics
experiments (see, e.g.\ \citet{polla94,schni00,king01}). While
astrophysicists in general do not have the luxury of being able to
control (or in many cases even determine) the initial conditions of
their experiments, simulations and observational data are achieving
levels of quality that will soon prove such post-processors
invaluable. One successful post-processor application in astrophysics
(for MHD simulations no less!)
has been the production of synthetic observations of jets from radio
galaxies, which are large enough to have detailed features resolved
spatially \citep{smith85,matth90,tregi01}.

Even without imaging observations of accretion disks, the results of
the post-processor
analysis are very useful in understanding the structure and
evolution of the MHD disks, as well as giving valuable insights into
X-ray timing measurements of accreting black holes, particularly {\it
  RXTE} observations of 
binary black holes. In this paper we
focus primarily on emission models that may track the disk's thermal
emission. In future work, we plan to develop other
models that will perform the analogous job for coronal emission
processes. 

The outline and major results of this paper are as follows: we begin
in Section \ref{model} with a detailed description of the MHD
simulations and the relativistic radiative transfer equation. We
describe three different emission models (optically thin line,
optically thick line, and thermal blackbody), each useful as a different
diagnostic of the accretion disk structure. For the thermal emission model, we
describe a method to convert from the dimensionless ``code
units'' of the simulation to more useful cgs units that appear in the
absorption and emission coefficients in the radiation transport. In
Section \ref{results} we apply these methods to a single MHD
simulation of an accretion flow around a Schwarzschild black hole and
present light curves and power spectra for the three different
emission models and a sample of observer inclination angles. We
estimate our sensitivity to detecting high-frequency QPOs in the
simulated light curves and discuss the significance and
robustness of a transient pair of QPOs with a 2:3 frequency ratio.

To gain a deeper understanding of the power spectra and physical
mechanism behind the QPOs, in Section
\ref{az_structure} we develop a number of new tools to analyze the
azimuthal structure of hot spot perturbations in the turbulent
magnetized disk. For a range of radii throughout the disk, we
find the characteristic hot spot shape to be self-similar with
constant extent in azimuth and radius. Furthermore, we find the hot
spot lifetimes to be distributed exponentially, much like the simple
model of \citet{schni05}, with a typical lifetime proportional to the
orbital period (i.e.\ the dynamical time) at that radius. Finally, in
Section \ref{discussion}, we discuss the implications of these results
and outline possibilities for more direct comparison with
observations. 

\section{DESCRIPTION OF THE MODEL}\label{model}

\subsection{Data from Global GRMHD Simulations}\label{grmhd_data}
Our basic data come from a single numerical simulation of accretion onto
a Schwarzschild black hole that uses the fully general relativistic
3-dimensional MHD code described in \citet{devil03a}. This code follows the
dynamical evolution of magnetized matter in the ideal MHD approximation
assuming that the fluid's internal energy evolves adiabatically except
for the creation of entropy in shocks. Magnetic field evolution
utilizes the constrained transport algorithm \citep{evans88}.

The calculation is done in Boyer-Lindquist coordinates, so the grid's
inner radial boundary is placed just outside the event horizon ($r =
2.1M$; throughout this paper, unless stated otherwise, we adopt units
in which $G=c=1$). The outer radial boundary is at $120M$. At both
radial boundaries, matter can leave the grid, but not enter it. To
avoid the (flat-space) coordinate singularity at the polar axis, there
are reflecting boundary conditions at polar angles $\theta=0.045\pi$
and $0.955\pi$. The full range of azimuthal angles is included in the
grid, with, of course, periodic boundary conditions linking $\phi = 0$
and $2\pi$. The $2\pi$ range in $\phi$ is divided into 256
equal-sized zones. However, the 192 zones in both $\theta$ and $r$ are
concentrated where we expect the sharpest gradients: near the
equatorial plane and at small radius (see \citet{devil03b} for
details).

In the initial state, all the matter is contained within a hydrostatic 
torus centered on the equatorial plane with pressure maximum at
$r=25M$ and inner edge at $r=15M$. It possesses a purely poloidal and
axisymmetric magnetic field whose fieldlines follow isodensity
contours; the plasma initially has a mean $\beta$ of 100. Outside the
torus, the density is fixed at $10^{-10}$ times the density at the
pressure maximum. As found in many simulations before (e.g.,
\citet{devil03b,mckin04}), within a few orbits
after the beginning of the simulation, the magneto-rotational
instability grows to nonlinear amplitude and magnetic stresses
redistribute angular momentum throughout the matter. The result is a
disk of moderate thickness ($h/r \sim 0.1$) and nearly-Keplerian
angular momentum distribution accreting onto the central black hole.

\begin{figure}[tp]
\caption{\label{plotone} Density-weighted averages of hydrodynamic
  properties of the 
  disk versus radius, for the MHD simulations ({\it solid}) and a
  steady-state Novikov-Thorne disk with a non-zero torque at the ISCO
  ({\it dashed}). The MHD disk behaves like an accretion disk only
  inside $r/M \approx 25$, where $\dot{M}(r)$ is roughly constant.}
\begin{center}
\scalebox{0.45}{\includegraphics{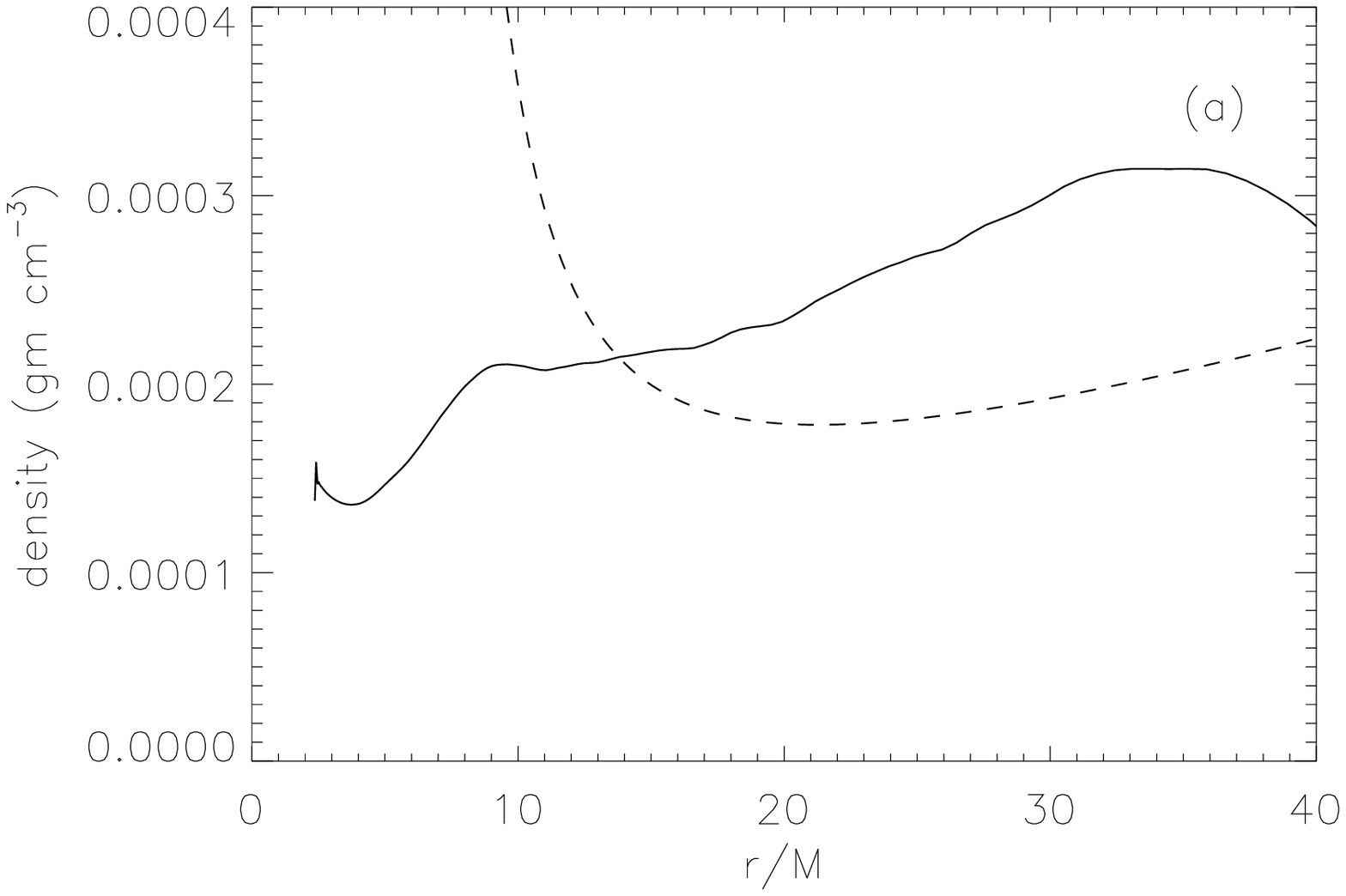}}
\scalebox{0.45}{\includegraphics{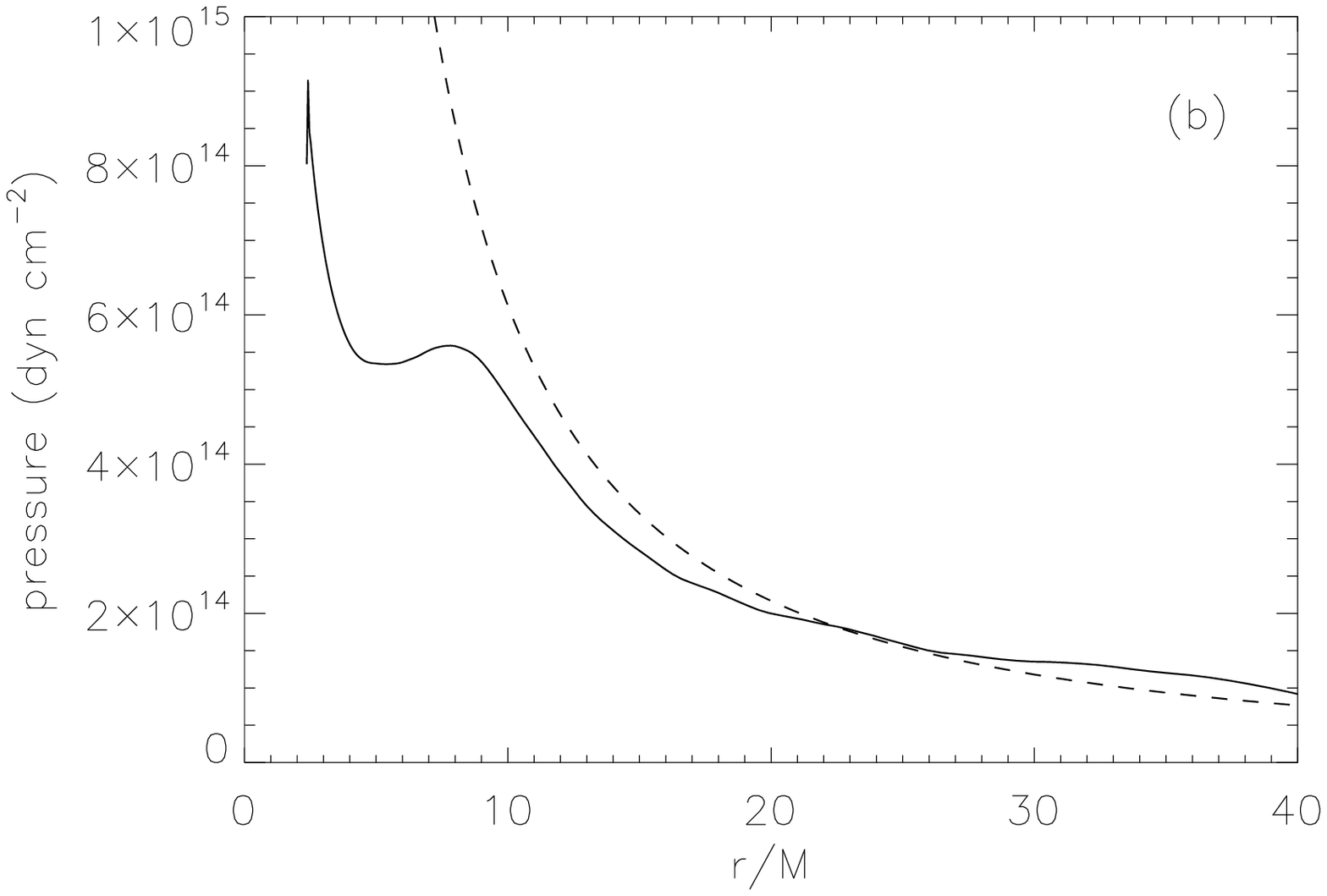}}\\
\scalebox{0.45}{\includegraphics{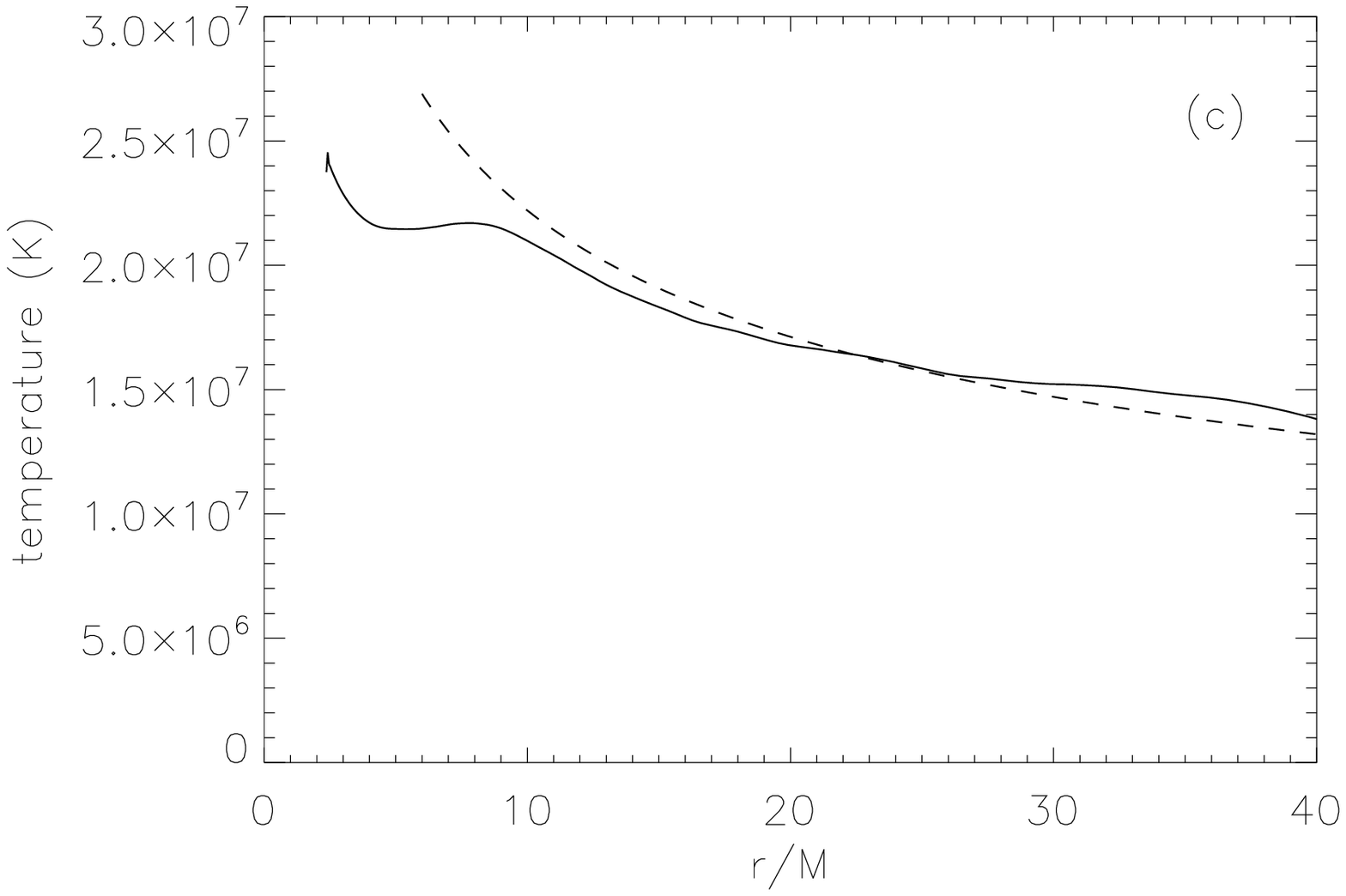}}
\scalebox{0.45}{\includegraphics{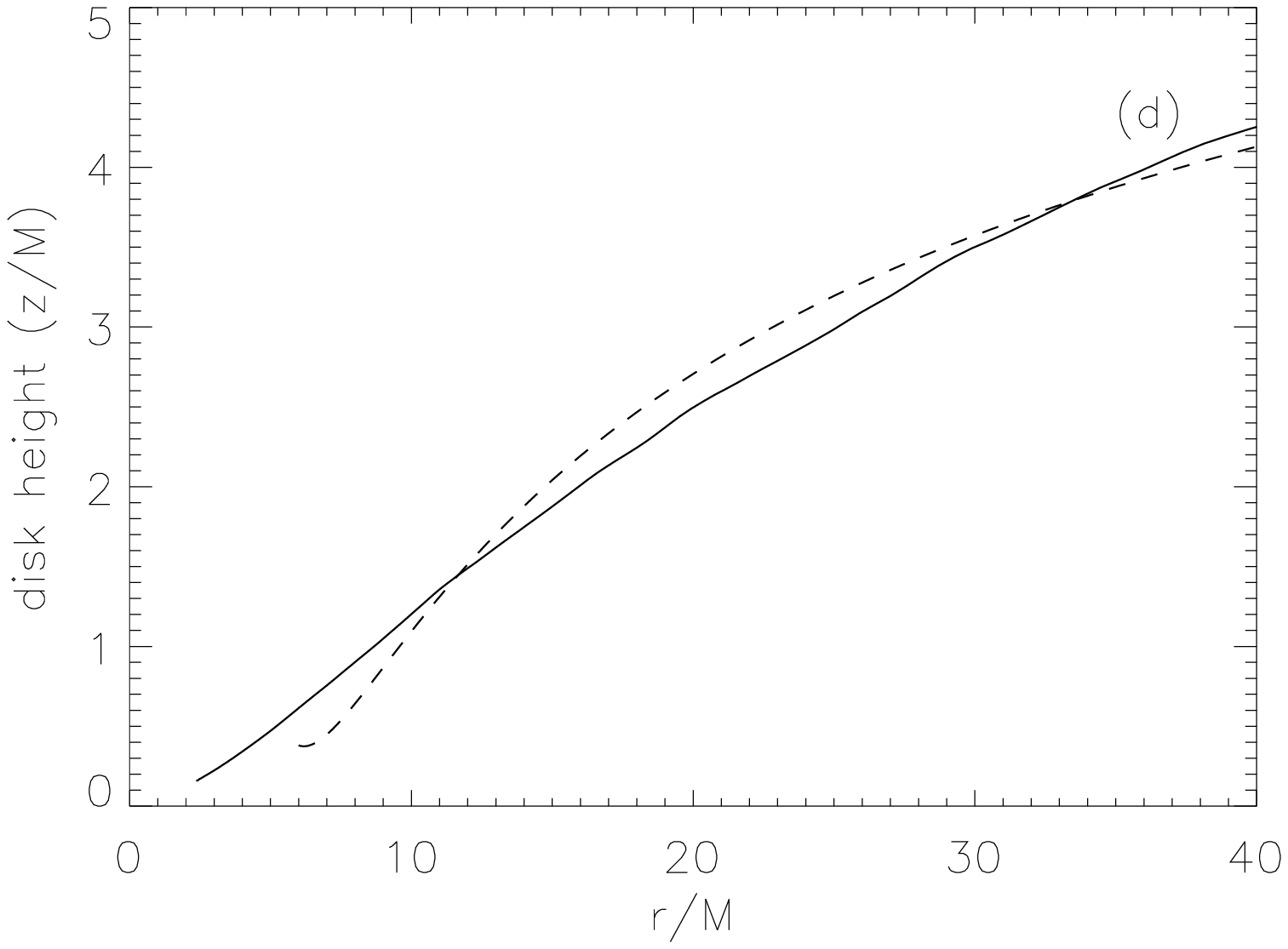}}\\
\scalebox{0.45}{\includegraphics{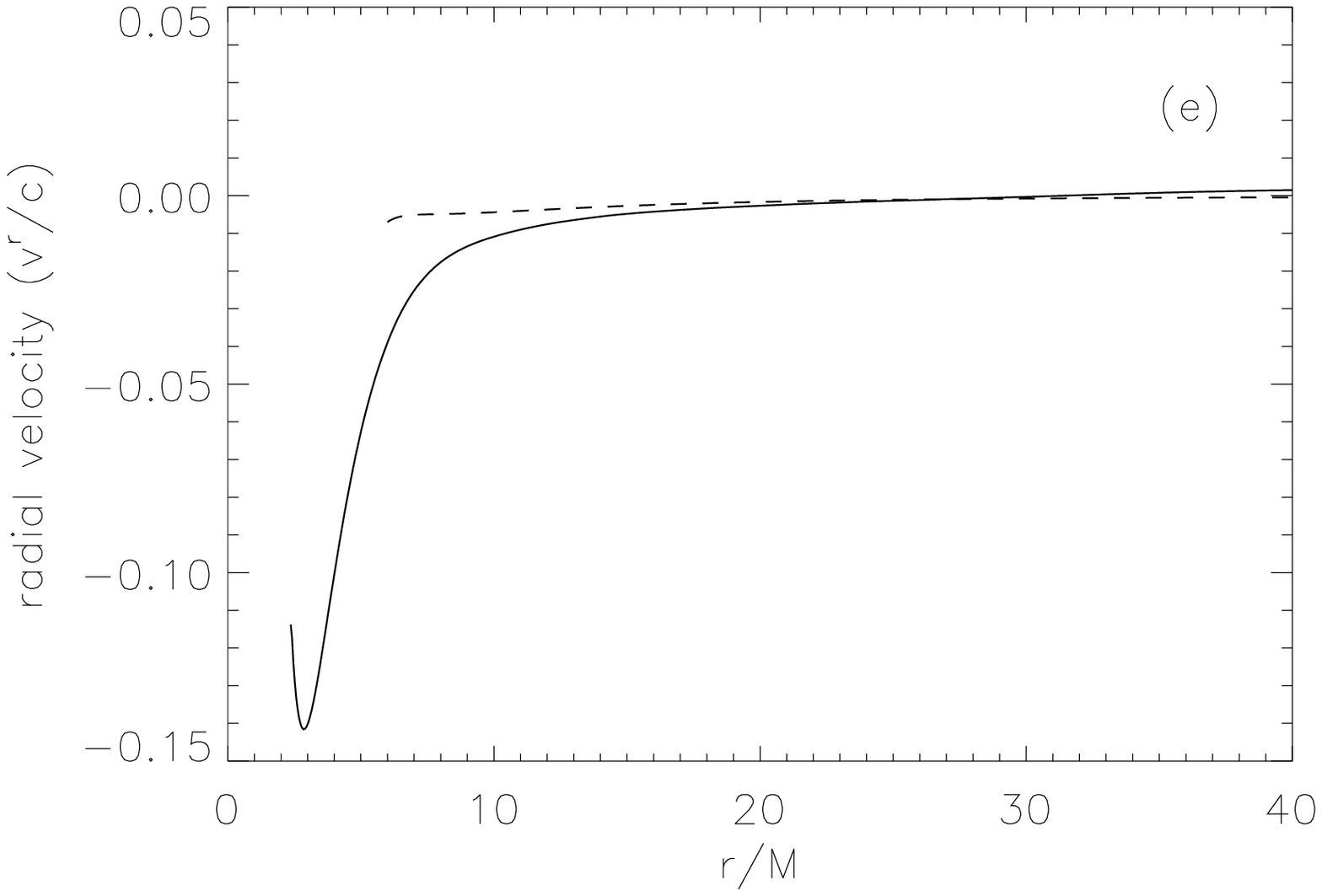}}
\scalebox{0.45}{\includegraphics{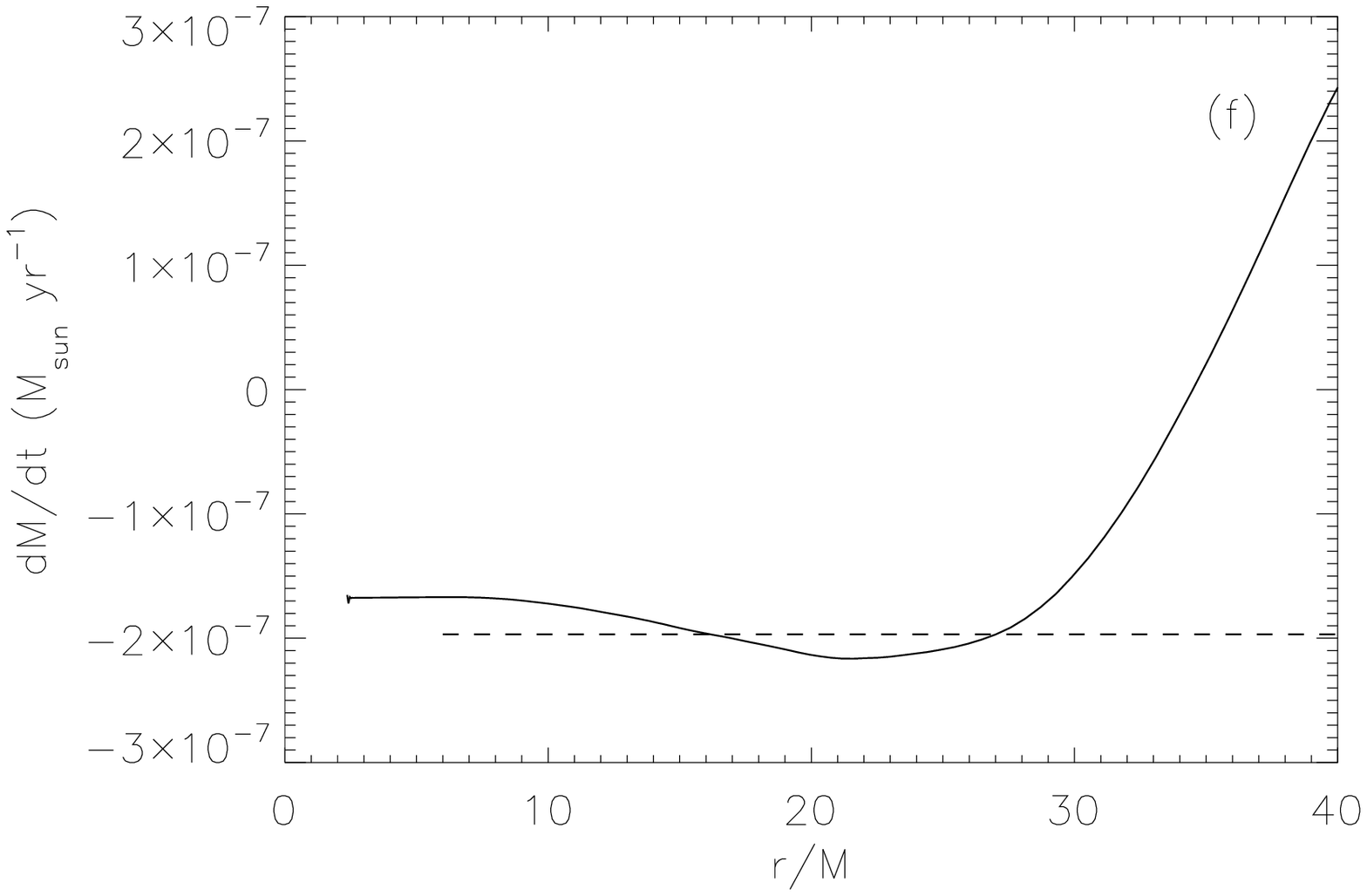}}
\end{center}
\end{figure}

Over the entire simulation run time $T_{\rm sim} = 6000 M$, at $5 M$
intervals in coordinate time, we write
out full 3-dimensional snapshots of the major hydrodynamic
variables: the density, pressure, and fluid 4-velocity (in practice we
do not actually save data from the first $\sim 1700 M$, the time when
the early transient features relax).
In Figure~\ref{plotone} we plot a selection of time- and
azimuth-averaged disk variables from the simulation data (solid
curves), and compare them with a steady-state Novikov-Thorne disk (dashed
curves). The conversion from code
units to cgs units follows the proceedure described in Section
\ref{emission_models}, assuming
an accretion rate of $0.5\dot{M}_{\rm Edd}$ onto a black hole of mass
$10M_\odot$. Additionally, for the steady-state disk, we have added a
small torque at the inner-most stable circular orbit
(ISCO). Following \citet{agol00}, we have parameterized this torque by
the extra efficiency it adds to the disk through increased dissipation
just outside the ISCO. Specifically, in Figure \ref{plotone} we use
$\delta\eta=0.007$ (chosen to best match the disk height from the
simulations; see Fig.\ \ref{plotone}d), while the no-torque efficiency
for a Schwarzschild black hole is $\eta=0.057$. 

The density, temperature, pressure, and radial velocity shown in
Figure~\ref{plotone} are mass-weighted averages, to provide a focus on
the most ``disk-like'' region of the MHD simulation near the
midplane. Outside $r/M \approx 25$, the simulation cannot really be
thought of as an accretion disk, since there is actually net mass flow
outwards in that region, the result of the
initial torus expanding as angular momentum is transferred from the
inner to the outer disk. In the inner-most regions ($r/M \lesssim
10$), the simulation diverges from the Novikov-Thorne disk as the gas
acquires a larger inward radial velocity. To maintain mass
conservation, this requires a 
significant decrease of the surface density compared to the
Novikov-Thorne disk, as is evident from Figures \ref{plotone}a and
\ref{plotone}e. 

We should note that it may be somewhat coincidental
that the agreement between the simulations and analytic model is so
good, as they use rather different equations of state. The simulations
include only gas pressure, and ignore any loss of internal energy to
radiation, while the Novikov-Thorne model is based on a
radiation-pressure dominated disk that balances energy generation by
means of viscous dissipation with energy loss by radiation flux from the
disk surface. In fact, using significantly different values of $(L/L_{\rm
 Edd})$ would change the scale height of the Novikov-Thorne disk,
while only modifying the cgs scale of the density and pressure of the
simulation data and leaving the disk thickness unchanged.

\subsection{Ray-tracing and Radiative Transport}\label{ray_tracing}
The ray-tracing method used in this analysis can be broken up into two
independent pieces: (i) the calculation of photon geodesic
trajectories through the black hole spacetime, and (ii) the
integration of the relativistic radiative transfer equation along
these geodesic paths. The first step, which is carried out using a
Hamiltonian formulation for the equations of motion, is described in
detail in \citet{schni04} for the general Kerr metric with
Boyer-Lindquist coordinates. The only major difference between our
current approach and the methods used in that paper is the manner in
which the photon positions
and momenta are tabulated. In \citet{schni04}, the photons passed
through a disk of finite thickness, crossing each surface of constant
coordinate $\theta$ exactly once (see Fig.\ 1 of that paper). While
computationally convenient, this approach limited the range of viewer
inclination angles (fully edge-on views were impossible as the
observer at infinity would be embedded inside the computational grid
and photon intersections with the $\theta$ surfaces would occur near
the observer and not the accretion disk).
Furthermore, since each photon path would terminate after
passing through all the $\theta$-surfaces once, we could not produce
multiple images via strong gravitational lensing. 

\begin{figure}[tp]
\caption{\label{plottwo} Schematic diagram of the ray-tracing
  procedure. Photon paths are integrated along geodesic trajectories
  from a distant observer to the region around the black hole. The
  positions and momenta are tabulated along the photon path and used
  to integrate the radiative transfer equation through the accretion
  disk ({\it shaded region}) towards the observer. Here the tabulation points
  are indicated schematically by diamonds; in practice much higher
  resolution is used. The photons either terminate at the black hole
  horizon ({\it thick circle}) or escape to infinity.
}
\begin{center}
\scalebox{0.8}{\includegraphics*[40,400][570,700]{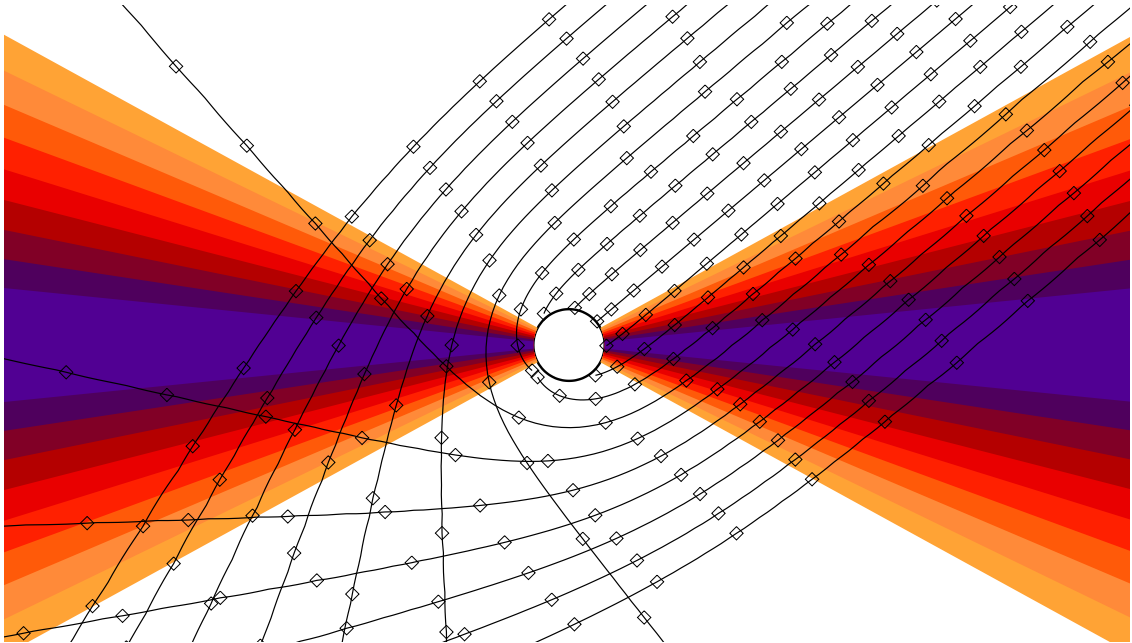}}
\end{center}
\end{figure}

We have now solved these problems by tabulating the photon paths
according to coordinate time $t$ (the independent variable in our
Hamiltonian equations of motion) instead of latitude $\theta$. For
each photon, we record the position and momentum at events equally
spaced in $t$ over a section of the trajectory centered on the point
of closest approach (minimum $r$) to the black hole (this value of $r$
will always
be unique for null geodesics). Figure \ref{plottwo} shows a
schematic of this approach for a Schwarzschild black hole, but the
discussion below can be applied equally well to the general
case of a spinning black hole. This method allows
for arbitrary viewing orientations and the modeling of multiply lensed
images. Instead of recording the coordinate momenta
$p_\mu(\mathbf{x})$, at each point along the path we convert to the
locally flat frame of a
zero-angular momentum observer (ZAMO), denoted by hatted indices:
$p_{\hat{\mu}}(\mathbf{x})$. Working in the ZAMO frame simplifies the
solution of the radiative transfer equation, described below.

From the tabulated set of photon positions and momenta we obtain a
series of null vectors $dx^{\hat{\mu}}_i$ parallel to
$p^{\hat{\mu}}$, defined in the ZAMO frames
at space-time coordinates $x^\mu_i$. The fluid properties at these
points are interpolated from the fixed computational grid that holds
the data output from the MHD simulation. Specifically, we record the
fluid 4-velocity $u^{\hat{\mu}}$, also defined in the ZAMO frame, and
the gas density and pressure as measured in the fluid frame. As
described below in Section \ref{emission_models}, the code units of
density and pressure are converted to cgs units and the temperature is
derived assuming a radiation pressure-dominated gas. 

With all these pieces in place, the problem is reduced to solving the
classical radiative transfer equation through a relativistic fluid
with (locally) uniform density, temperature, and velocity. We first
write the nonrelativistic radiative transfer equation as
\begin{equation}\label{rad_trans_eq}
\frac{dI_\nu}{ds} = j_\nu -\alpha_\nu I_\nu\ ,
\end{equation}
where $ds$ is the differential path length and $I_\nu$, $j_\nu$, and
$\alpha_\nu$ are respectively the spectral intensity, emissivity, and
absorption coefficient of the fluid medium at a frequency $\nu$.
The absorption coefficient is related to the opacity
$\kappa_\nu$ through the density $\rho$: $\alpha_\nu =
\rho\kappa_\nu$. In this form, true emission and absorption are
included, but not electron
scattering. The inclusion of scattering adds an integral over
$I_\nu(\cos\theta)$ to the rhs of equation (\ref{rad_trans_eq}),
greatly complicating its solution. 

Defining the optical depth $\tau_\nu$ through
\begin{equation}\label{tau_nu}
d\tau_\nu \equiv \alpha_\nu ds,
\end{equation}
the transfer equation can be written as
\begin{equation}\label{rad_trans_eq2}
\frac{dI_\nu}{d\tau_\nu} = S_\nu - I_\nu,
\end{equation}
where the source function is defined as $S_\nu \equiv
 j_\nu/\alpha_\nu$.
Over regions of constant source function $S_\nu$,
equation (\ref{rad_trans_eq2}) has the simple solution
\begin{equation}\label{rad_solution}
I_\nu(d\tau_\nu) = S_\nu + e^{-d\tau_\nu}[I_\nu(0)-S_\nu].
\end{equation}

Both $I_\nu$ and $S_\nu$ have the same properties under Lorentz
transformations, namely $I_\nu/\nu^3$ and $S_\nu/\nu^3$ are both
invariant. Another Lorentz invariant is the optical depth, since the
fraction of
photons passing through a finite medium is given by $e^{-\tau}$, which
is just a number, and thus the same in any reference frame. Following
\citet{rybic79}, we can calculate the absorption coefficient in a relativistic
medium by considering a small volume of matter flowing in the
$\mathbf{e}_{\hat{x}}$ direction with respect to the lab frame.
Since the motion is in the $x$
direction, the fluid volume thickness $l$ in the $y$-dimension is the
same in the lab (unprimed) and fluid (primed) frames. For a photon
propagating at angle
$\theta$ with respect to the fluid velocity in the lab frame, the
total optical depth in the $y$-dimension can be written
\begin{equation}\label{tau_nu1}
\tau_\nu = \frac{l\alpha_\nu}{\sin\theta} =
\frac{l}{\nu\sin\theta}\nu\alpha_\nu = \mbox{Lorentz invariant}.
\end{equation}
Since $\nu\sin\theta$ is proportional to the $p_y$ component of the
photon 4-momentum, it must be the same in both frames because the
boost is in a perpendicular direction. Thus $\nu\sin\theta$ is another
Lorentz invariant, and so is $\nu\alpha_\nu$.
From the definition of the source function $j_\nu = \alpha_\nu S_\nu$,
we find that $j_\nu/\nu^2$ is also Lorentz invariant, or
\begin{equation}\label{j_nu2}
j_\nu = \left(\frac{\nu}{\nu'}\right)^2 j_\nu'.
\end{equation}

In the ZAMO frame (here it can be thought of as the lab frame), the
spatial path length of the photon trajectory is given by $ds^2 =
\eta_{\hat{j}\hat{k}}dx^{\hat{j}} dx^{\hat{k}}$. The photon's
4-momentum and the 4-velocity of the fluid in the ZAMO basis give the
angles $\theta$ and $\theta'$, and the absorption $\alpha_\nu'$ and
emissivity $j_\nu'$ are given in the fluid frame by the specific
emission model being used. Now we have enough
information to solve the radiative transfer equation in a relativistic
flow: 
\begin{equation}\label{rad_trans_eq3}
\frac{dI_\nu}{ds} = \left(\frac{\nu}{\nu'}\right)^2 j_\nu' -
\left(\frac{\nu'}{\nu}\right) \alpha_\nu' I_\nu.
\end{equation}
The special relativistic redshift between
the photons in the ZAMO frame and fluid frame is
\begin{equation}
\frac{\nu}{\nu'} = \gamma(1+\beta\cos\theta'),
\end{equation}
where $\gamma \equiv 1/\sqrt{1-\beta^2}$ as usual. 

The above analysis, while quite useful for special relativistic flows
in the locally flat ZAMO basis, ignores all general relativistic effects of curved
spacetime around the black hole. To include these effects, we need
only consider the invariant $I_\nu/\nu^3$ along the geodesic path of
the photons. This is particularly straightforward from a computational
point of view, where the spectrum is stored as a discrete array $I^j$,
evaluated at the frequencies $\nu^j$. These frequencies are redshifted
from one point on the trajectory to the next due solely to
gravitational effects:
Let $U^\mu_i$ be the 4-velocity of a ZAMO at position $x^\mu_i$. We
can define the inner product with the photon 4-momentum as 
\begin{equation}\label{alpha_pv}
\chi_i \equiv p_{\mu, i}U^\mu_i.
\end{equation}
Then the array of frequencies is redshifted along the photon path
according to 
\begin{equation}\label{nu_j_ip1}
\nu^j_{i+1} = \nu^j_i \left(\frac{\chi_{i+1}}{\chi_i}\right).
\end{equation}
Similarly, the spectral intensity defined at each frequency point
scales as
\begin{equation}\label{I_j_ip1}
I^j_{i+1} = I^j_i \left(\frac{\chi_{i+1}}{\chi_i}\right)^3.
\end{equation}
While these methods are well-suited for our implementation of the
ray-tracing code, it should be noted that purely covariant approaches
also exist for solving the radiative transfer equation in curved
spacetime \citep{fuers04}. 

\subsection{Emission Models}\label{emission_models}

As in \citet{schni06}, here we consider three different emission
models for the disk. The simplest model, optically thin line emission,
assumes that the gas is emitting monochromatic radiation
isotropically in the fluid rest frame. The emissivity is proportional
to rest mass density, and there is no absorption: $j_\nu \propto
\rho\delta(\nu-\nu_{\rm em})$ and $\alpha_\nu=0$. While not entirely
realistic (geometrically thin accretion disks are generally
optically thick), this model is extremely useful as a diagnostic
tool. The observed photons serve as tracers for the disk's surface
density, relativistic beaming, and the gravitational redshift of
the fluid near the black hole. 

The second model considered is one of optically thick line
emission and absorption, with both $\alpha_\nu$ and $j_\nu$
proportional to $\rho\delta(\nu-\nu_{\rm em})$. Just as the optically
thin model served as a diagnostic of surface density perturbations,
the optically thick model effectively maps out the Doppler and
gravitational red-shifts of
the disk's photosphere. In this regard, it can be thought of as a
model for the fluorescent iron line seen in many active galactic
nuclei and galactic black holes (modulo a specified emissivity profile in
$r$). However, unlike the many popular thin-disk models
such as \citet{laor91}, our optically thick model samples the upper
layers of a vertically extended disk and corona, somewhat complicating
the simple iron line profiles predicted by a flat disk.

The third emission model combines a Kramers opacity law with its
thermodynamic inverse emissivity:
\begin{equation}\label{bb_emission}
j_\nu \propto \rho^2 T^{-1/2}e^{-x}
\end{equation}
and
\begin{equation}\label{kramers}
\alpha_\nu \propto \rho^2 T^{-7/2} \left(\frac{1-e^{-x}}{x^3}\right),
\end{equation}
giving the classical blackbody distribution in an optically thick
region of constant temperature: 
\begin{equation}\label{bbI_nu}
I_\nu = j_\nu/\alpha_\nu \propto T^3 x^3 \frac{e^{-x}}{1-e^{-x}},
\end{equation}
where $x \equiv h\nu/kT$. We believe this final model gives the most
accurate description of the ``Thermal Dominant,'' or ``High-Soft''
State for X-ray binaries.

For the thermal emission model, we must first convert the hydrodynamic
variables as provided by the MHD simulations in code units to
more useful physical (cgs) units of density, temperature and
pressure (as described in \citet{devil03a}, the fluid variables used
in the code are the transport velocity $V^i$,
matter density $\rho$, pressure $P$, and Lorentz factor $W$). To do so
requires choosing a few specific model parameters. First, we
set the overall time and distance scales of the problem by picking a
mass for the central black hole. For the Boyer-Lindquist coordinates
used in the simulation, this sets
\begin{subequations}\label{cgs_units}
\begin{equation}
dt_{\rm cgs} = 4.9 \times 10^{-5} dt_{\rm code} M_{10} \mbox{ sec},
\end{equation}
\begin{equation}
dl_{\rm cgs} = 1.45 \times 10^6 dl_{\rm code} M_{10} \mbox{ cm},
\end{equation}
\end{subequations}
where $M_{10}$ is the black hole mass in units of $10 M_\odot$. 
To convert code rest-mass density to g/cm$^3$, we need to set an average mass
accretion rate $\dot{M}$ in cgs units:
\begin{equation}
\dot{M}_{\rm cgs}=\dot{M}_{\rm Edd}\left(\frac{L}{L_{\rm Edd}}\right)
= \frac{(L/L_{\rm Edd})}{\eta}\frac{L_{\rm Edd}(M)}{c^2}.
\end{equation}
We then use the code-calculated accretion rate $\dot{M}_{\rm code}$ to
get
\begin{equation}
\rho_{\rm cgs} = 2.3\times 10^{-5} \frac{(L/L_{\rm Edd})}{\eta}
\rho_{\rm code} \dot{M}^{-1}_{\rm code} M_{10}^{-2} \mbox{ g/cm}^3,
\end{equation}
where the Eddington luminosity is $L_{\rm Edd}(M) = 1.3\times 10^{39}
M_{10}$ erg s$^{-1}$, and the efficiency $\eta$ gives the bolometric
luminosity through
$L=\eta\dot{M}c^2$. The Novikov-Thorne model predicts $\eta=
0.057$ for the Schwarzschild black hole, and the non-zero torque at
the ISCO provides an additional efficiency factor of $\delta\eta=0.007$.

The cgs pressure and temperature of the gas are determined by assuming
a radiation-pressure dominated equation of state, which is quite
reasonable for the inner regions of a black hole
accretion disk with an appreciable fraction of the Eddington luminosity.
For such a gas, the isothermal sound speed is $c_s =
(p_r/\rho)^{1/2}$, giving 
\begin{equation}
p_{\rm cgs} = \rho_{\rm cgs}c_s^2 = p_{\rm
  code}\left(\frac{\rho_{\rm cgs}}{\rho_{\rm code}}\right) c^2,
\end{equation}
from which the temperature is easily derived:
\begin{equation}
T_{\rm cgs} = \left(\frac{3p_{\rm cgs}}{a}\right)^{1/4},
\end{equation}
where $a$ is simply the radiation density constant ($7.6\times
10^{-19}$ erg cm$^{-3}$ ${\rm K}^{-4}$).

As mentioned above in Section \ref{grmhd_data}, this method of
determining the gas
temperature is not entirely consistent with the assumptions made in
the MHD simulation. Namely, the code assumes a gas pressure-dominated
adiabatic equation of state, with no explicit radiation pressure and
the gas entropy increased only by shock heating. Yet perhaps this is a
case of the ``ends justifying the means:'' even though we assume a
radiation-dominated pressure to derive the temperature from a calculation that
actually ignores radiation pressure, the resulting values are actually
in close agreement with the analytic predictions of a standard
alpha-disk for the same black hole mass and accretion
rate. Alternatively, instead of using the mass accretion rate to determine
the code-to-cgs conversion factors, we could attempt to match any of
the other Novikov-Thorne disk variables, such as surface density, central
pressure, or temperature. These all give somewhat different scaling
factors for the density, but we find the behavior of the light
curves for the thermal emission model are ultimately very similar in all
cases. 

\begin{figure}[tp]
\caption{\label{plotthree} Snapshots of the inner regions of the MHD
  simulation, considering only emission inside of $r<25M$. The three
  different emission models used are optically thin line ($a$, $b$),
  optically thick line ($c$, $d$), and thermal ($e$, $f$). The panels
  $a$, $c$, and $e$ have viewer inclination 
  angle of $i=0^\circ$ and the panels $b$, $d$, and $e$ have
  $i=70^\circ$. For each panel, the color scheme represents a
  logarithmic scale, normalized to the peak intensity for that frame.}
\begin{center}
\scalebox{0.45}{\includegraphics*[50,370][415,730]{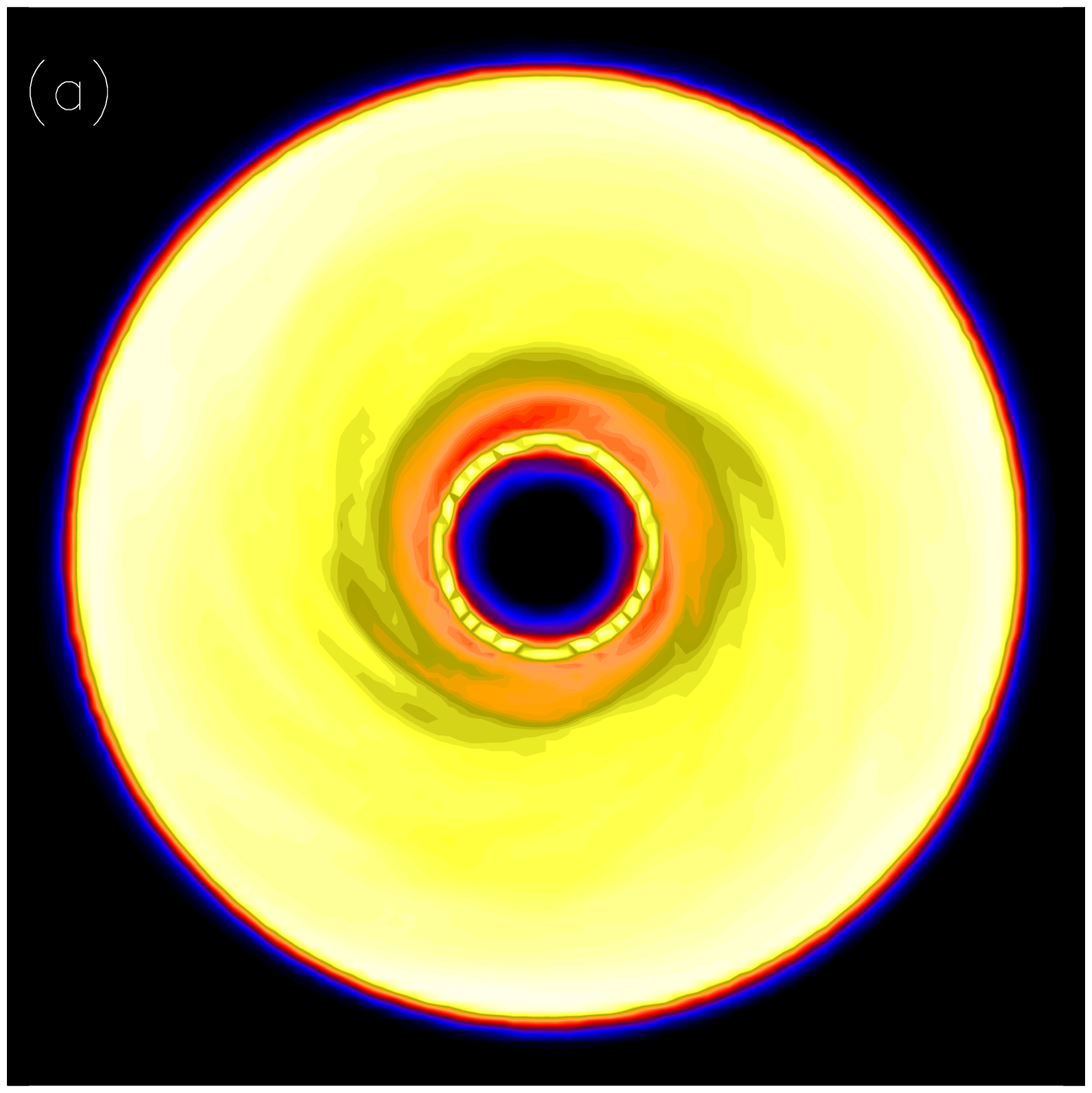}}
\scalebox{0.45}{\includegraphics*[50,370][415,730]{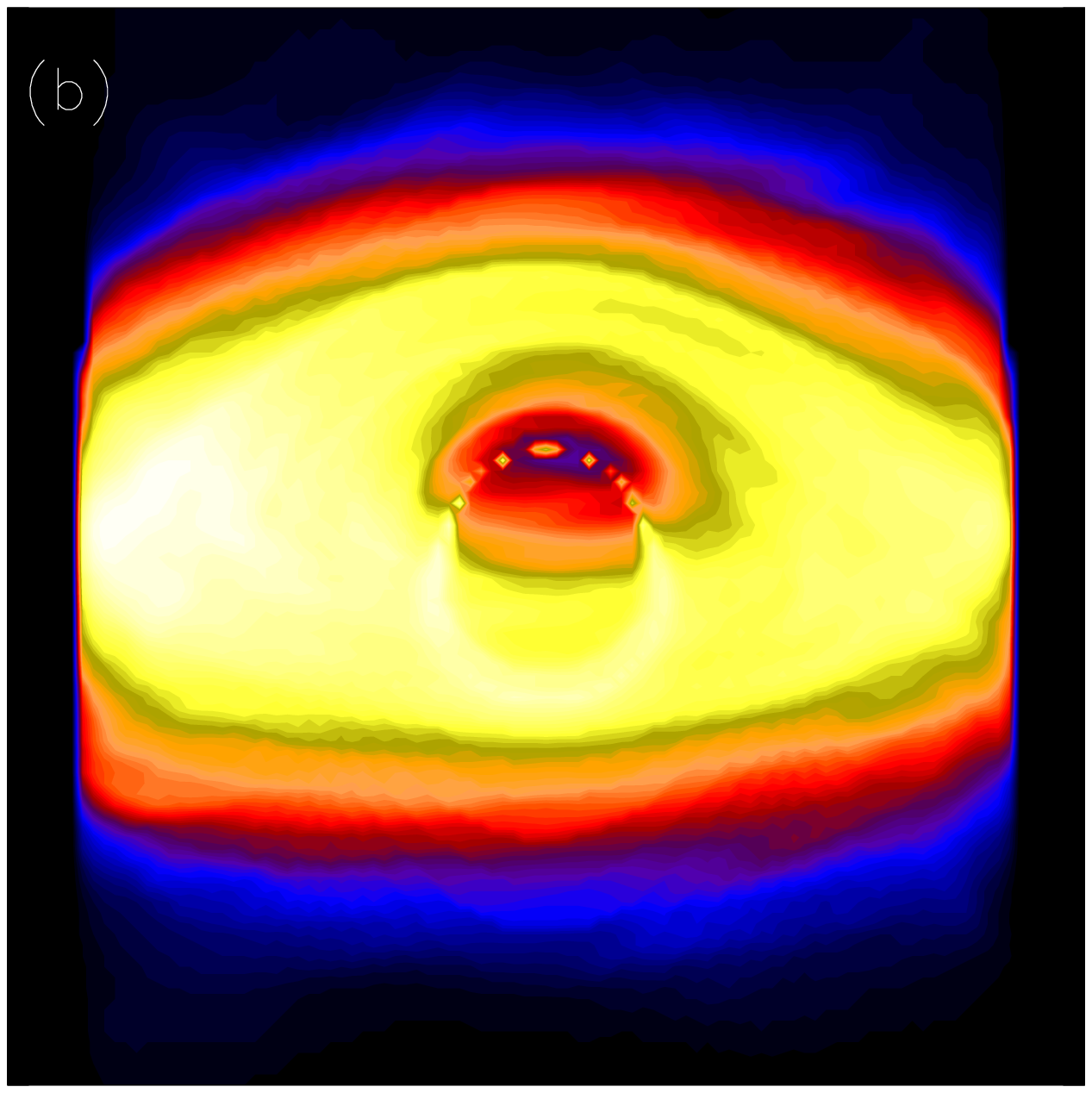}}\\
\vspace{0.1cm}
\scalebox{0.45}{\includegraphics*[50,370][415,730]{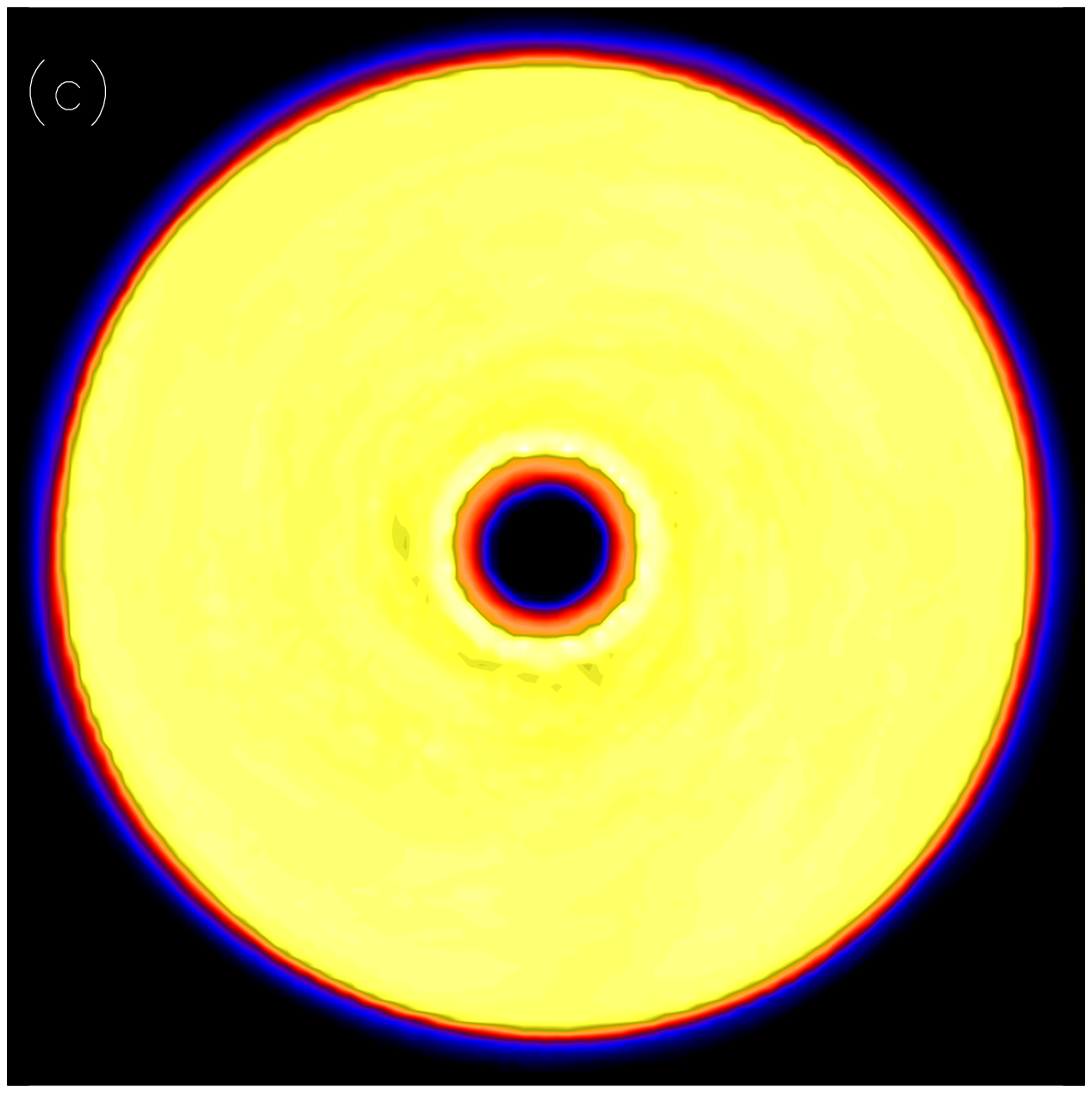}}
\scalebox{0.45}{\includegraphics*[50,370][415,730]{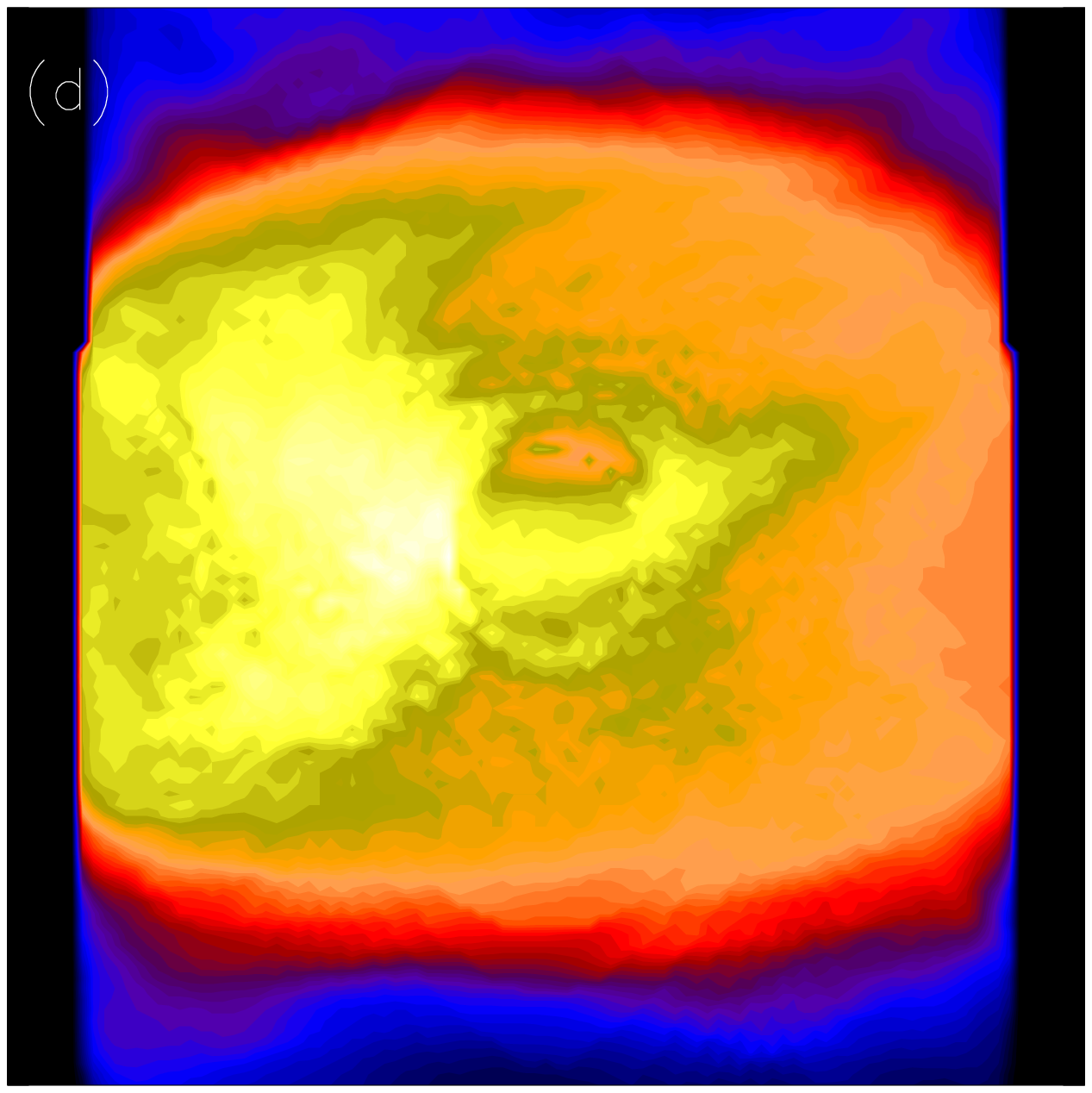}}\\
\vspace{0.1cm}
\scalebox{0.45}{\includegraphics*[50,370][415,730]{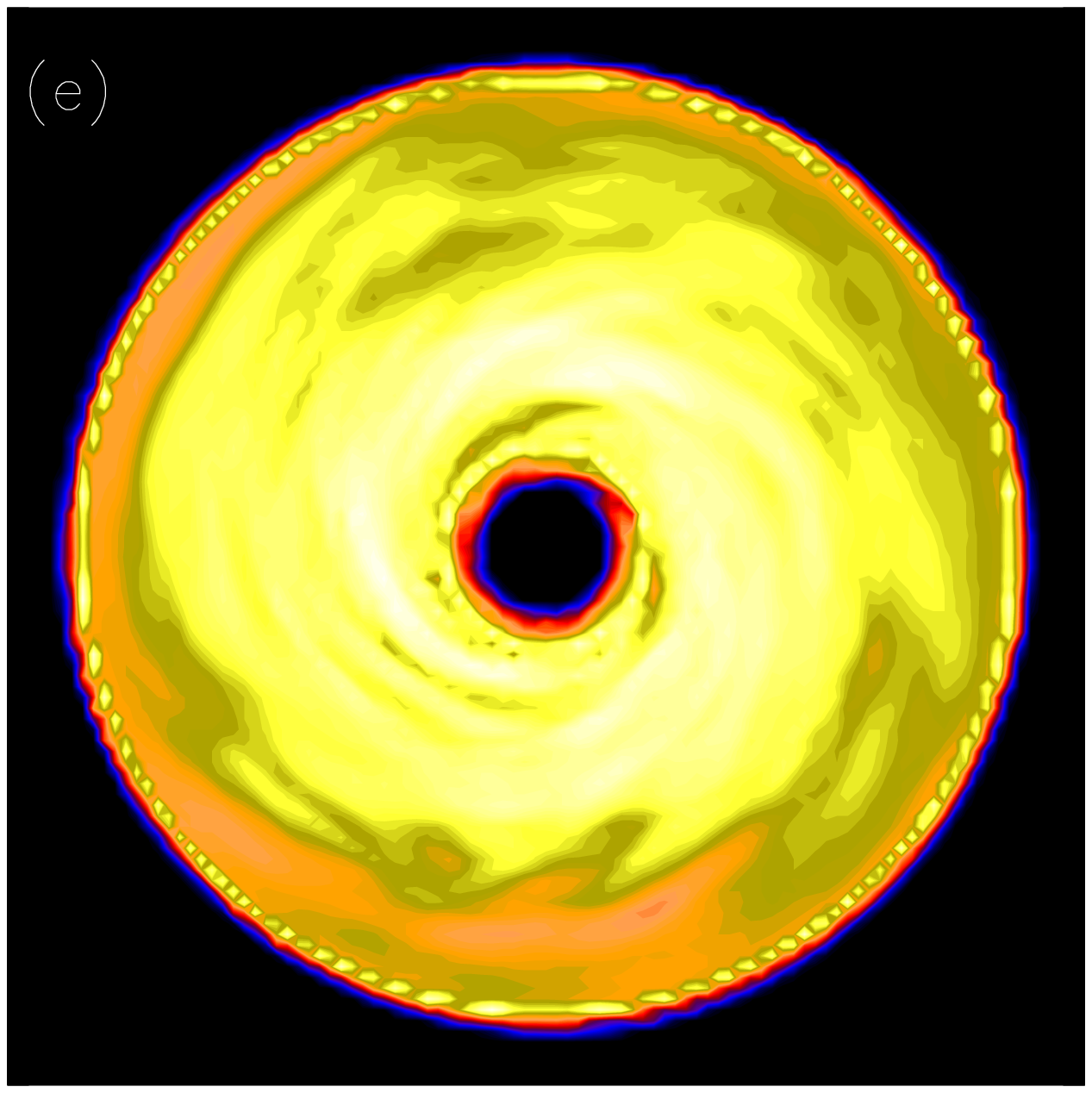}}
\scalebox{0.45}{\includegraphics*[50,370][415,730]{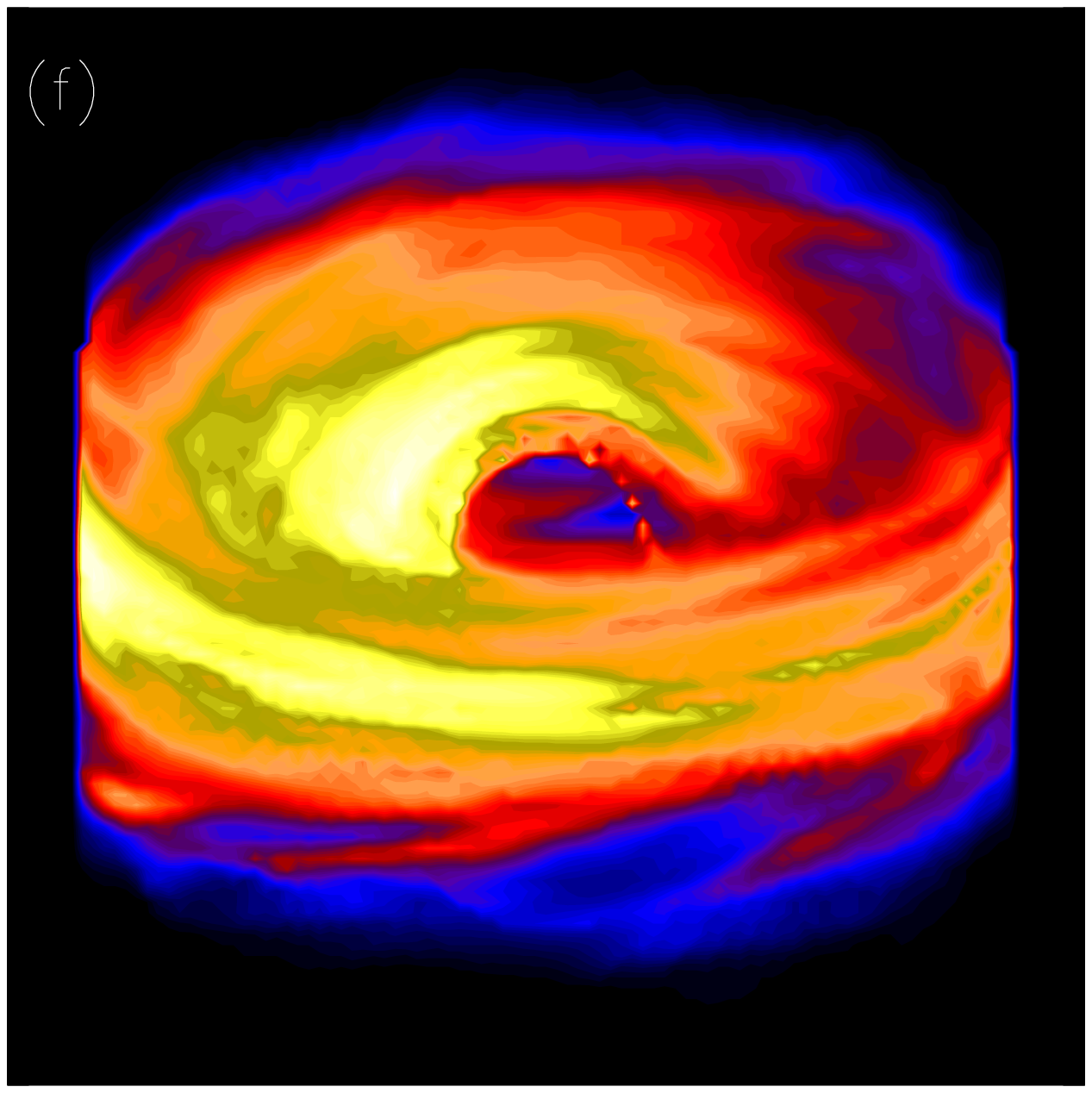}}\\
\scalebox{0.65}{\includegraphics*[30,300][580,420]{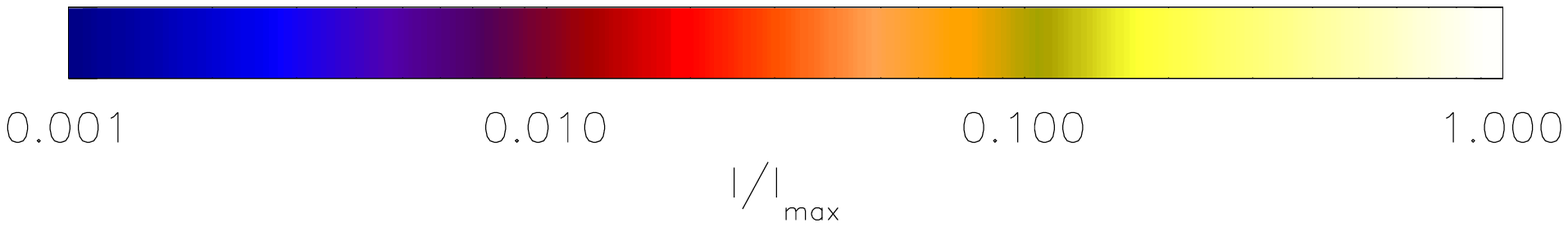}}
\end{center}
\end{figure}

As a demonstration of how these different models can be used to probe
the structure of the accretion disk, we show in Figure \ref{plotthree}
a snapshot of the inner disk, as seen by observers at inclinations of
$i=0^\circ$ and $70^\circ$ ($90^\circ$ is edge-on) for the three
emission models. For these figures and throughout the majority of this
paper, we include only emission from the inner region of the accretion
flow with $r<25M$, where the gas behaves most like a classical
accretion disk. In all panels, the observed intensity is represented by a
color-coded logarithmic scale normalized to the peak intensity of that
frame. In the high-inclination frames, the gas is moving towards the
observer on the left-hand side of the disk, i.e.\ in the positive
$\hat{\phi}$ direction. 

In Figures \ref{plotthree}a and \ref{plotthree}b, the optically thin
nature of the gas is
clear from the Einstein ring seen in the inner-most region of the
disk: this ring is formed by the photons that pass around the
black hole and then through the disk and to the observer. With the
exception of the emission from the Einstein ring, the intensity
clearly falls off with decreasing radius as the surface density
decreases inward with radius and the gravitational redshift further reduces
the observed emission. In Figure \ref{plotthree}b, the special
relativistic beaming of photons is clear from the higher intensity
seen from the gas moving towards the observer on the left. 

In Figures \ref{plotthree}c and \ref{plotthree}d, we show the
optically thick line emission model, demonstrating the nearly uniform,
monochromatic emission from the face-on disk. For the face-on view,
the only variations in intensity come from the gravitational
redshift and the transverse Doppler shift of gas orbiting more rapidly
in the inner-most disk. Small
local pertubations are also caused by the Doppler shift due to turbulent
motion of the photosphere perpendicular to the plane of the disk. As
in Figure \ref{plotthree}b, the edge-on view in Figure
\ref{plotthree}d shows the beaming 
of photons towards the observer, but produced from the upper layers of
the optically thick disk and surrounding atmosphere. 

Finally, in Figures \ref{plotthree}e and \ref{plotthree}f, we show the
image of the disk using the thermal emission and absorption
model. Unlike the optically thin model where emission traces surface
density, here the emission is greatest in the inner disk where the
temperature is highest. Also, the non-linear dependence of emission
and absorption on
the hydrodynamic variables produces larger amplitude modulations,
clearly seen in the structure of the disk. Because we include
emission and absorption only inside of $r=25M$, the interior of the disk
can be seen in Figure \ref{plotthree}f as a bright band at the outer
edge. While this effect is not physical (in a real accretion
disk it would be blocked by the outer portion of the disk), it
nonetheless serves as a useful imaging device and we believe it does
not significantly affect the qualitative behavior of the light
curve. One piece of evidence in support of this claim is the lack of
any significant excess power at the orbital frequency of $r=25M$
in the high-inclination light curves.

Just as the optically thin emission model can be used to track surface
density in the disk, the thermal model is able to measure the local
density and temperature at different depths of the disk. In this way,
the ray-tracing analysis provides us with both {\it literal} and {\it
  figurative} X-ray vision of the disk. The regions of highest
temperature (typically closer to
the disk midplane) produce photons in the high-energy tail that can
pass directly through the disk to the observer ($\alpha_\nu \sim
x^{-3}$ for $x\gg 1$). Conversely, since the
disk is optically thick to the low-energy Rayleigh-Jeans tail of the
thermal spectrum, the low-energy photon emission is a
direct measurement of the temperature at the surface of the disk,
where the intensity is proportional to $\nu^2 T$. 

\begin{figure}[tp]
\caption{\label{plotfour} X-ray images of the inner regions of the MHD
  simulation, considering only emission inside of $r<25M$. For a
  thermal emission model with $T_{\rm in} \sim 2-3$ keV, the three
  images are at photon energies of 1, 10, and 100
  keV. The inner-most region, where the disk has the highest surface
  temperature, emits most of the low-energy X-rays (panel $a$). On the
  other hand, since the disk is optically thin to the highest-energy
  X-rays, which are produced in the regions of largest overdensity
  near the midplane, they can be seen equally well in the outer
  regions (panel $c$). In between (10 keV; panel $b$), the thermal
  peak produces the majority of the observed emission, and thus
  resembles the total intensity shown in Figure \ref{plotthree}e. For each panel,
  the color scheme represents a logarithmic scale, normalized to the
  peak intensity for that frame.}
\begin{center}
\scalebox{0.5}{\includegraphics*[50,390][415,710]{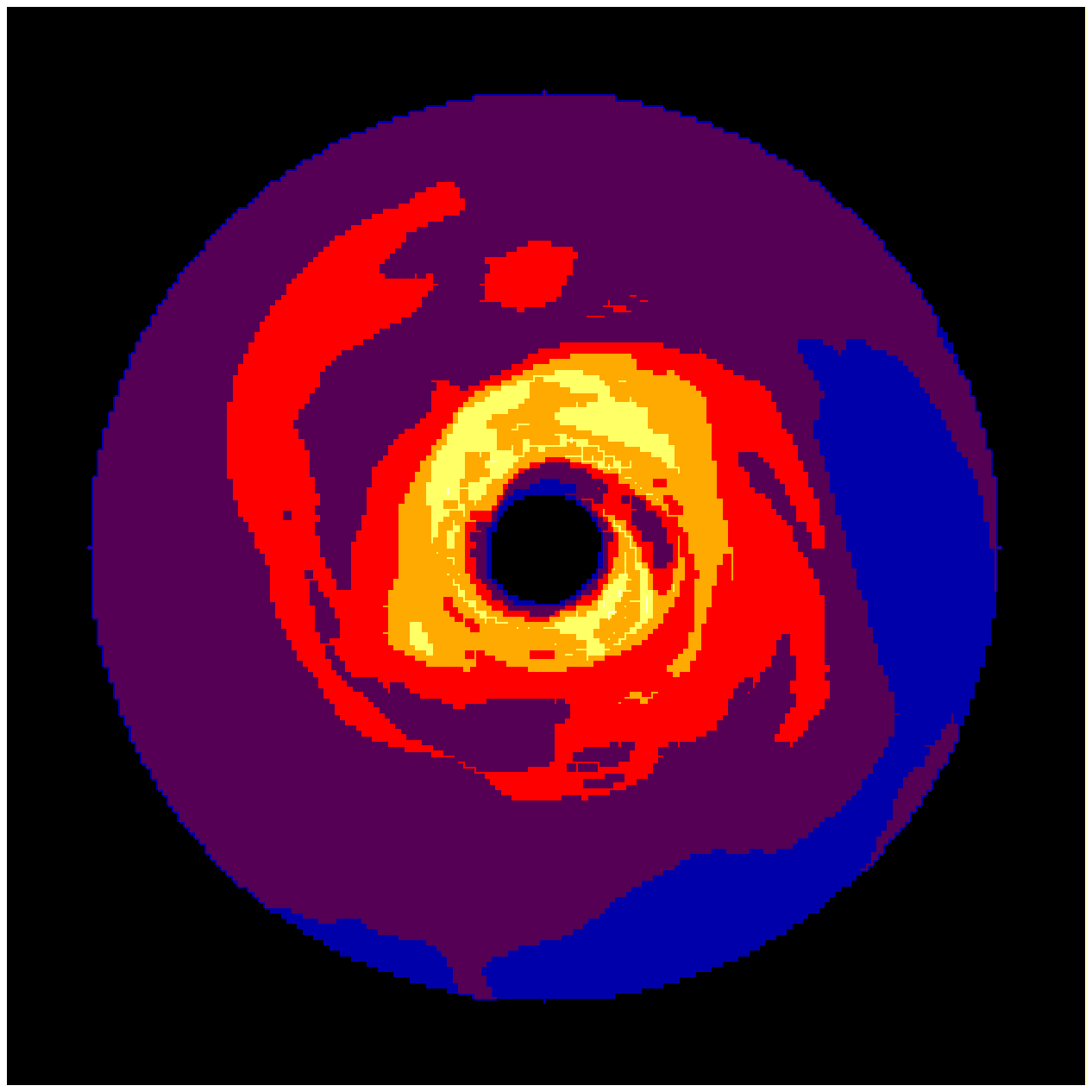}}
\scalebox{0.5}{\includegraphics*[50,390][415,710]{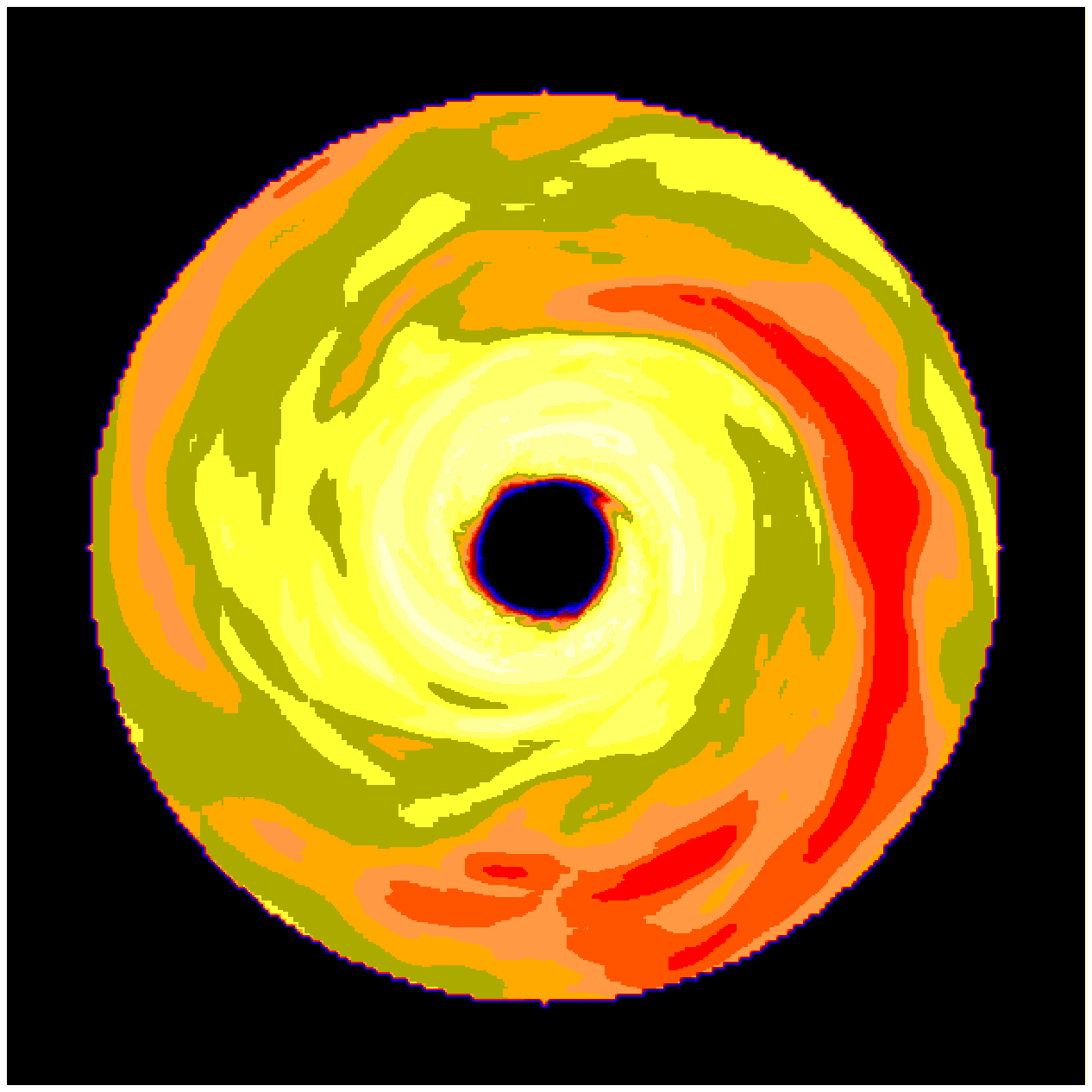}}\\
\vspace{0.2cm}
\scalebox{0.5}{\includegraphics*[50,390][415,710]{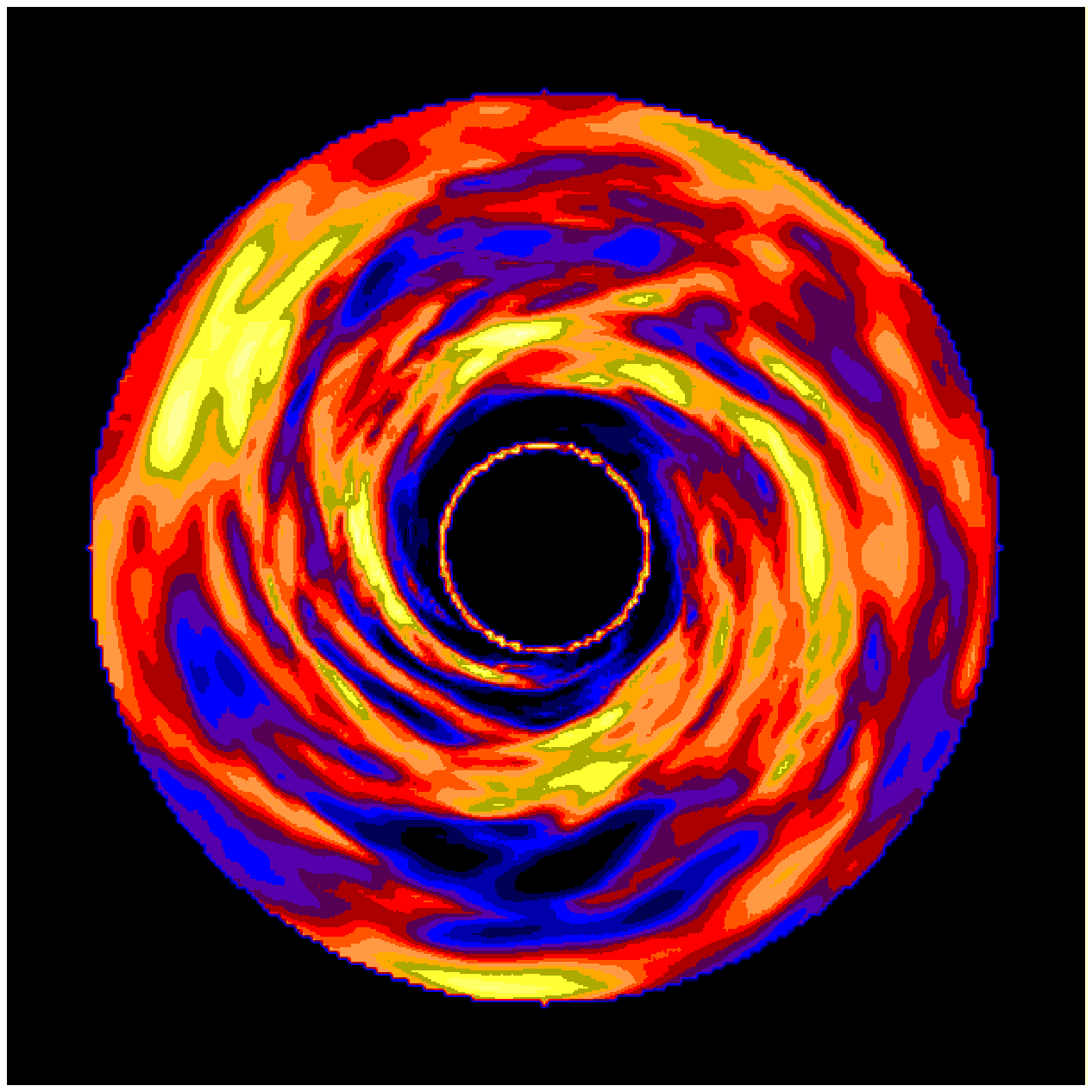}}
\scalebox{0.65}{\includegraphics*[30,300][580,420]{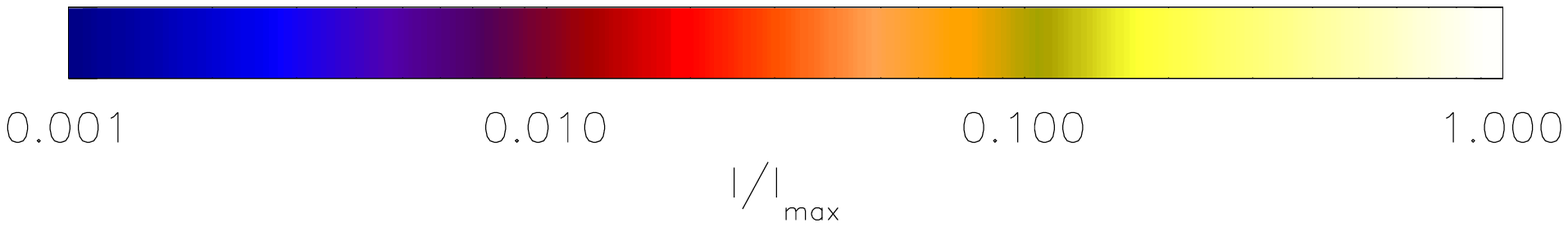}}
\end{center}
\end{figure}

This 3-dimensional imaging method is demonstrated in Figure
\ref{plotfour} for the same disk snapshot shown in Figure
\ref{plotthree}e, now divided into three different energy bands: 1,
10, and 100 keV. The thermal emission model has an inner disk
temperature $T_{\rm in}\sim 2-3$ keV, so the three plots sample the
low- and high-energy tails of the spectrum and the thermal peak. As described
above, the 1 keV emission shown in Figure \ref{plotfour}a probes the
surface temperature of the disk, highlighting the inner-most region
where the gas is hottest. The peak of the thermal spectrum near
$3kT_{\rm in} \approx 10$ keV (Fig.\
\ref{plotfour}b) of course produces the 
majority of disk emission and thus more closely resembles the integrated
emission shown in Figure \ref{plotthree}e. Since the disk is actually
optically thin to
the high-energy photons at 100 keV, they can be used to trace the
regions of greatest temperature and over-density near the midplane
(Fig.\ \ref{plotfour}c). We find that these regions are distributed
nearly uniformly in radius throughout the disk. 

\section{LIGHT CURVES AND POWER SPECTRA}\label{results}
\begin{figure}[tp]
\caption{\label{plotfive} Integrated light curves for the
  three different emission models: ($a$) optically thin line, ($b$)
  optically thick line, and ($c$)
  thermal emission/absorption. The {\it solid} lines have inclination
  $i=0^\circ$, the {\it dashed} lines have $i=45^\circ$, and
  the {\it dot-dashed} lines have $i=70^\circ$. The light curves are
  all normalized to their mean amplitude, linear secular trends have
  been removed, and the higher inclination light curves have been
  shifted vertically in order to facilitate comparison. The time scale
  is set by assuming a black hole mass of $10 M_\odot$.}
\begin{center}
\scalebox{0.5}{\includegraphics{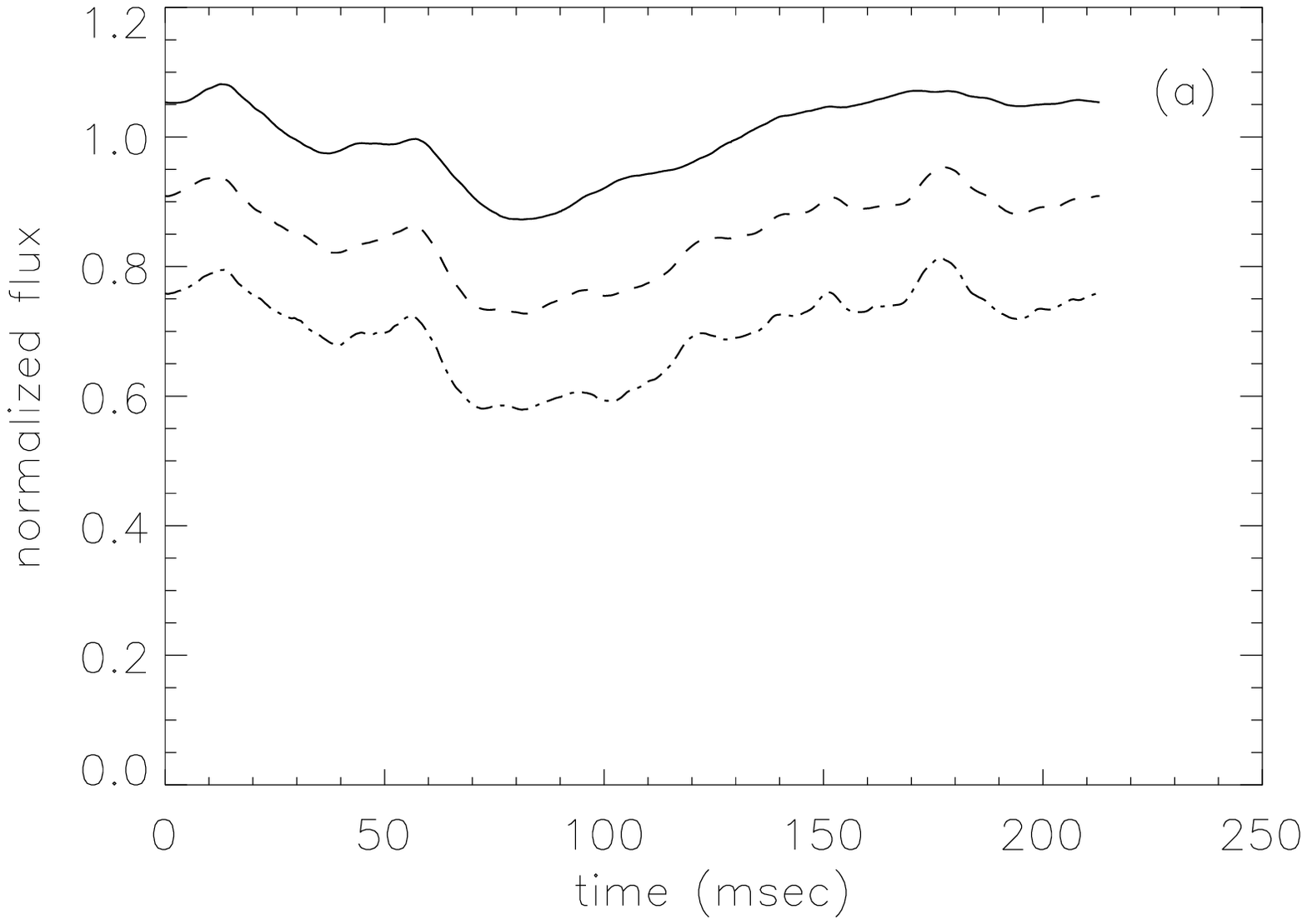}}
\scalebox{0.5}{\includegraphics{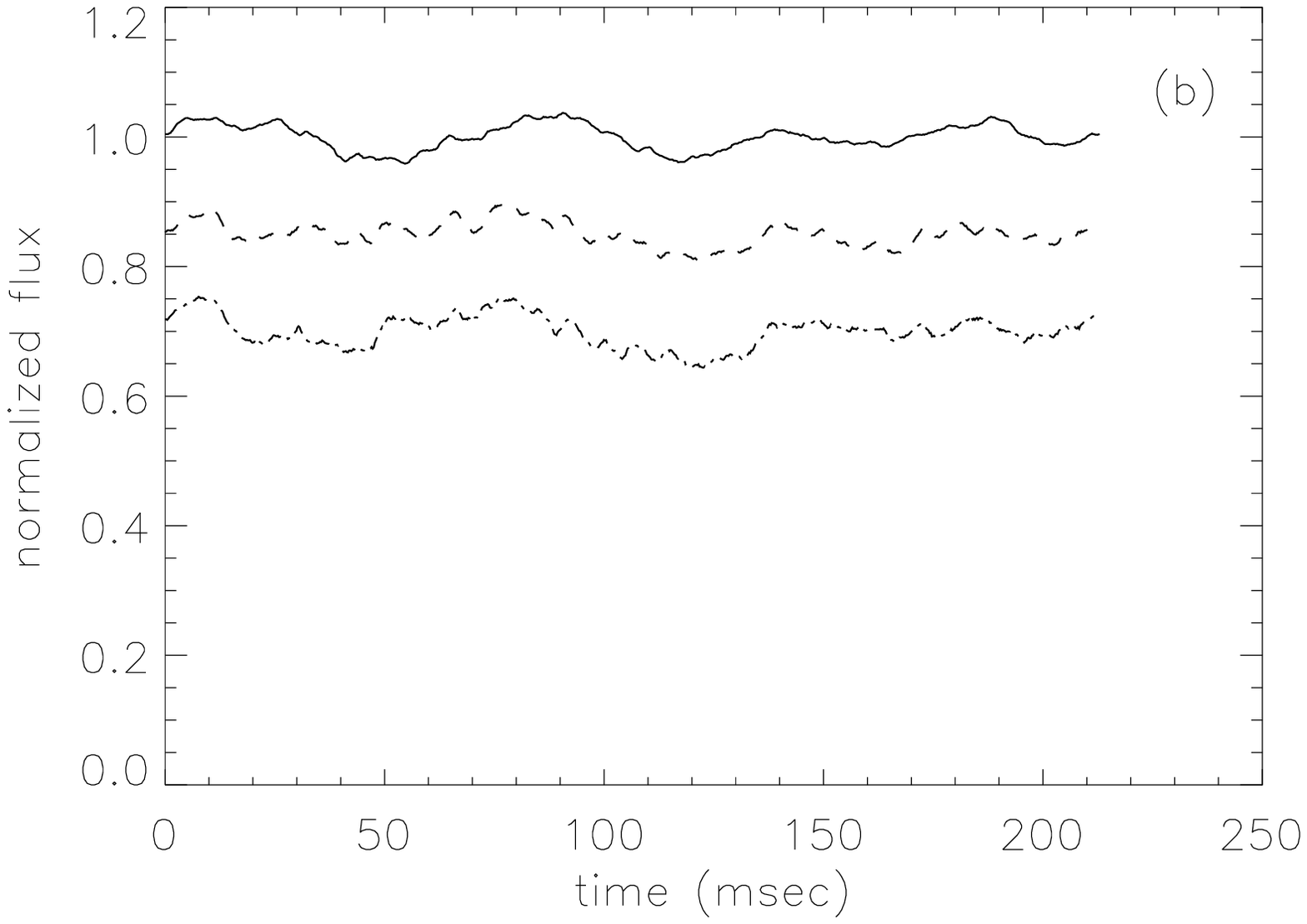}}
\scalebox{0.5}{\includegraphics{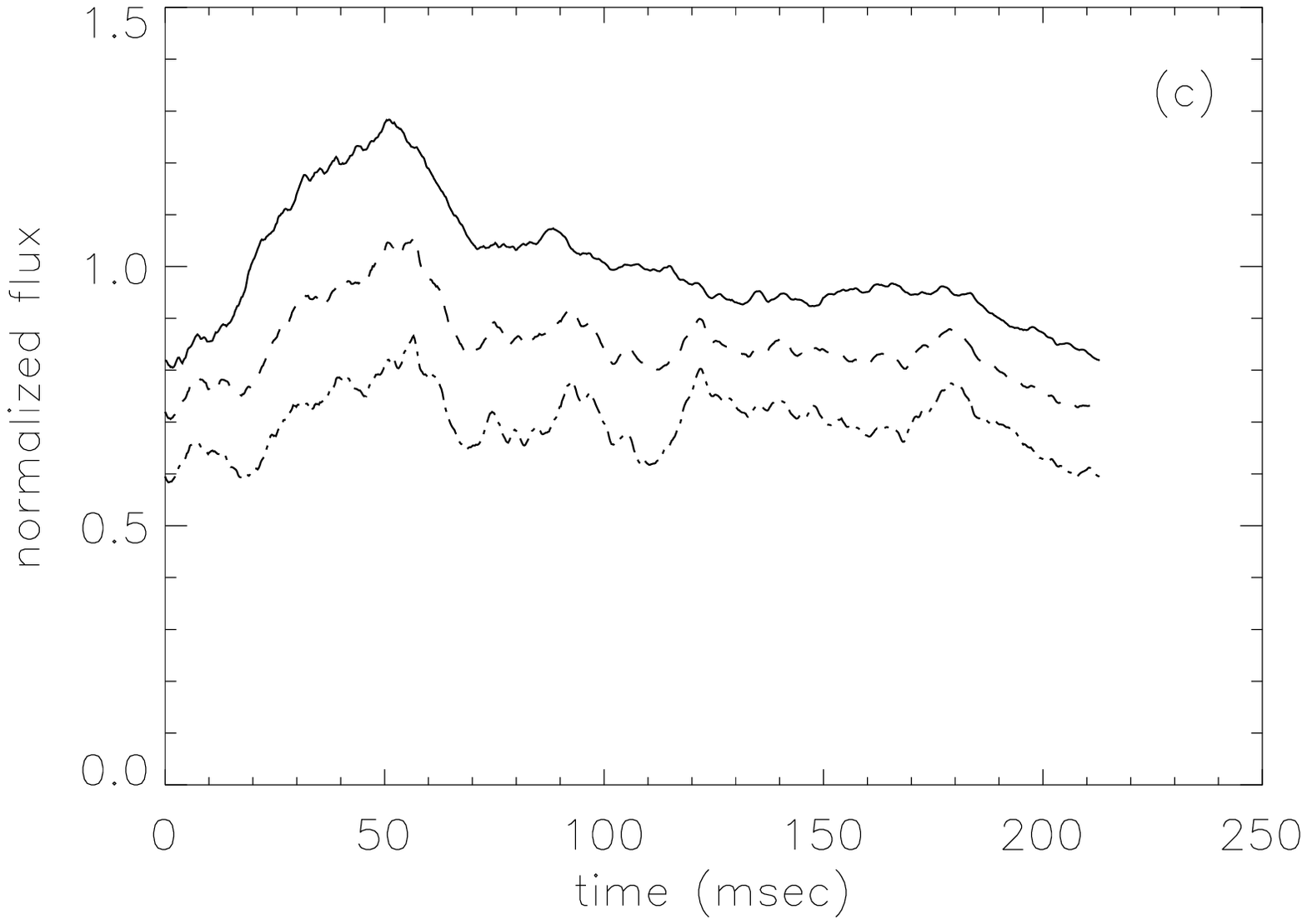}}
\end{center}
\end{figure}

Merging the ray-tracing methods described above with the
time-dependent output from
the MHD simulations, we are able to produce simulated light
curves of the accretion disk. For each of the three emission models,
we plot in Figure \ref{plotfive} the light curves for three
different inclination angles. A total of 870 data frames have been
sampled, spaced at coordinate time steps of $5M$, corresponding to a
total simulation time of just over 200 ms for a $10M_\odot$ black
hole, or about fifty orbits at the ISCO. This is a remarkably small
amount of ``real'' time for a
simulation that takes hundreds of hours to run on a major
supercomputer. Nonetheless, we believe it is still sufficiently long enough to
characterize the broad-band features in the power spectrum, and
additionally, as we show below, to
provide interesting constraints on the possible presence of
high-frequency QPOs.

Each of the light curves in Figure \ref{plotfive} has been
normalized and, for inclinations $i=45^\circ$ and $75^\circ$, shifted
vertically in order to facilitate comparison. Additionally, we have
applied a ``pre-whitening'' filter to each light curve, removing
linear secular trends and fixing $I(t_i)=I(t_f)$ in order to avoid
spurious high-frequency noise in the Fourier power spectra.
For the optically thin
line emission model, shown in Figure \ref{plotfive}a, it is clear
that the same large-amplitude, low-frequency features appear in phase
for all inclinations. These features most closely represent the total
mass within $r<25M$, a variable that is primarily driven by the mass
flux in the outer regions of the disk, and of course is independent of
inclination angle. The higher-frequency variability is due to regions
of overdensity moving through the gravitational potential and also
getting beamed towards and away from the observer for higher
inclinations. Thus the short-term variability is greater for
inclinations of $45^\circ$ and $70^\circ$, which is evident from
inspection of Figure \ref{plotfive}a. 

For the other emission models, the low-frequency correlation between
different inclinations is still present, but weaker than in the
optically thin case. This is quite reasonable, as the optically thick
models are more sensitive to localized perturbations that appear
different to observers at different inclinations. Similarly, there is
clearly more fractional variation on short time scales, again due to the
increased importance of local perturbations: if a hot spot rises to
the top of the disk and then disappears again on a dynamical time
scale, the effect will be greater for an optically thick disk than a
transparent one. Furthermore, the light curves from different
emission models do not appear to be correlated with each other,
confirming the presumption that they probe fundamentally different
properties of the disk. 

Moving beyond simple visual inspection of the light curves, we plot in
Figure \ref{plotsix} the power spectral density, in units of
$({\rm rms}/{\rm mean})^2{\rm Hz}^{-1}$, for the same light curves plotted
in Figure \ref{plotfive}. As suggested by the discussion above, we
find significantly more high-frequency power in the optically thick
and thermal emission models. Within each model, we find more
high-frequency power for the systems with higher inclinations. The
total power for each light curve is listed in Table \ref{table_power}, as
well as the amount of high-frequency $(f>100 \mbox{ Hz})$
power.

In all the plots shown in Figure \ref{plotsix}, the power
spectra can be described by a steep power-law in frequency. There is
some evidence of a flatter slope below $\approx 50$
Hz, but this is quite possibly a spurious effect due to the short
duration of the simulation, so that we have poor statistical sampling of
low-frequency features in the light curves. 
We have listed in Table \ref{table_power} the best-fit power-law indices
for each power spectrum above 50 Hz, mostly in the range $\Gamma \sim
3-4$. While the spectra are somewhat
flatter for the optically thick and thermal models, all appear steeper
than those seen in {\it RXTE} spectra, which are more typically around
$\Gamma \sim 1$ \citep{mccli05} (however it should be noted that the
data become increasingly poor above $\sim 100$ Hz for most
observations in the Thermal state). At the same time, the simulations of
\citet{armit03}, where emission traces the vertically integrated
stress, produce power spectra with slopes of $\Gamma \approx 2$.

In all cases the slope appears to be independent of the
inclination angle, and a function only of the emission model. For each
emission model, the total rms amplitude of the light curve
fluctuations increases with inclination, a result that was also
found by \citet{armit03}. In fact, this may be the most robust measure
with which to compare the simulated light curves with
observations. For a broad range of possible emission mechanisms, we
find that the MHD turbulent disks consistently appear more
variable when viewed at high inclinations, a prediction that should be
testable with archived {\it RXTE} data. 

\begin{figure}[tp]
\caption{\label{plotseven} Power spectra for light curves produced
  by different annuli of the accretion disk, using the thermal
  emission model and an inclination of $i=0^\circ$. The inner-most regions
  of the disk produce larger amplitude fluctuations and contribute
  relatively more power at higher frequencies (flatter spectrum above
  $\sim 100$ Hz).}
\begin{center}
\scalebox{0.8}{\includegraphics{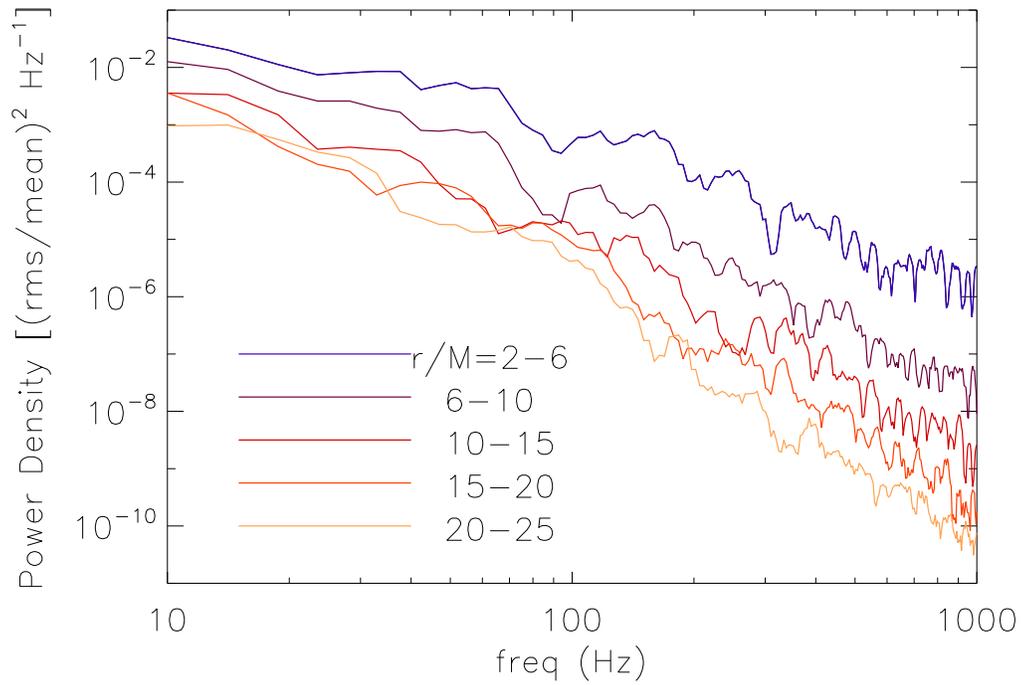}}
\end{center}
\end{figure}

As shown in \citet{armit03}, the power spectrum from a
single radius (i.e.\ annulus of the disk) should show a clear
break frequency at roughly the orbital frequency at that
radius. Following that approach, we show in Figure
\ref{plotseven} the power spectra from five different radii in
the disk. Again we use the thermal emission model and a face-on
inclination $i=0^\circ$ so as not to introduce spurious features from
the direct emission from the disk interior (artificially visible due
to cutting off absorption outside the given annulus). The radii
sampled in Figure \ref{plotseven} have $r/M=(2-6)$, $(6-10)$,
$(10-15)$, $(15-20)$, and $(20-25)$. There is some visible
evidence for broken power laws, with lower break frequencies for
annuli at larger radius, but again the short simulation time severely
limits our confidence in the behavior of the power spectra at
low frequencies. More statistically significant is the difference
between the power-law slopes at high frequency, where the inner
regions of the disk clearly have flatter power spectra. The inner
regions also produce much more total power, dominating the overall
variance in the integrated flux. These results are summarized in Table
\ref{annuli_power}, which lists the best-fit power-law indices above
and below 100 Hz, as well as the integrated power for each annulus. 

Even if we were to include emission only from the
inner-most regions of the disk, the corresponding power spectra would
still be significantly steeper than those seen in the {\it RXTE}
observations. The discrepancy with the the $\Gamma=2$ result of
\citet{armit03} should also be investigated in future work in order to
understand the relative importance of the simulation details compared
to the post-processor emission models in producing a given power-law
slope. In fact, these discrepancies provide one of the strongest
arguments demonstrating the need of an 
accurate radiation transfer post-processor for the MHD simulations. By
looking solely at the variation of the code variables, as many
previous works have done, one cannot hope to reproduce the light
curves and power spectra that are actually observed from the
disk.

To agree quantitatively with the data will require either
new physics in the MHD simulations or new emission models for the
post-processors, presumably with an increased focus on the emission
from the corona. Additionally, we know electron scattering plays an
important role in the radiation transport for hot accretion disks, and
therefore geodesic photon paths are not completely realistic. 
As yet further motivation for understanding the emission processes
that produce these power spectra, recent observations of GRS 1915+105
actually show a much steeper power-law, with $\Gamma=2.8-3.0$
\citep{bello06}. However, those observations were taken during a
relatively hard state (as opposed to the soft Thermal state), further
complicating direct comparison with our emission models.

\subsection{Sensitivity to QPOs}\label{upper_limits}
\begin{figure}[tp]
\caption{\label{plotsix} Power spectra for the
  light curves plotted in Figure \ref{plotfive}. As in that
  Figure, the {\it solid} lines have inclination
  $i=0^\circ$, the {\it dashed} lines have $i=45^\circ$, and
  the {\it dot-dashed} lines have $i=70^\circ$.}
\begin{center}
\scalebox{0.5}{\includegraphics{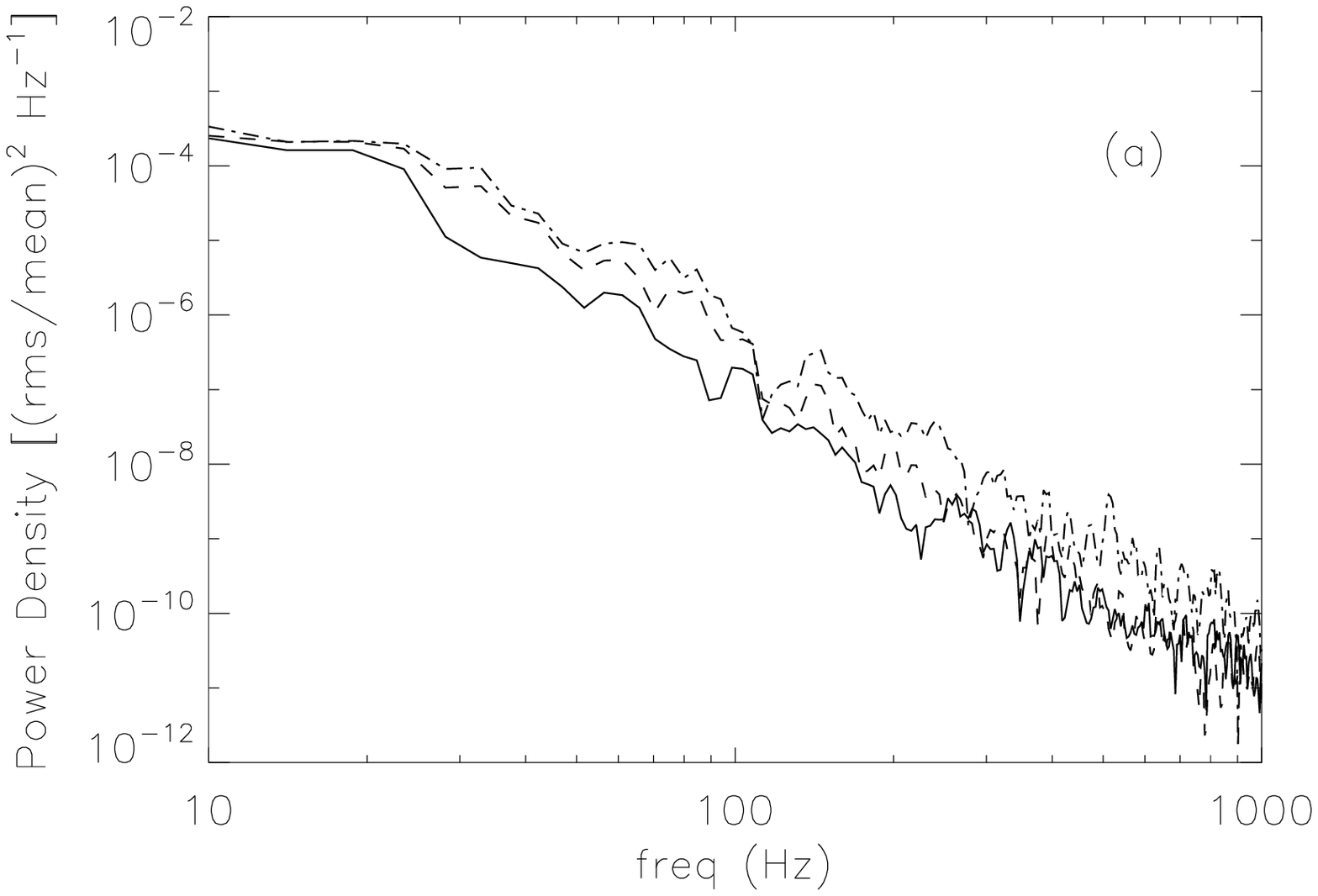}}
\scalebox{0.5}{\includegraphics{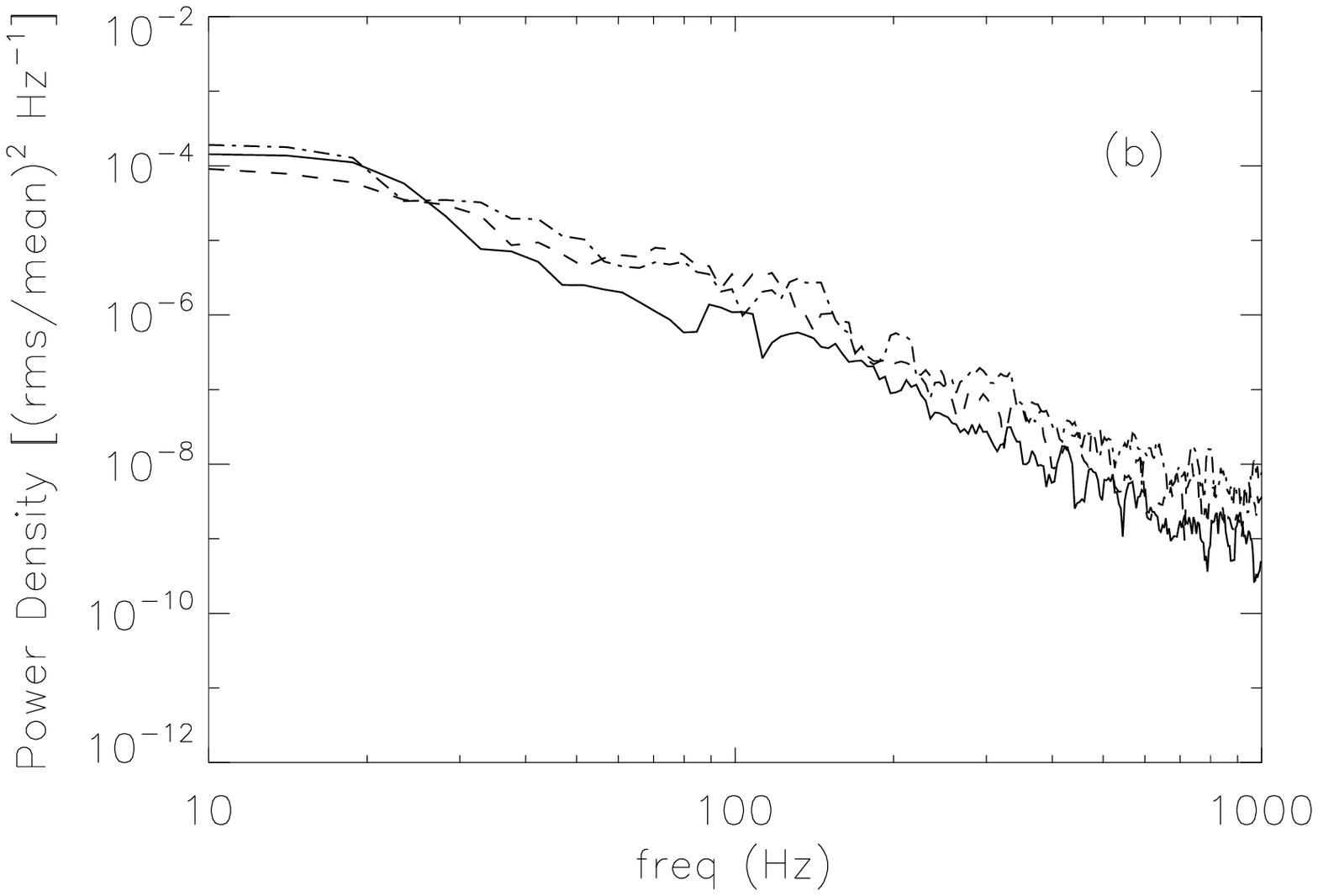}}
\scalebox{0.5}{\includegraphics{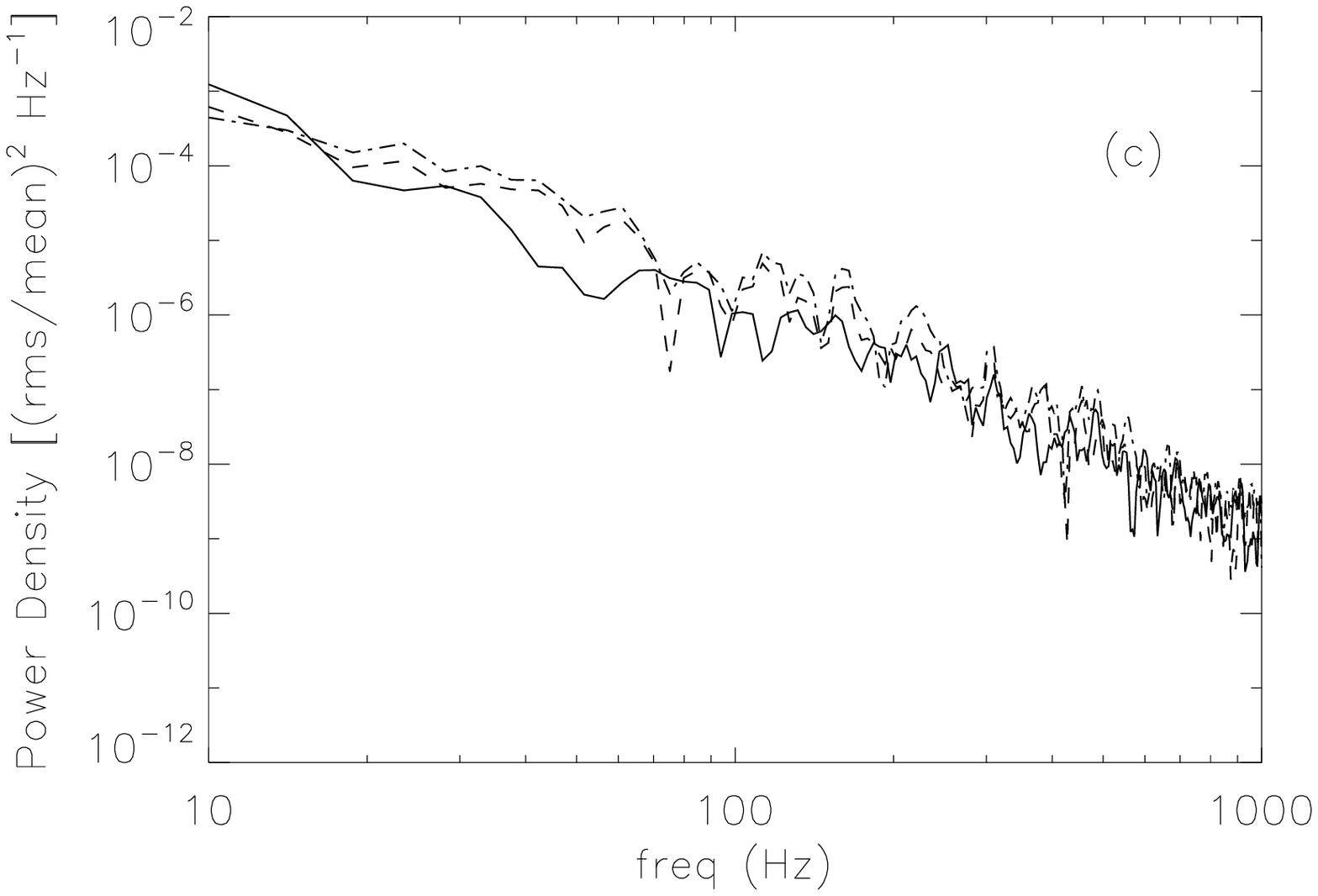}}
\end{center}
\end{figure}

From visual inspection of the power spectra in Figure
\ref{plotsix}, no clear features (e.g.\ peaks, power-law breaks,
etc.) are immediately evident. However, the short simulation time
makes inherent statistical fluctuations unavoidable, which could
either hide true features or falsely imitate absent ones. We would
therefore like to derive a more quantitative understanding of our
sensitivity to timing features in the simulated light curves.
From the power spectra plotted in Figure
\ref{plotsix}c, we can derive an estimate for the theoretical
upper limits on the possible existence of high frequency QPOs in the MHD
simulations.

With the high spatial resolution of our ray-tracing
post-processor, we can safely assume the power-law behavior of the
power spectrum is
due to the inherent turbulence of the disk, as opposed to any sort of 
``detector'' or finite sampling statistics. Further assuming that the power
in neighboring frequency bins is uncorrelated, we can treat the
variations (jitter) in the power spectrum as behaving locally like
Poisson variants. In the limit of an infinite length simulation, the
frequency bins would become infinitesimally small, and thus could be
averaged to get an arbitrarily smooth power spectrum, i.e., the
``inherent'' turbulent power, which we call $P_\infty(\nu)$, modeled
as a simple power-law:
\begin{equation}
P_\infty(\nu) = P_0 \nu^{-\Gamma}.
\end{equation}
Then the probability distribution function for the power at a given
frequency is given by that of a Poisson variant:
\begin{equation}\label{fPnu}
f(P_\nu) = \frac{1}{P_0\nu^{-\Gamma}}
\exp\left(\frac{-P_\nu}{P_0\nu^{-\Gamma}}\right),
\end{equation}
which has mean $\mu = P_0\nu^{-\Gamma}$ and variance
$\sigma^2=\mu^2$. We have checked this relation empirically for the
power spectra in Figure \ref{plotsix} and find that the
measured variations in $P_\nu$ do indeed behave according to equation
(\ref{fPnu}). 

For this distribution, the observed power in a single frequency bin
is expected to be $P_\infty(\nu)$, plus or minus 100\% (hence the
critical importance of frequency binning). To identify a potential
QPO, we would
integrate over some range in frequency $\nu-\Delta\nu$ to
$\nu+\Delta\nu$. For a simulation duration of time $T_{\rm sim}$, the
number of frequency bins in this range is 
\begin{equation}
N_{\rm bin} = 2\Delta\nu T_{\rm sim}.
\end{equation}
Assuming $P_\infty(\nu)$ is roughly constant over this range (i.e.\
$\Delta\nu \ll \nu$), the
total power $P_N$ in the $N_{\rm bin}$
frequency bins should have a distribution than is approximately normal
(central limit theorem for large $N_{\rm bin}$):
\begin{equation}\label{f_Ptot}
f(P_N) = \frac{1}{\sqrt{2\pi N_{\rm bin}\sigma^2}}
\exp\left[\frac{-(P_N-N_{\rm bin}\mu)^2}{2N_{\rm
      bin}\sigma^2}\right].
\end{equation}

A potential QPO in the power spectrum would appear as a
small amount of excess power above the ``background noise''
$P_\infty(\nu)$. A QPO centered around $\nu_0$ with a FWHM of
$2\Delta\nu$ and a Lorentzian profile
\begin{equation}
P_{\rm QPO}(\nu) = \frac{A^2_{\rm
    QPO}\pi/\Delta\nu}{1+\left(\frac{\nu-\nu_0}{\Delta \nu}\right)^2}
\end{equation}
has total (rms/mean) amplitude of
\begin{equation}
 A_{\rm QPO}(\mbox{rms \%}) = \sqrt{\int d\nu P_{\rm QPO}(\nu)}.
\end{equation}
For our significance estimate, we will consider only the power within
$\nu_0\pm \Delta\nu$, which has an rms departure from the background
power-law of $A_{\rm QPO}/\sqrt{2}$. From
equation (\ref{f_Ptot}), we see that the total measured power
\begin{equation}
P_{\rm tot}(\nu_0,\Delta\nu)= \int_{\nu_0-\Delta\nu}^{\nu_0+\Delta\nu}
P_\nu d\nu
\end{equation}
can be used to give a ``1-sigma'' confidence limit on the QPO power by
\begin{equation}
P_{\rm tot} = 2\Delta\nu P_\infty(\nu_0)+P_{\rm QPO} < 
(N_{\rm bin}\mu+\sqrt{N_{\rm bin}\sigma^2})d\nu,
\end{equation}
where $d\nu=1/T_{\rm sim}$ is the size of the frequency bin in
Hertz. More generally, at the ``n-sigma'' significance level, we can
rule out any QPO with rms amplitude greater than
\begin{eqnarray}\label{A_significance}
A_{\rm QPO}(n_\sigma) &=& \left[2(N_{\rm bin}\mu d\nu+n_\sigma \sqrt{N_{\rm
      bin}\sigma^2}d\nu-2\Delta\nu P_0\nu_0^{-\Gamma})\right]^{1/2}\nonumber\\
&=& \left[2n_\sigma d\nu\sqrt{N_{\rm bin}\sigma^2}\right]^{1/2} =
    \left[n_\sigma P_0\nu_0^{-\Gamma}\sqrt{\frac{8\Delta\nu}{T_{\rm
      sim}}}\right]^{1/2}.
\end{eqnarray}
In terms of the oscillator quality factor
$Q=\nu_0/FWHM=\nu_0/(2\Delta\nu)$, the amplitude limit can be written
\begin{equation}\label{A_significance2}
A_{\rm QPO}(n_\sigma) = \left(2n_\sigma P_0
\nu_0^{-\Gamma+1/2}\right)^{1/2} (T_{\rm sim} Q)^{-1/4}.
\end{equation}
Plugging in some typical numbers from our simulations (e.g., the power
spectra in Fig.\ \ref{plotsix}c), we have $P_0 = 10$,
$\Gamma=3$, and $T_{\rm sim}=200$ msec. For a typical high-frequency
QPO with $\nu_0=200$ Hz and $\Delta\nu=20$ Hz ($Q=5$;
\citet{remil02}), we should thus be able to rule out the presence of a
QPO with rms amplitude 1\% at the 3-sigma significance level.

A number of important relationships can be seen from equation
(\ref{A_significance}). First, the dependence on $\nu_0$ shows that due to
the steep power-law behavior of the background turbulence, with the
current simulations it is very difficult to detect or rule out QPOs
at {\it low} frequencies. For example, a similar QPO with quality
factor $Q=5$ at $\nu_0=50$ Hz would require an amplitude of $\sim 6\%$
to be detected at the same significance level. Narrower QPOs (large
$Q$, small $\Delta\nu$) are easier to detect, but not by much, due to
the $\Delta\nu^{1/4}$ dependence in equation
(\ref{A_significance}). Similarly, a much longer simulation would also
not improve sensitivity that much, due to the same weak dependence on
$T_{\rm sim}$. 

\subsection{Identification of High-frequency QPOs}\label{hfqpos}
The discussion above provides a simple way to estimate {\it a priori}
the amplitude of any potential QPO at a given significance
limit. Turning the problem around, and without assuming anything about
the shape or amplitude of the QPO, one can
simply calculate the significance of any power excess
found in a localized group of $N_{\rm bin}$ frequency bins. In this
approach, we
set a fixed value for the quality factor $Q$, which in turn gives
$N_{\rm bin}(\nu)$ and thus $P_N(\nu)$, from which the significance
of any region of excess (or deficit) power can be determined by
\begin{equation}\label{n_sigma}
n_\sigma(\nu) = \frac{P_N(\nu)-N_{\rm bin}\mu}{\sqrt{N_{\rm
      bin}\sigma^2}}.
\end{equation}
Note that this definition of $n_\sigma$ allows for negative
significance values (i.e.\ deficit power) and has a mean value of
$0$. If the distribution of power in individual frequency bins follows
an exponential distribution, which is what we find for our
power spectra, the distribution of $n_\sigma$ will have a chi-squared
distribution with $2N_{\rm bin}$ degrees of freedom. For large values of
$N_{\rm bin}$, this approaches a normal distribution with variance
$N_{\rm bin}\sigma^2$.

\begin{figure}[tp]
\caption{\label{ploteight} Excess power (in units of significance
  $\sigma$) above the
  ``inherent'' shape of the power spectrum, as defined by equation
  (\ref{n_sigma}). The different plots correspond to different viewing angles of
  the same disk, all using the thermal emission model. The small ticks
  at the top of each plot
  correspond to integer multiples of a fundamental frequency $\nu_f$: $2\nu_f$,
  $3\nu_f$, $5\nu_f$, and $6\nu_f$, where $3\nu_f =230$ Hz, close to
  the geodesic orbital frequency at the ISCO (220 Hz).}
\begin{center}
\scalebox{0.45}{\includegraphics{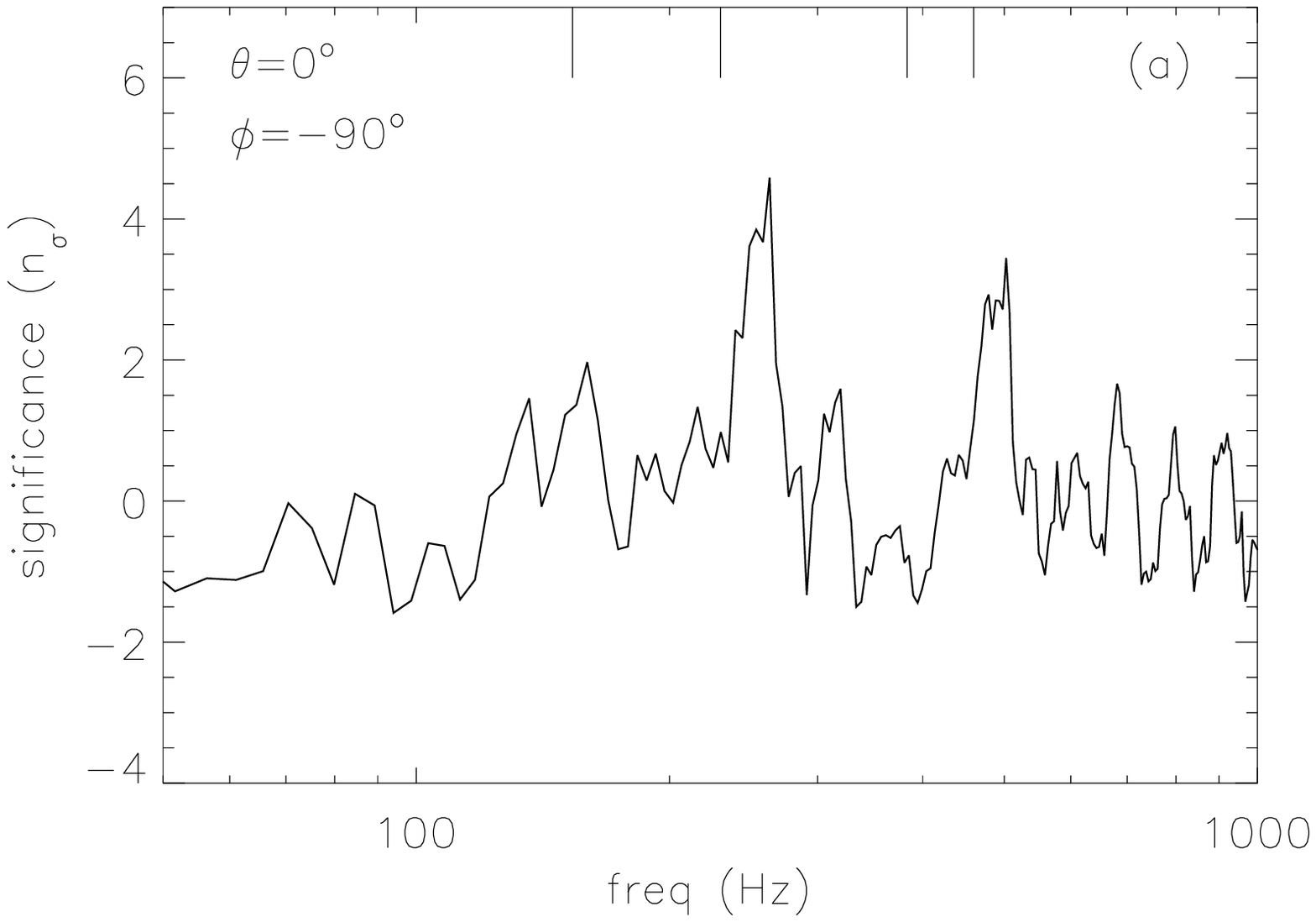}}
\scalebox{0.45}{\includegraphics{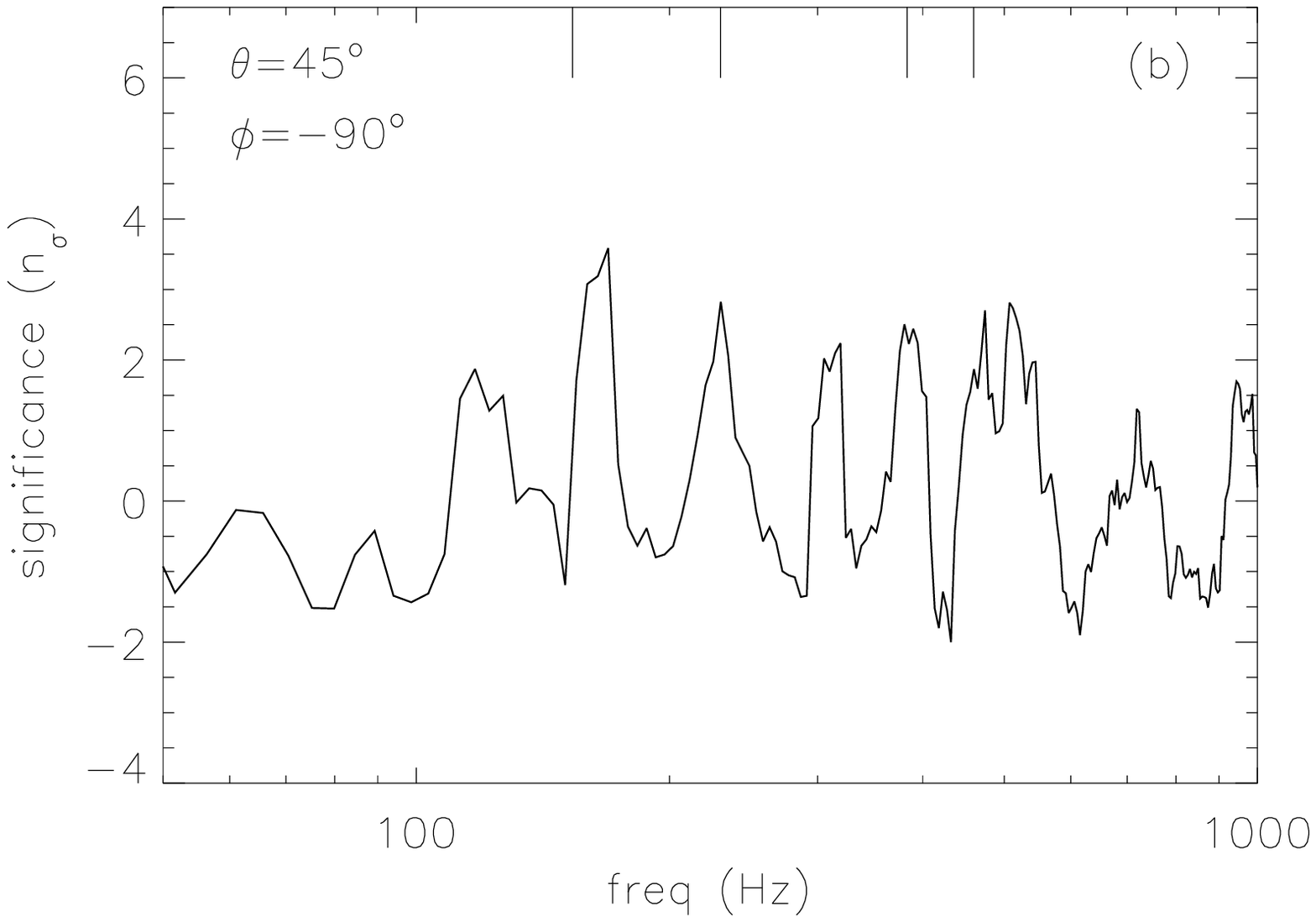}}\\
\scalebox{0.45}{\includegraphics{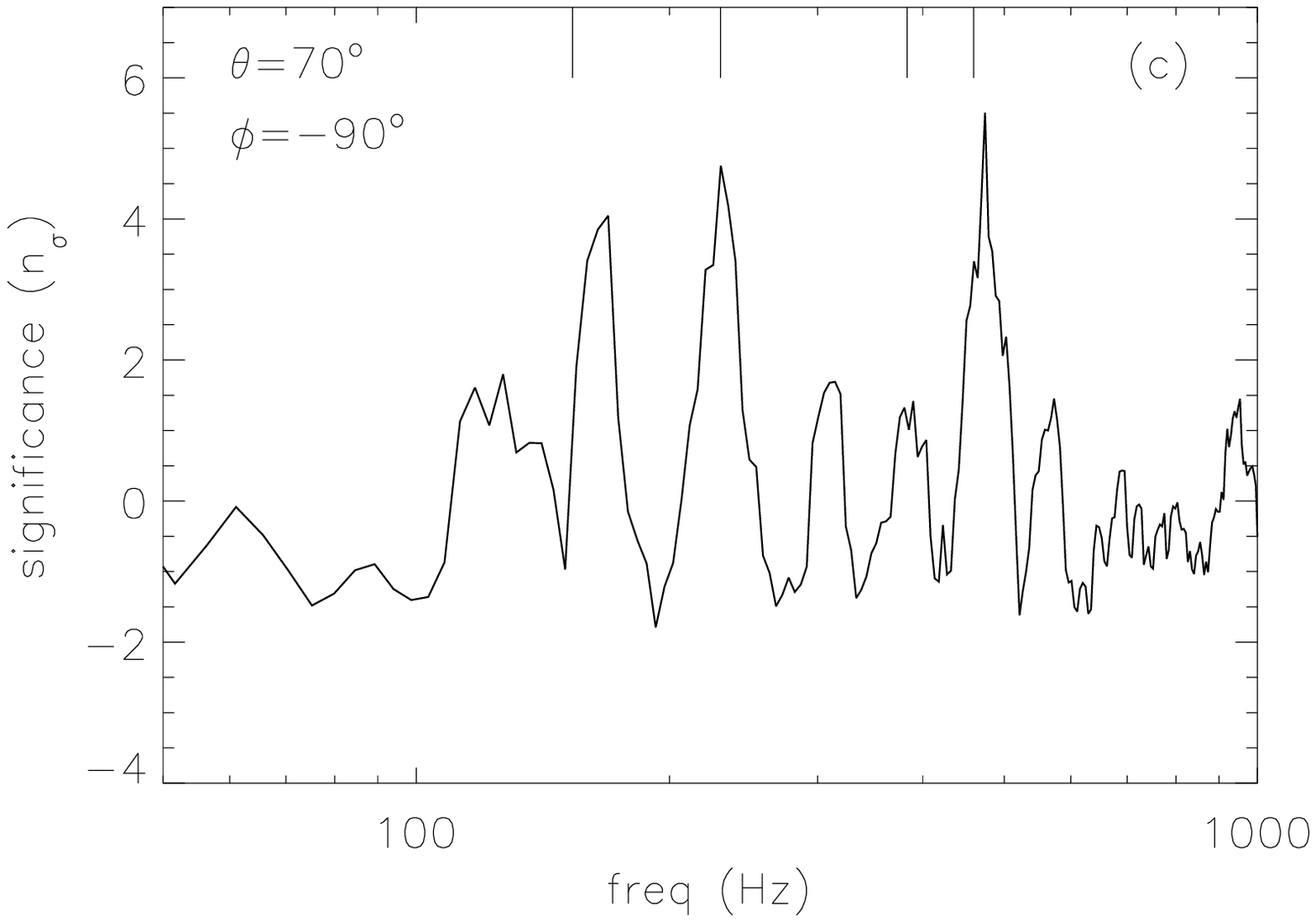}}
\scalebox{0.45}{\includegraphics{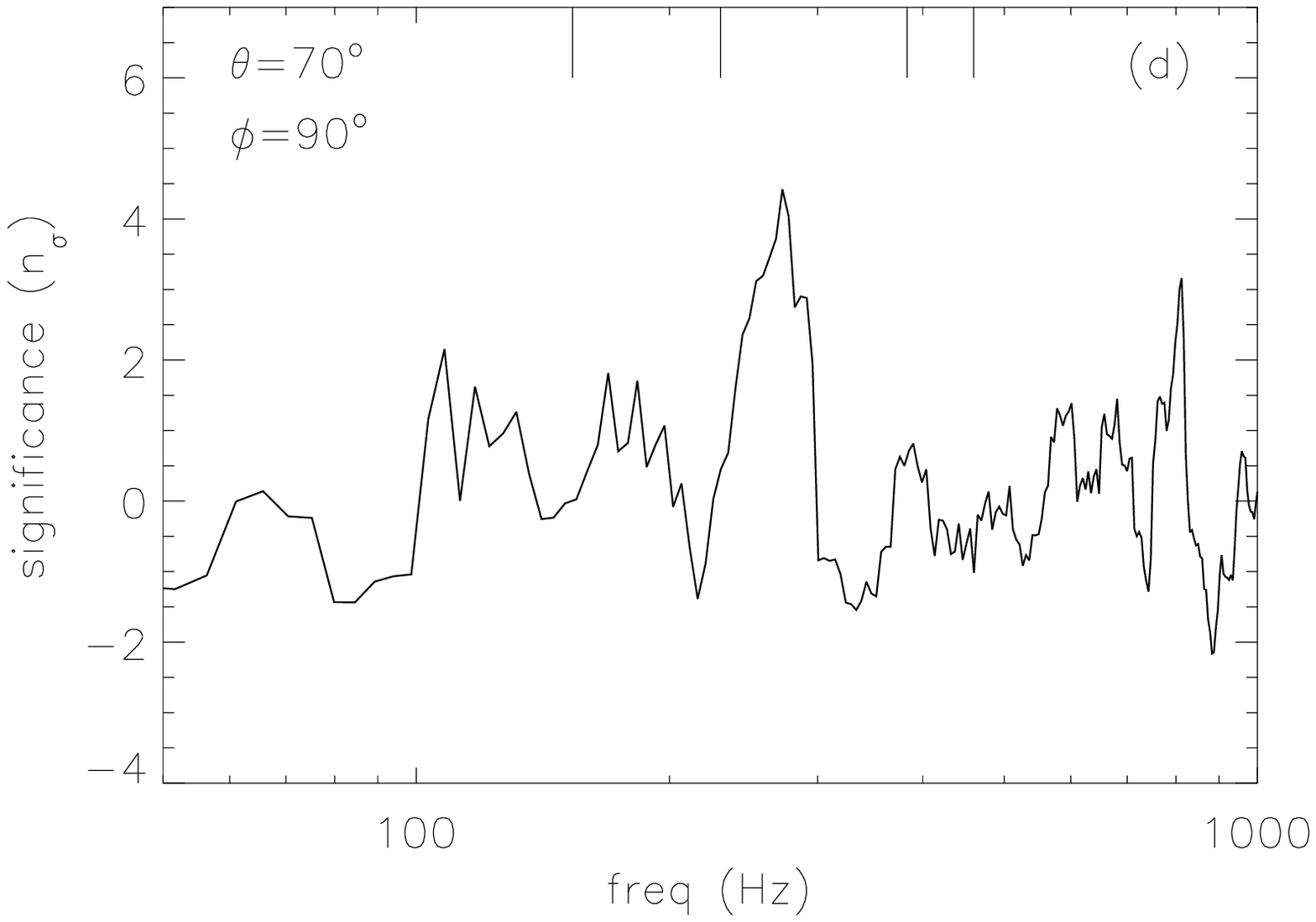}}\\
\scalebox{0.45}{\includegraphics{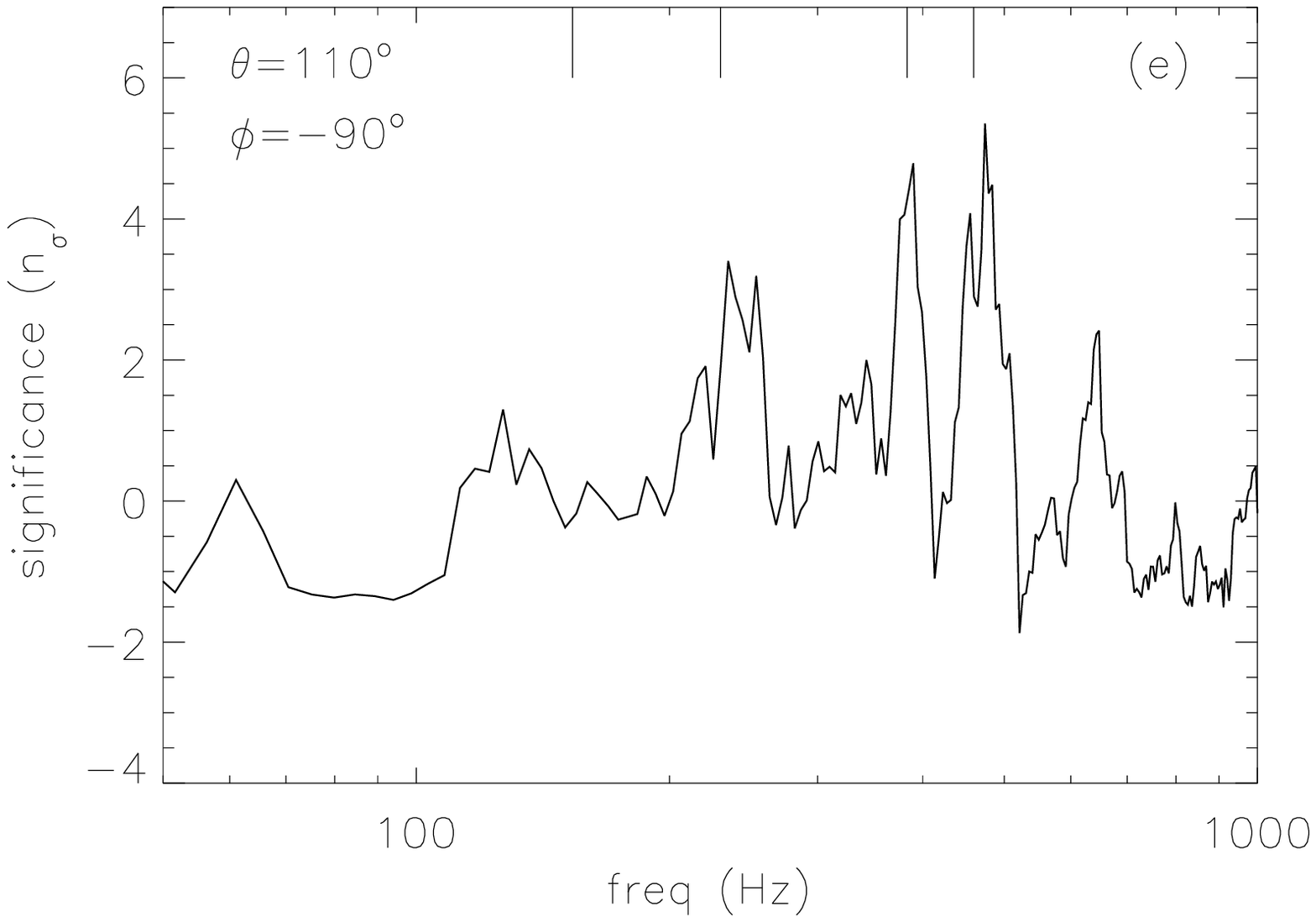}}
\scalebox{0.45}{\includegraphics{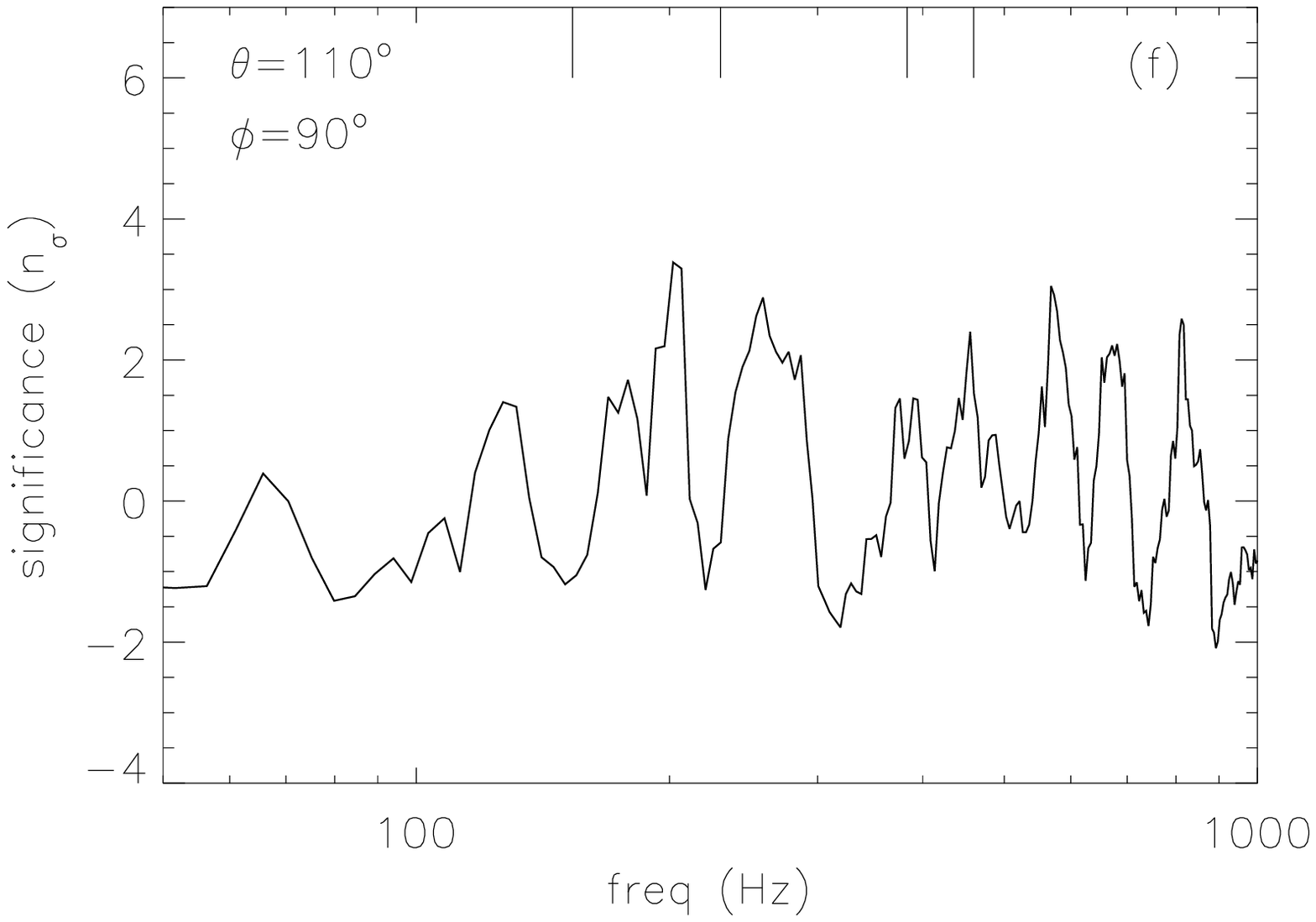}}\\
\end{center}
\end{figure}

In Figure \ref{ploteight} we plot a collection of these significance
curves for the thermal emission model with a range of inclination
angles and orientations, fixing $Q=10$. For each power spectrum, we
determine the power-law background $P_\infty(\nu)$ with a
least-squares fit to the simulated data. While not immediately obvious
in the original power spectra, a number of interesting features appear
in the significance analysis. Most striking are the peaks in Figure
\ref{ploteight}c that appear close to the orbital frequency at the ISCO,
as well as $2/3\nu_\phi({\rm ISCO})$ and the
first harmonic $2\nu_\phi({\rm ISCO})$. The significance of these
peaks increases with inclination (compare with the smaller peaks at
the same frequencies in Fig.\ \ref{ploteight}b, where $i=45^\circ$), as
predicted by the hot spot model described in \citet{schni04}.

This result is very intriguing, especially in light of a number of recent
observations of black hole binaries that find multiple QPO peaks with
a 2:3 frequency ratio \citep{mille01,remil02,homan05}. Thus it is quite
important to understand the robustness of such a result, if possible
without requiring hundreds of hours more of supercomputer
simulations. As shown below in Section \ref{az_structure}, the
average density pertubation in the disk has a lifetime of less than a
third of an orbital period. Therefore two different light curves of
the same disk
viewed from $\phi \pm 90^\circ$ should be uncorrelated when seen at high
inclination, where the emission is dominated by special relativistic
beaming from the half of the disk moving towards the
observer. Similarly, for an optically thick disk, an observer looking
at the ``underside'' of the disk ($i\to 180^\circ-i$) might also see
an uncorrelated light curve. We calculated light curves and power
spectra with $\theta=70^\circ,110^\circ$
and $\phi=\pm 90^\circ$, and show the significance plots
in Figures \ref{ploteight}d-f.  The light curves from $\phi = \pm 90^{\circ}$
are indeed uncorrelated.  On the other hand, the top-bottom views appear to be
strongly correlated. From this we infer that the largest hot spot
perturbations extend vertically through the entire disk, and are thus
visible from both sides simultaneously. 

As is clear from Figures \ref{ploteight}d and \ref{ploteight}f, the 2:3
QPO peaks are not always present, while other peaks may appear
at other frequencies.  Although the transience of this result is somewhat
disappointing, it is
actually consistent with the data from {\it RXTE}, which detects the
subtle HFQPOs in only a small fraction of all observations
\citep{mccli05}. On the other
hand, if the simulations most accurately represent the Thermal state
(which is by no means certain, as evidenced by the large discrepancy in
theoretical vs.\ observed power-law slope),
one would not expect to see high-frequency QPOs at all, because they
are not generally observed in the state. Yet another
caveat is that around 230 Hz, for $T_{\rm sim}=200$ msec, $N_{\rm
  bin}\approx 5$, so the distribution of $n_\sigma$ is not quite
normal. Whereas a random variable with a Gaussian distribution would
have $4\sigma$ outliers only $0.006\%$ of the time, the actual
distribution for $N_{\rm bin}=5$ (roughly chi-squared with 10 degrees
of freedom) gives $4\sigma$ results with probability 
$0.2\%$ and $5\sigma$ outliers $0.03\%$ of the time. Thus is it
unlikely, but not impossible, to form a few large ($>4\sigma$) peaks
from purely random fluctuations.

\begin{figure}[tp]
\caption{\label{plotnine} Probability of randomly producing a QPO at a given
  significance $(n_\sigma)$, based on Monte Carlo calculations of a
  power-law power spectrum with fluctuations of individual frequency
  bins defined by an exponential probability distribution. As
  predicted by the analytic theory, we find $4\sigma$ results roughly
  $0.1-0.2\%$ of the time.}
\begin{center}
\scalebox{0.8}{\includegraphics{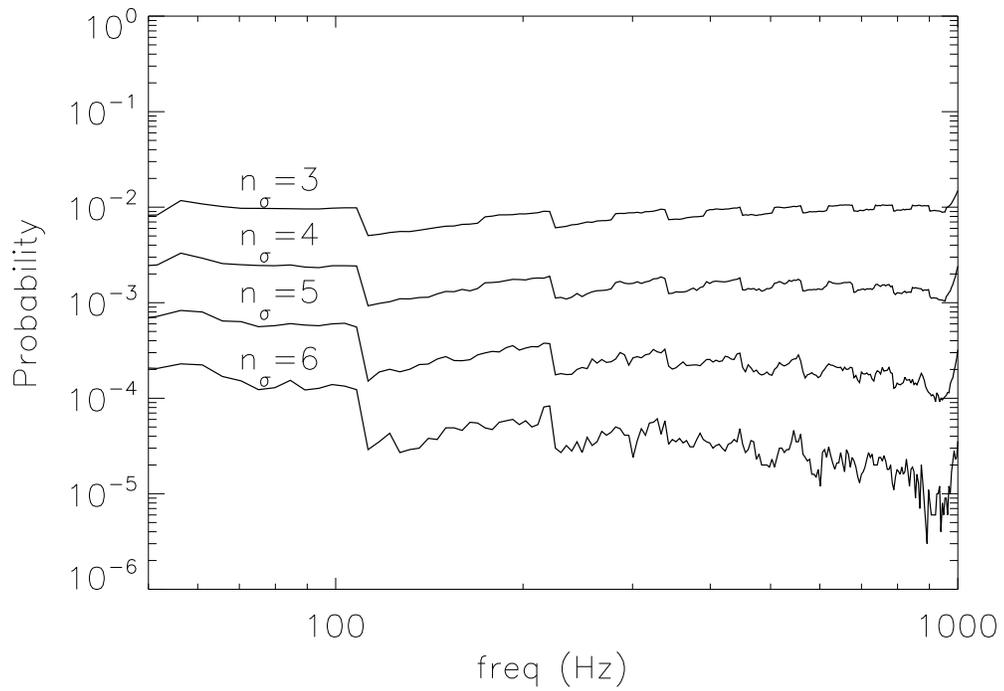}}
\end{center}
\end{figure}

In order to understand better the likelihood of these results, we
performed a Monte Carlo calculation of $10^6$ simulated power-law power
spectra. This is a similar approach to that taken by people
searching for QPOs in X-ray observations of Sgr A$^\ast$ and AGN
\citep{benll01,belan06}. For fixed values of $P_0$ and $\Gamma$, we generate a random
value for the power at each frequency according to the exponential
distribution in equation (\ref{fPnu}). Then we analyze this simulated
power spectrum with the exact same methods described above, generating
a plot of significance versus frequency like the ones shown in Figure
\ref{ploteight}. Repeating this procedure many times gives the
probability of producing a random QPO signal at a given frequency and
significance. For $Q=10$, Figure \ref{plotnine} plots this
probability for $n_\sigma=$ 3, 4, 5, and 6. The results of the Monte
Carlo calculation appear to be very similar to the analytic estimates
presented above, confirming the non-normal nature of the
distribution. The ``steps'' or ``edges'' seen in Figure
\ref{plotnine} are simply due to the finite number of frequency
bins used in each QPO: for a fixed value of $Q$, $N_{\rm bin}$
increases with frequency, and every integer step corresponds to a
small change in the significance, as can be seen from equation
(\ref{n_sigma}). 

In the end, we
are still not quite sure if the 2:3 ratio QPOs are ``real,'' and if so,
whether they represent the same QPO pairs seen in some
observations. Analogous to the observations, the only way really to
improve our confidence in these results may be to run more simulations
and see if the 2:3 peaks are reproducible and if so, with what range of
QPO parameters. With more simulated data, we should also be better
able to identify the exact mechanism that is causing the QPOs. At this
point, it seems somewhat unlikely that they are caused by a simple
hot spot model as in \citet{schni05}, which requires much longer
lifetimes for the quality factors seen here. To understand the
robustness of these QPO pairs, in addition to longer
simulation times, we must also explore the effects of black hole spin,
one of the major ingredients in most HFQPO models. This effort, too, will
require significant computational resources, but should be more
productive than simply running the current Schwarzschild simulation
longer [see eqn.\ (\ref{A_significance2})]. 

\section{AZIMUTHAL STRUCTURE OF ACCRETION DISK}\label{az_structure}

To gain a deeper understanding of the source of the light curve
fluctuations, we must first characterize the behavior of the
hydrodynamic fluctuations in the accretion disk. For all three
emission models, and particularly the optically thin model, the
variations in brightness are closely related to the
surface density fluctuations. Thus, to analyze the structure and
evolution of the hot spots in the disk, we will primarily focus on the
surface density $\Sigma(r,\phi,t)$ defined as
\begin{equation}
\Sigma(r,\phi,t) = \int d\theta \sqrt{g_{\theta\theta}} \rho(r,\theta,\phi,t).
\end{equation}
For each annulus of constant radius and time, we subtract out the mean
and normalize the surface density such that
\begin{subequations}\label{normal_sigma}
\begin{equation}
\int d\phi \bar{\Sigma}(r,\phi,t)=0
\end{equation}
and
\begin{equation}
\int d\phi \bar{\Sigma}^2(r,\phi,t)=1.
\end{equation}
\end{subequations}

To estimate the distribution of hot spot lifetimes, we calculate the
correlation between the surface density maps at two different
times. The time-symmetric overlap function is defined as
\begin{equation}
c(r,t,\Delta t) \equiv \frac{1}{2}\left[\int d\phi \bar{\Sigma}(r,\phi,t)
  \bar{\Sigma}(r^+,\phi^+,t+\Delta t)\right]^{1/2}
+\frac{1}{2}\left[\int d\phi \bar{\Sigma}(r,\phi,t)
  \bar{\Sigma}(r^-,\phi^-,t-\Delta t)\right]^{1/2}
\end{equation}
so that $c(r,t,0)=1$. Here the advanced and retarded coordinates
$(r^+,\phi^+)$ and $(r^-,\phi^-)$ are the spatial positions where a
clump of matter at $(r,\phi,t)$ would be at times $t+\Delta t$ or
$t-\Delta t$. These coordinates are determined by integrating the
time- and azimuth-averaged fluid velocities at each radius in the disk. For
example, 
\begin{subequations}
\begin{equation}
r^+ = r+\int_0^{\Delta t} \bar{v}^r[r(t)] dt
\end{equation}
and
\begin{equation}
\phi^+ = \phi+\int_0^{\Delta t} \bar{v}^\phi[r(t)] dt,
\end{equation}
\end{subequations}
where $\bar{v}^i(r)$ is the average fluid transport
velocity $(dx^i/dt)$ in the disk midplane. 
This more detailed approach is necessary for the inner disk where the
radial velocities are not negligible and the orbital velocities change
rapidly with $r$.

Now the complete correlation function at each radius can be given by
averaging over the duration of the simulation:
\begin{equation}\label{C_rt}
C(r,\Delta t)=\frac{1}{t_f-t_i}\int_{t_i}^{t_f} c(r,t,\Delta t) dt.
\end{equation}
The correlation is normalized so that if there is no evolution
of the disk perturbations in the local fluid frame, then $C(r,\Delta
t)=1$. Of course, in practice there {\it is} significant evolution, and therefore
we expect $C(r,\Delta t\to \infty) \to 0$ as the turbulent disk at
late times is completely uncorrelated with the initial
conditions. Then the characteristic decay time for $C(r,\Delta t)$ is
a measure of the lifetime of a typical hot spot at that radius. 

\begin{figure}[tp]
\caption{\label{plotten} Cross-correlation of surface density
  fluctuations at different radii in the disk, as defined in equation
  (\ref{C_rt}). Outside of the ISCO, 
  the typical density perturbation lasts a characteristic lifetime of
  about a quarter-orbit.}
\begin{center}
\scalebox{0.8}{\includegraphics{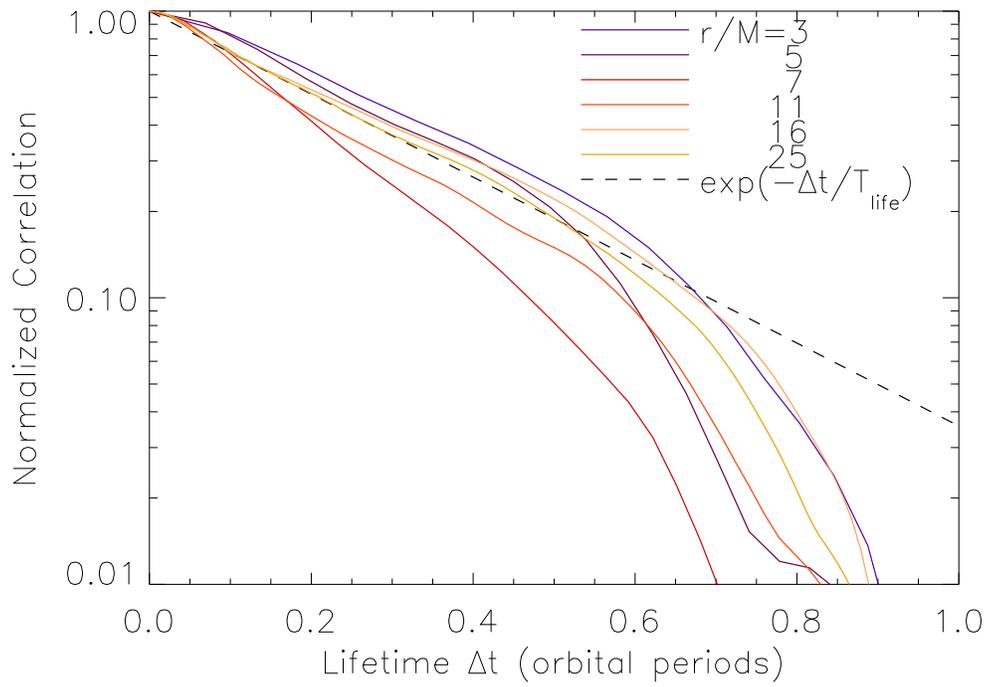}}
\end{center}
\end{figure}

To compare lifetimes at different radii, we rescale the
time delay $\Delta t$ by the orbital period $(2\pi/\Omega)$ at
each radius. These correlation functions are plotted in Figure
\ref{plotten} for a range of $r$, including the region inside the ISCO.
Remarkably, we find very similar behaviors for hot spots
over a large range of radii $(3M \le r \le 25M)$, with nearly exponential
decay at early times with a time constant of 
\begin{equation}
T_{\rm life} \approx 0.3 T_{\rm orb}.
\end{equation}

This behavior is quite similar to the simple hot spot model developed
by \citet{schni05} where hot spots were created and destroyed around a
single radius with random phases and exponentially-distributed
lifetimes, naturally giving Lorentzian peaks in the power spectrum at
the geodesic frequencies corresponding to the chosen ``resonant''
radius. However, for the MHD simulations, there appears to be no
special radius, and even for the potential QPOs discussed above in
Section \ref{hfqpos}, the geodesic hot spot model may not be
appropriate. 

In addition to the typical lifetimes for the density perturbations, we
have also analyzed the characteristic {\it spatial} dimensions of the
hot spots in $r$ and
$\phi$. We would ideally like to follow each
hot spot in its own locally non-rotating frame to remove the shearing
effects of differential rotation in the disk. This would give a better
estimate for the ``true'' size and shape of the local perturbations
without including the global dynamics of the disk. In fact, there is a
simple analytic transformation that can be applied to remove the shear
effects from the entire disk simultaneously (at least the region that behaves in a
roughly Keplerian way; $6M \lesssim r \lesssim 25M$ in our simulation).

To construct this transformation, consider a sheared blob extending in
$r$ from $r_0-dr$ to $r_0+dr$. The orientation of the blob at its creation
can be estimated by shifting each point backwards in time by $T_{\rm
  life}(r_0)$. This amounts to a shift in azimuth of
$\Delta\phi(r_0 \pm dr)=-T_{\rm life}(r_0)\Omega(r_0 \pm dr)$ or
\begin{equation}\label{ddphi_dr}
\frac{d}{dr}\Delta\phi(r) = -T_{\rm life}(r)\frac{d}{dr}\Omega(r)
\approx 0.9\left(\frac{\pi}{r}\right),
\end{equation}
where we have assumed a Keplerian orbital velocity of
$\Omega(r)\sim r^{-3/2}$ (recall this relation holds even for
relativistic orbits in the Schwarzschild metric). Integrating equation
(\ref{ddphi_dr}) gives the azimuthal coordinate transformation
$\varphi(r)=\phi(r)-\Delta\phi$, with 
\begin{equation}
\Delta\phi(r) = 0.9\pi \ln\frac{r}{r_{\rm in}}.
\end{equation}

\begin{figure}[tp]
\caption{\label{ploteleven} Surface density perturbations
  in the accretion disk at a single moment in time. In ($a$), clear shearing
  patterns are seen, caused by the differential rotation of the
  disk. In ($b$), we show the same perturbations, after ``unshearing''
  the disk by translating each annulus of the disk by $\Delta\phi(r) =
  0.9\pi \ln(r/r_{\rm in})$. The contours are spaced linearly and
  cover a range of $0.1-1.0$ times the maximum surface density.}
\begin{center}
\scalebox{0.7}{\includegraphics{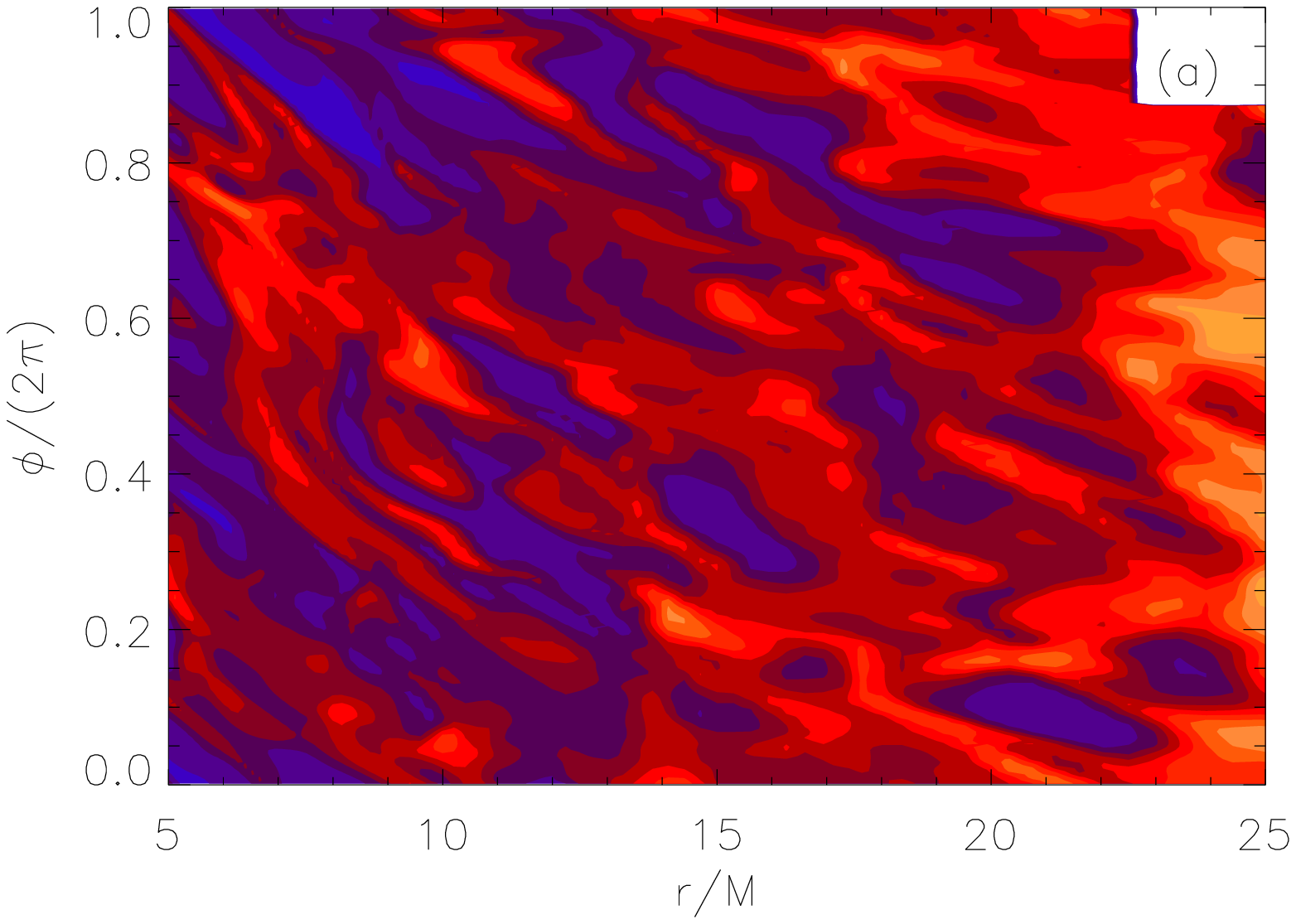}}
\scalebox{0.7}{\includegraphics{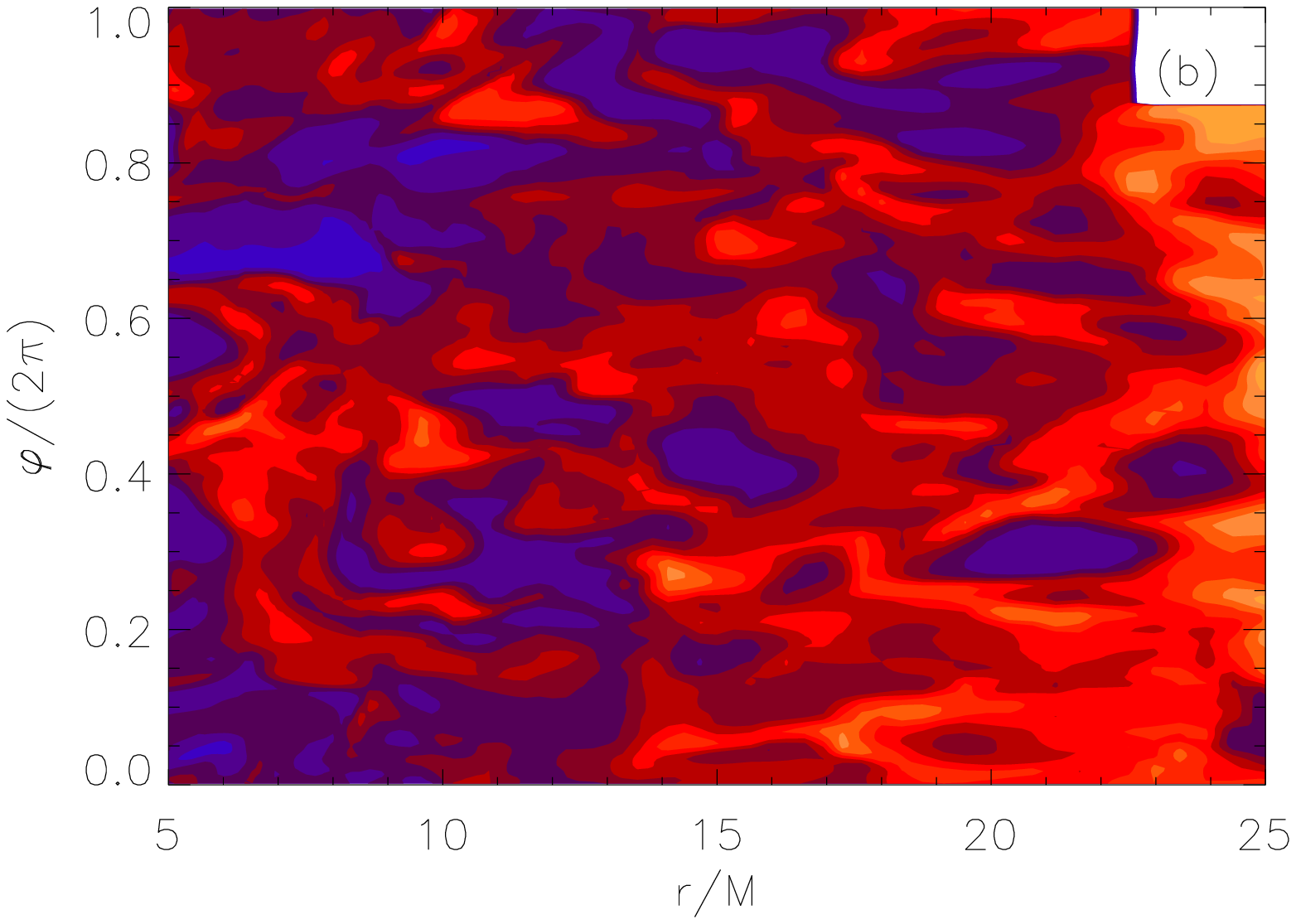}}
\scalebox{0.6}{\includegraphics*[30,300][580,420]{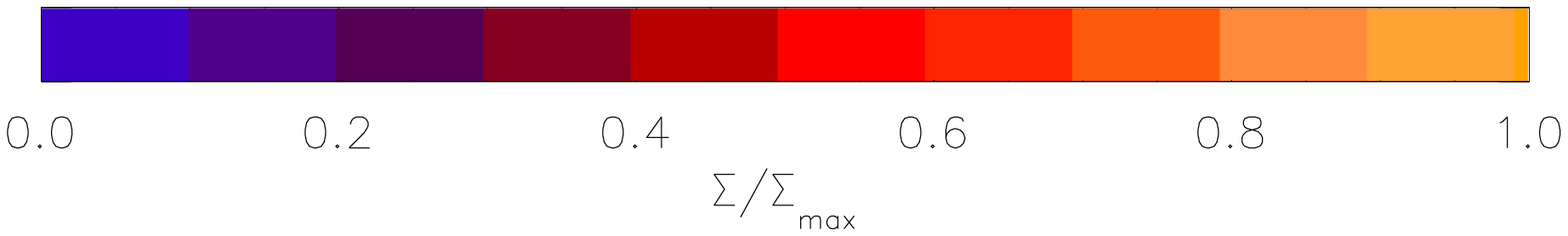}}
\end{center}
\end{figure}

Figure \ref{ploteleven} shows the effect of applying this
transformation to a snapshot of the accretion disk surface density. In
the ``before'' image (Fig.\ \ref{ploteleven}a), the differential rotation is clearly present, as
material in the inner disk gets swept along faster, and shears the
overdensities into diagonal arcs in the $(r,\phi)$ plane. After
performing the transformation defined above, the blobs appear rather
uniform in shape and orientation (Fig.\ \ref{ploteleven}b). Finally, to compare the aspect
ratios of blobs at different radii in the disk, we perform a simple
coordinate transformation in the radial direction as well:
\begin{equation}
\eta(r) \equiv\ln(r/r_{\rm in}).
\end{equation} 
As we will show, these coordinates highlight the invariance of $\delta
r/r$ for
density perturbations throughout the disk.

\begin{figure}[tp]
\caption{\label{plottwelve} Variance in the normalized surface density as a
  function of wavenumber for (a) azimuthal modes and (b) radial
  modes. The characteristic size of a
  density perturbation is determined by the wavenumber of the
  power-law break.}
\begin{center}
\scalebox{0.7}{\includegraphics{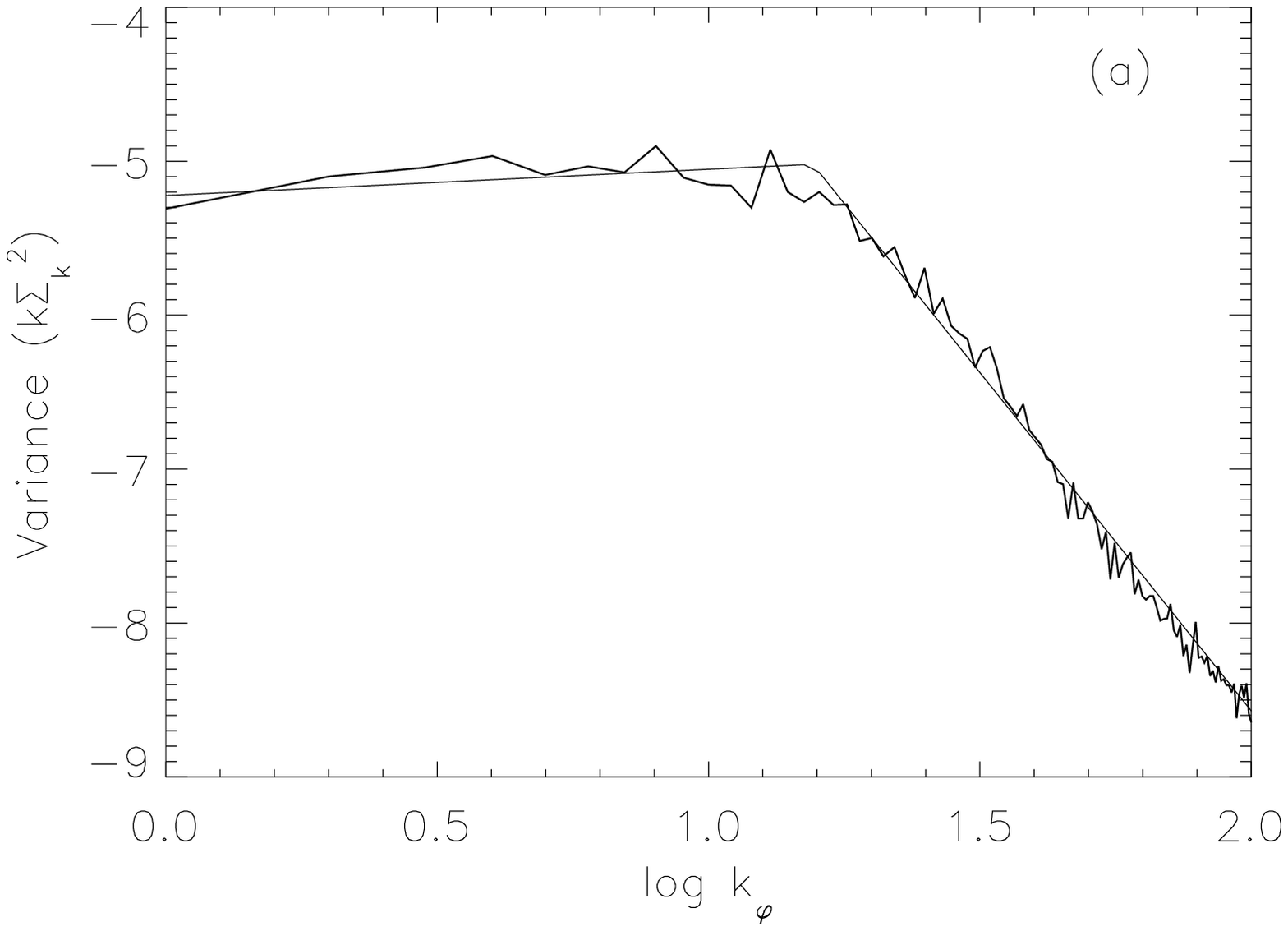}}
\scalebox{0.7}{\includegraphics{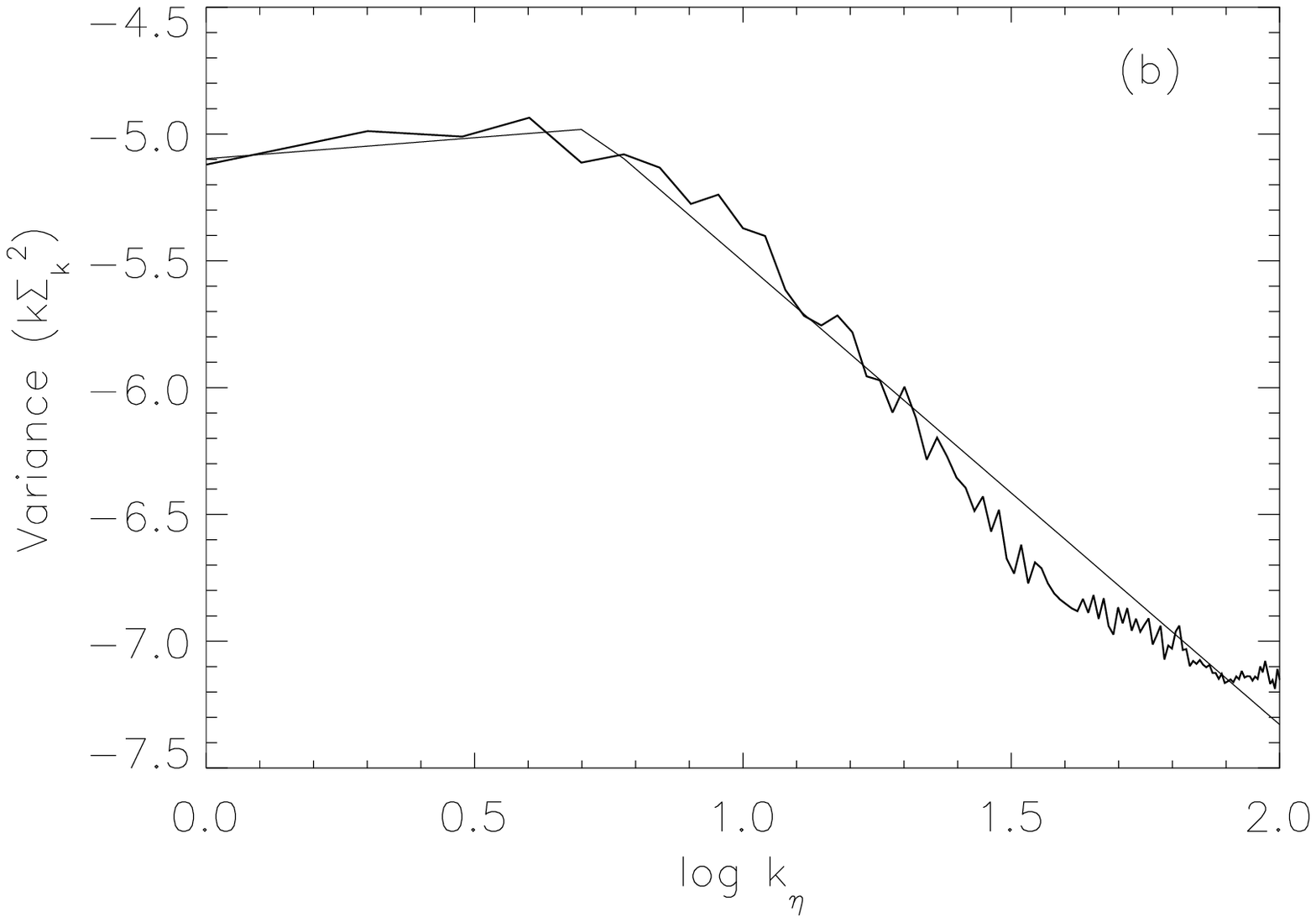}}
\end{center}
\end{figure}

Given a surface density map $\Sigma(\eta,\varphi,t)$ for each frame in
time of the simulation, we normalize to get $\bar{\Sigma}$ as in equation
(\ref{normal_sigma})
for all values of $\eta$ and $t$. Then, for each time step we create two
partial Fourier transforms, one in each spatial direction:
\begin{subequations}
\begin{equation}
\tilde{\Sigma}(k_{\eta},\varphi,t)=\frac{1}{2\pi}\int d\eta
\bar{\Sigma}(\eta,\varphi,t) e^{-ik_{\eta}\eta}
\end{equation}
and
\begin{equation}
\tilde{\Sigma}(\eta,k_{\varphi},t)=\frac{1}{2\pi}\int d\varphi
\bar{\Sigma}(\eta,\varphi,t) e^{-ik_{\varphi}\varphi}.
\end{equation}
\end{subequations}
Averaging (in quadrature) over the duration of the simulation gives
\begin{subequations}
\begin{equation}
\tilde{\Sigma}^2(k_{\eta},\varphi)=\frac{1}{N_t}\sum_{j=1}^{N_t}
\tilde{\Sigma}^2(k_{\eta},\varphi,t_j)
\end{equation}
and
\begin{equation}
\tilde{\Sigma}^2(\eta,k_{\varphi})=\frac{1}{N_t}\sum_{j=1}^{N_t}
\tilde{\Sigma}^2(\eta,k_{\varphi},t_j).
\end{equation}
\end{subequations}
With this construction, $\tilde{\Sigma}^2(k_{\eta},\varphi)$ is the
time-averaged power spectrum of radial variations at a fixed azimuth
$\varphi$. Similarly, $\tilde{\Sigma}^2(\eta,k_{\varphi})$ is the
average azimuthal power for a fixed radius $\eta$. Figure \ref{plottwelve}
shows examples of these power spectra for representative values of
$\eta$ and $\varphi$. It is clear from these plots that the variance
$k\tilde{\Sigma}^2(k)$ can be well approximated by a broken power law
for variations in both the radial and azimuthal directions. The
``break wavenumber'' gives the characteristic maximum size of a density
perturbation. Since we find the peak amplitude of the perturbations
increases with their spatial extent, it is reasonable that the
variance peaks at this break wavenumber. 

For each value of $\varphi$ we fit the power spectrum
$\tilde{\Sigma}^2(k_{\eta},\varphi)$ to a broken power-law in
$k_{\eta}$ and find an average value for $k_{\eta}^{\rm
  break}=5.5$. Applying a similar method to
$\tilde{\Sigma}^2(\eta,k_{\varphi})$ gives an average azimuthal break
wavenumber of $k_{\varphi}^{\rm break}=15$. These values correspond to
wavelengths of $\lambda_{\eta} = 0.26$ (giving
$\delta r/r=e^{\lambda_{\eta}}-1=0.3$) and $\lambda_{\varphi}=25^\circ$, or
an aspect ratio for the hot spots of $a/b=r \delta\varphi/dr \approx 1.5$. 
As we saw above in Section \ref{hfqpos}, the light curves from the top-
and bottom-views of the optically thick disk are highly correlated. This
suggests that the vertical extent of the density perturbations is limited
by the disk thickness, which has a roughly constant value of $\delta z/r
\approx 0.15$ (see Fig.\ \ref{plotone}d). Thus the density 
perturbations throughout the disk are self-similar ellipsoids with
principle axes in the ratio $(r\delta\varphi:\delta r:\delta z)=(3:2:1)$.

\begin{figure}[tp]
\caption{\label{plotthirteen} Best-fit parameters for the broken power-law
  curves plotted in Figure \ref{plottwelve}, where $\tilde{\Sigma}^2(k) \propto
  k^{-\Gamma_1}$ for $k<k^{\rm break}$ and $\tilde{\Sigma}^2(k) \propto
  k^{-\Gamma_2}$ for $k>k^{\rm break}$.}
\begin{center}
\scalebox{0.45}{\includegraphics{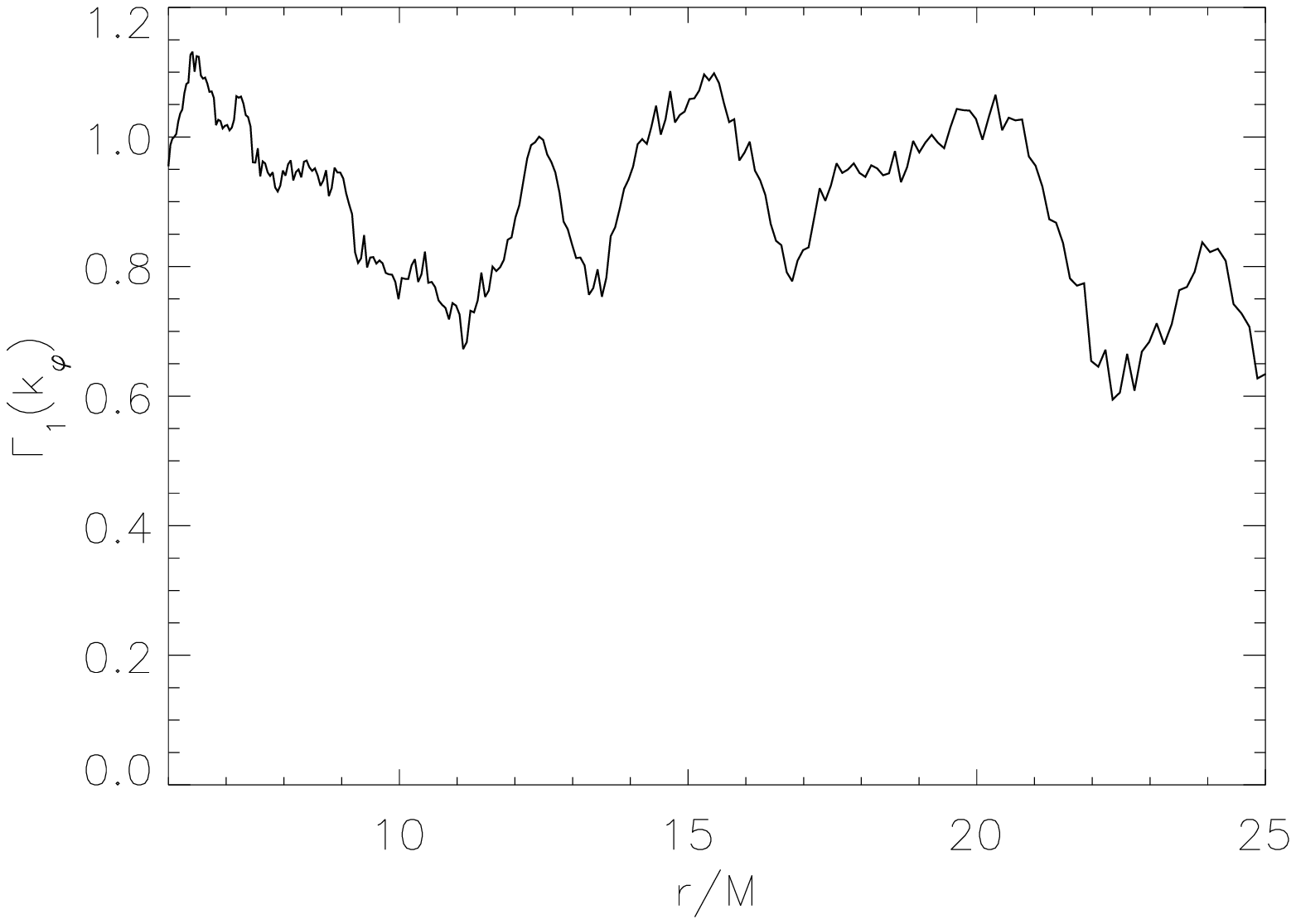}}
\scalebox{0.45}{\includegraphics{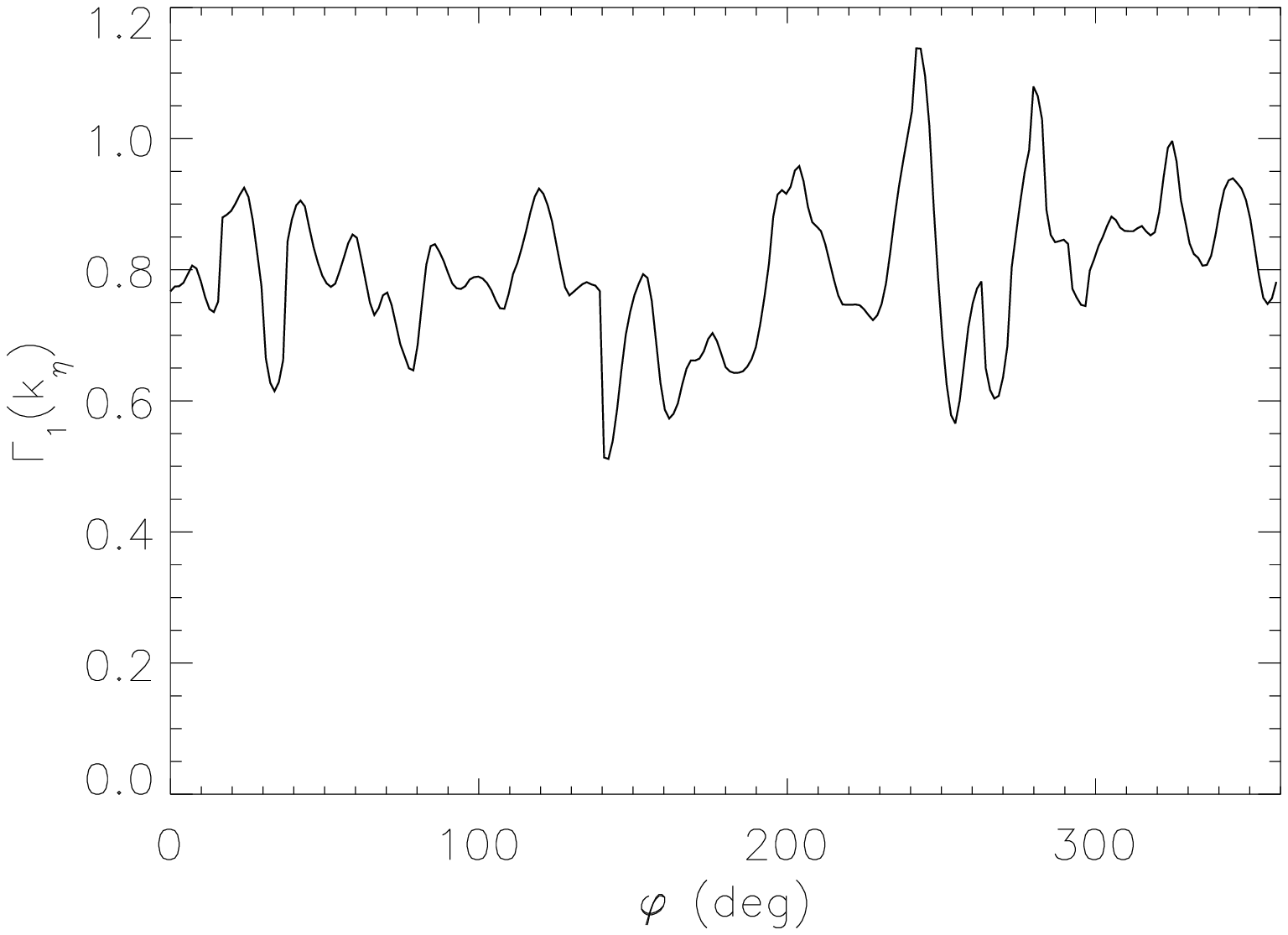}}\\
\scalebox{0.45}{\includegraphics{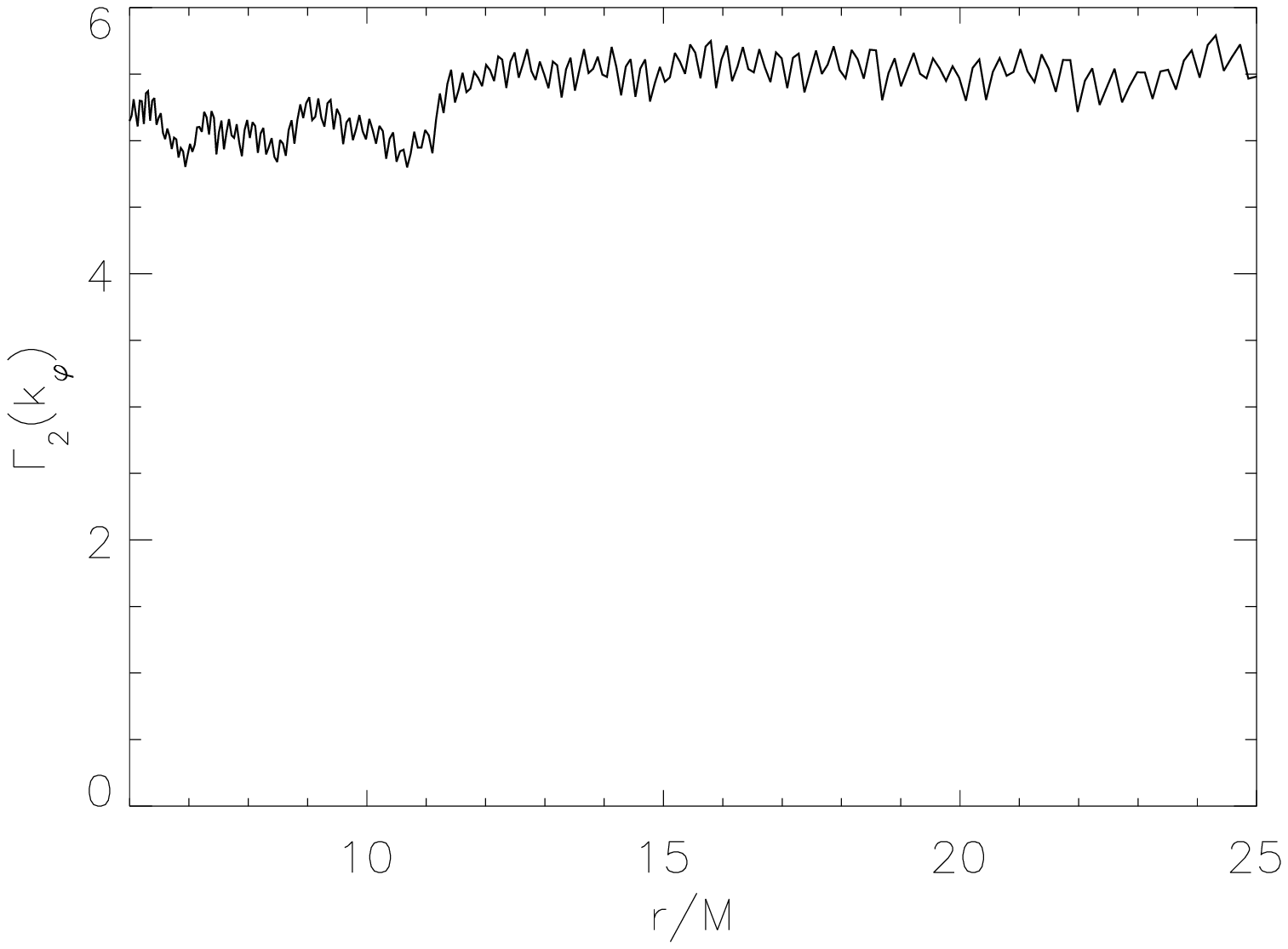}}
\scalebox{0.45}{\includegraphics{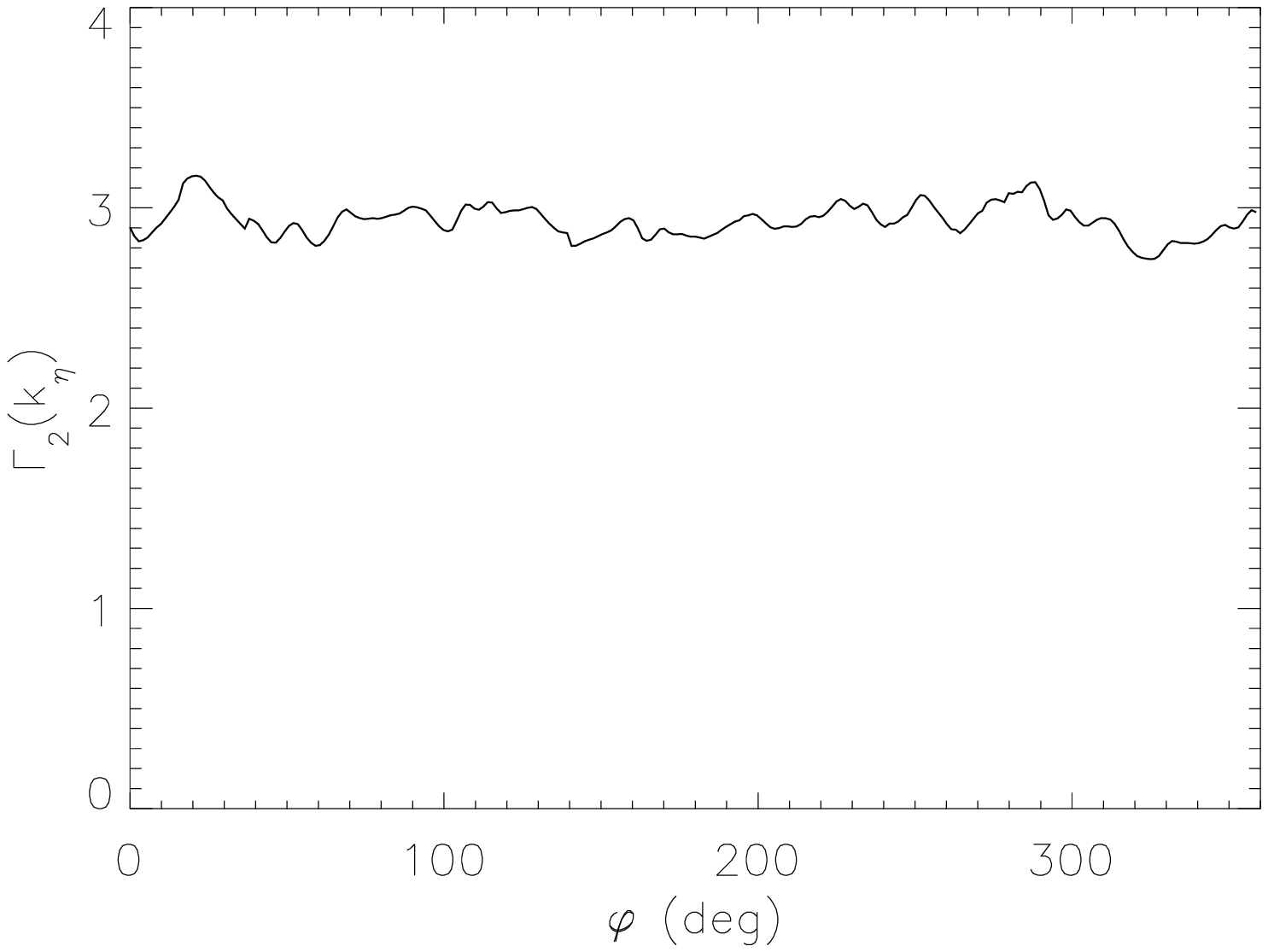}}\\
\scalebox{0.45}{\includegraphics{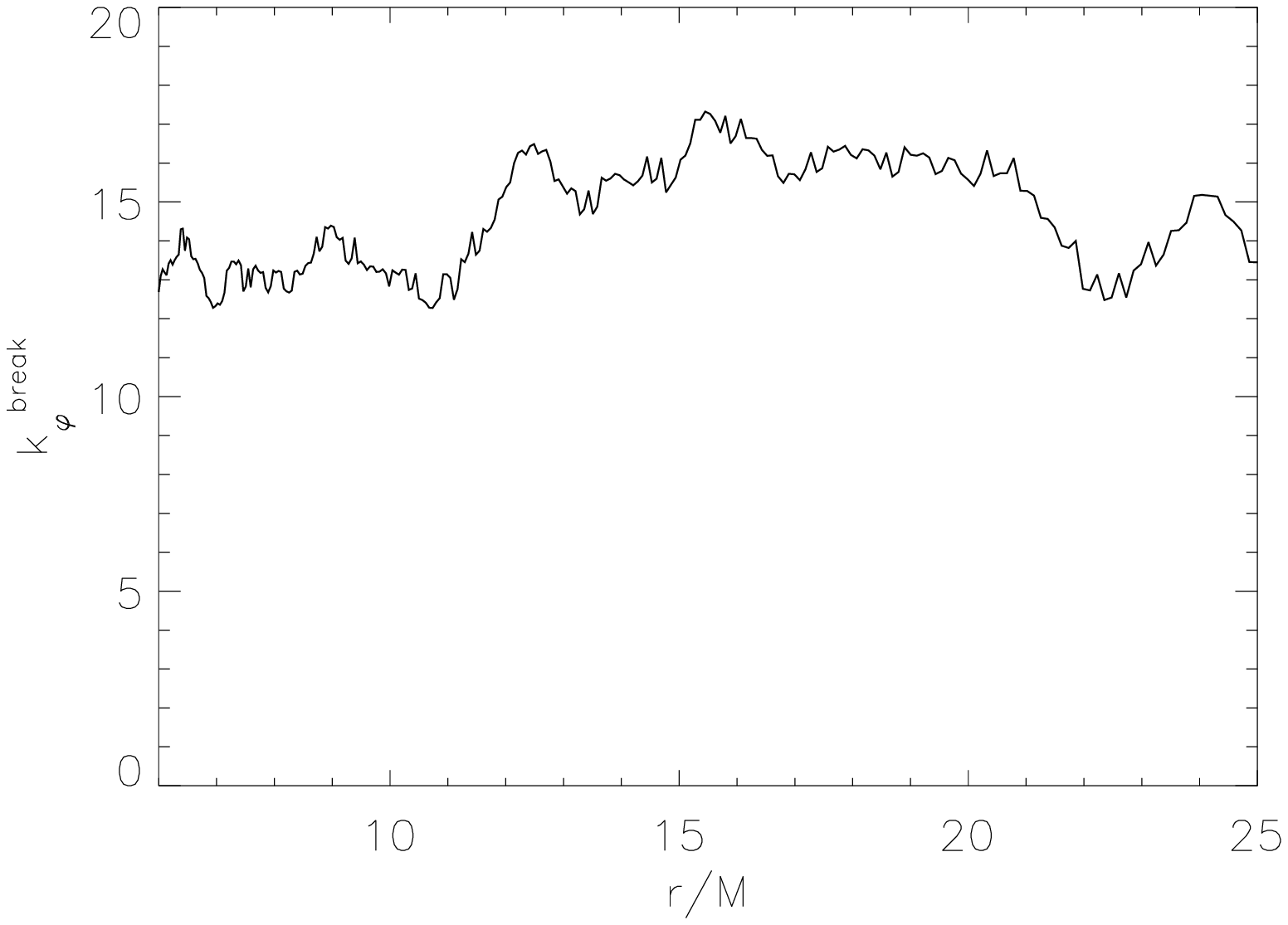}}
\scalebox{0.45}{\includegraphics{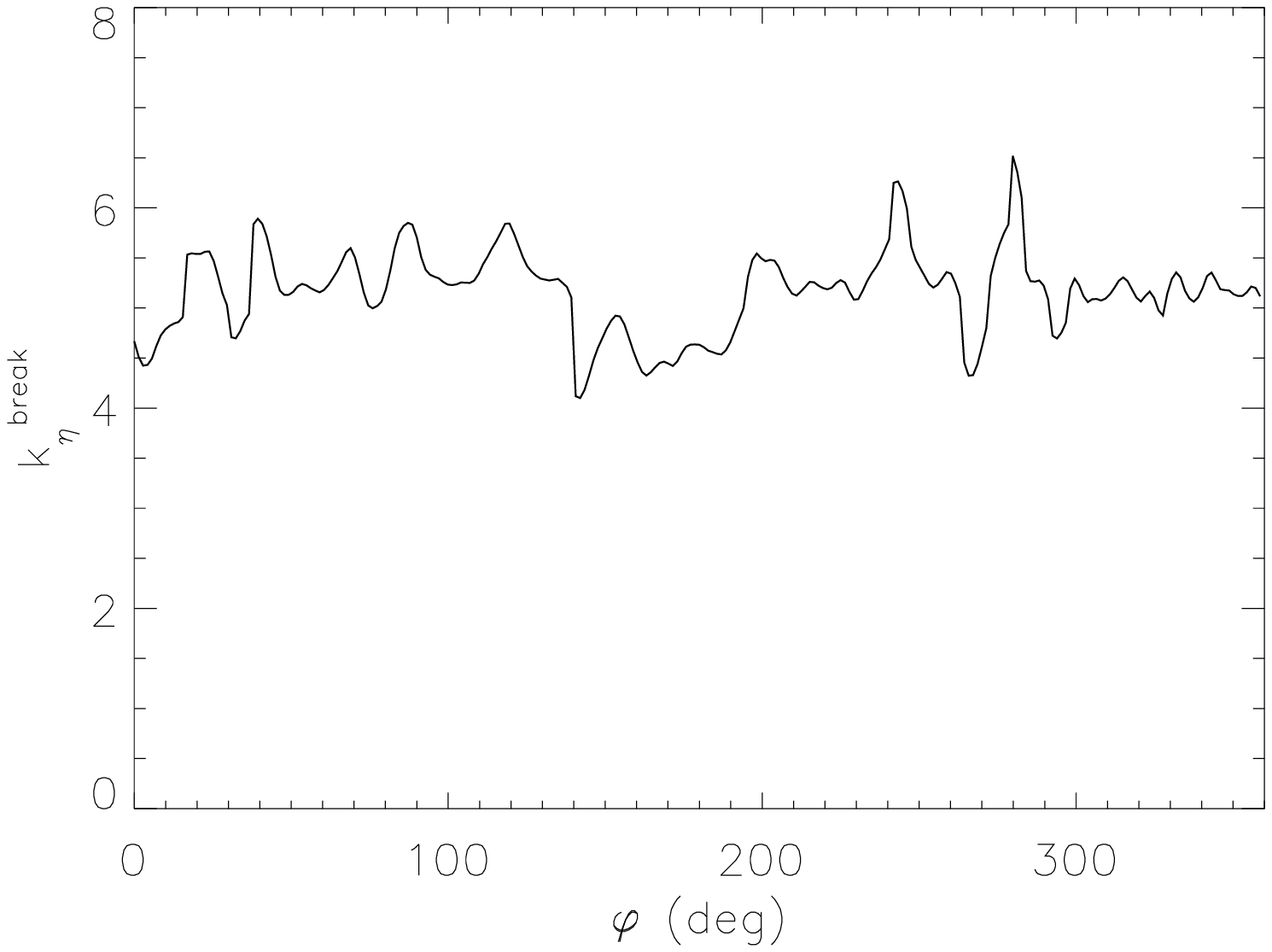}}
\end{center}
\end{figure}

To get a feeling for the stability of this wavenumber analysis, we plot
in Figure \ref{plotthirteen} the best-fit parameters for the broken
power-law at each slice in $r$ or $\varphi$. Because of the rotational
symmetry of the accretion system, it is expected that the fit
parameters should be roughly constant in $\varphi$, so the curves in
Figures \ref{plotthirteen}d,e,f really give an estimate of the typical errors in
our analysis. Somewhat more surprising is the near constancy of
the fit parameters as a function of $\eta(r)$, emphasizing
the fact that density perturbations at all radii have similar extent
in azimuth. Lastly, the fact that the curves in Figure \ref{plotthirteen}
do not appear as purely random fluctuations around the mean highlights
the level of correlation between neighboring points in
radius/azimuth.

\section{DISCUSSION AND CONCLUSIONS}\label{discussion}

\subsection{Summary of Findings}

We have developed a generalized post-processor analysis tool that
couples ray-tracing in the Kerr metric with global GRMHD simulations
to produce light curves and power spectra of accreting black
holes. Using a variety of emission models, we probe different
regions of the accretion disk and are better able to understand the
underlying causes of time variability in the observed flux. The
optically thick blackbody emission/absorption model is particularly
useful for understanding the behavior of stellar-mass black holes in
the Thermal Dominant state, which is characterized by a broad peak in
the photon energy spectrum and very low levels of variability. 

By fixing the black hole mass and accretion rate, we can
convert from dimensionless code units to physical units
for local fluid density. Then, assuming a radiation-pressure dominated
gas in the inner disk, we can derive a physical temperature, and thus
emissivity and absorption coefficients. The temperature and scale
height of the disk compare 
well with the Novikov-Thorne predictions for a Schwarzschild black
hole accreting at 50\% $\dot{M}_{\rm Edd}$ with a small torque at the
inner boundary. Despite the very different assumptions used in the
analytic model and the computer simulations, the
agreement is close enough to provide reasonable confidence in our
conversion factors for density and temperature, and thus the
conclusion that the simulations can be used to understand the Thermal
Dominant state. 

We have also developed new methods for analyzing the azimuthal
structure of the accretion flow, determining the characteristic sizes
and lifetimes of density perturbations. Over a large range of radii,
we found the perturbations have a nearly exponential distribution of
lifetimes, with $T_{\rm life}\approx 0.3T_{\rm orb}$. Also, the
characteristic shapes of the hot spots appears to be self-similar
throughout the Keplerian regions of the disk, with $\delta\phi \approx
25^\circ$ and $\delta r/r \approx 0.3$. 

From these short coherence times, it seems clear that the hot spots
formed by MHD turbulence 
cannot survive long enough to produce the QPOs with quality factors of
$Q\sim 5-10$ described in \citet{schni05}. Thus, if the transient high
frequency QPOs with 2:3 frequency ratios identified in Section \ref{hfqpos}
are in fact robust signatures of
GRMHD disk dynamics, they are most likely {\it not} produced by
geodesic hot spots.  This conclusion is supported by the fact that the
QPOs are seen only from a single azimuthal viewing angle at any one
time, perhaps suggesting some form of localized stationary wave in the disk.
Finally, as part of our search for QPOs in the simulation data, we
have also developed a general formalism for quantifying the
significance of such features in simulations. These analytic results
have been combined with Monte Carlo calculations to estimate our
confidence limits for the QPOs at $\gtrsim 99.9\%$. 

\subsection{Comparison with Observations}
As argued throughout this paper, we believe the MHD simulations
most closely resemble the Thermal state of black hole X-ray
binaries. This high-luminosity state is dominated by a broad thermal
peak in the
energy spectrum around $E \sim 0.7-1.5$ keV and a relatively small
contribution from a steep power-law tail at higher energies
\citep{mccli05}. The timing
properties are characterized by a featureless power spectrum with
$P(\nu) \sim \nu^{-\Gamma}$, with typical power-law index $\Gamma \approx
1$. The total power in this state is also small, with (rms/mean)$^2$
Hz$^{-1} \lesssim 10^{-3}$ above 1~Hz.  Our simulated power spectra
predict $\Gamma \approx 3-4$, which is rather greater than the
typical slope.  On the other hand, just this sort of steep power
spectrum has been seen in the hard state observations of \citet{bello06}.

%However, there remain some quantitative differences between our
%simulated power spectra and the data, namely our larger values for
%$\Gamma \approx 3-4$. While this actually agrees closely with the
%results of \citet{bello06}, those observations were made during a
%hard state, and also showed significant high frequency QPOs. 

One robust prediction of the MHD/ray-tracing simulations is that,
independent of the emission mechanism, the integrated rms power
increases with disk inclination. In principle, this should be an
observable prediction that could be tested with {\it RXTE} data from
black hole binaries in the Thermal state. Of course, there are
complications when comparing two different binary systems, including
different relative contributions to the thermal peak or power-law
tail, different total luminosities, and different black hole
masses. However, with a sufficiently large number of observations,
it should be possible to account for these variables and extract a
correlation between integrated spectral power and inclination.

Other observational trends that might be explored with future MHD
simulations include the observed linear relation between the X-ray
flux and rms, which appears to extend to AGN as well
\citep{uttle04,uttle05}. The timing properties of black hole binaries
also vary greatly between their different spectral states, a
general feature that is still not understood.
To explore either of these
relationships, we will require more sophisticated emission models and,
quite likely, more simulation data with a broader range of disk
parameters. 

However, even with improved simulations and light curve models, there
still exist some inherent difficulties in comparing theory with
observation. {\it RXTE} typically requires thousands of seconds of
observation to detect high-frequency QPOs at rms amplitudes of a few
percent (limited largely by photon counting statistics and ``dead time''
corrections). To achieve a similar sensitivity, our MHD simulations
require hundreds of hours corresponds to generate only 0.2 seconds of
real time for a $10 M_\odot$ black hole (we have perfect photon
statistics, but are limited by the inherent
variance in the power spectra of short time series). 
This is more closely analogous to the time scales of AGN
observations. If we were to scale the simulation to an
AGN of mass $10^8 M_\odot$, it would correspond to a sampling rate of
every 40 minutes for roughly a month, assuming {\it perfect} photon
counting statistics!
Thus it should be no surprise that QPOs are still so difficult to detect
with high significance from supermassive black holes \citep{benll01,vaugh05}. 

\acknowledgements{JDS acknowledges helpful discussions with Chris
  Reynolds and Tod Strohmayer. This work was supported by NSF grant
  PHY-0205155 and NASA grant NNG04GK77G (JFH),
  and by NSF grants AST-0205806 and AST-0313031
  (JHK).  The simulation described here was carried out on the
  DataStar system at SDSC.}

\newpage

\begin{table}
\caption{\label{table_power} RMS power for different emission models
  and inclination angles}
\begin{tabular}{lcccc}
\hline
\hline
Emission Model & Inclination & Power $>10$ Hz & Power $>100$ Hz &
$\Gamma(>50\mbox{ Hz})$ \\
               & (degrees)   & (\% rms)           & (\% rms) & \\
\hline
thin line & 0  & 5.74 & 0.22 & 3.8 \\
          & 45 & 6.96 & 0.36 & 4.5 \\
          & 70 & 7.79 & 0.54 & 4.1 \\
thick line& 0  & 4.94 & 0.78 & 3.1 \\
          & 45 & 4.51 & 1.33 & 3.2 \\
          & 70 & 5.89 & 1.39 & 2.7 \\
thermal   & 0  & 16.2 & 1.54 & 3.0 \\
          & 45 & 13.6 & 2.22 & 3.3 \\
          & 70 & 14.0 & 2.87 & 3.3 \\
\end{tabular}
\end{table}

\begin{table}
\caption{\label{annuli_power} RMS power from different annuli of the
  accretion disk with $i=0^\circ$ for the thermal emission model}
\begin{tabular}{rcccc}
\hline
\hline
radii & Power $>10$ Hz & Power $>100$ Hz & $\Gamma(10-100\mbox{ Hz})$ &
$\Gamma(100-1000\mbox{ Hz})$ \\
$(r/M)$   & (\% rms) & (\% rms) & & \\
\hline 
2-6       & 83.2 & 26.2 & 1.8 & 2.9 \\
6-10      & 43.8 &  6.6 & 2.7 & 3.6 \\
10-15     & 22.4 &  2.8 & 2.8 & 3.9 \\
15-20     & 18.3 &  1.6 & 2.4 & 4.4 \\
20-25     & 12.7 &  1.0 & 2.6 & 4.5 \\
\end{tabular}
\end{table}

\end{document}